\documentclass[traditabstract, longauth]{aa}

\usepackage[nonamebreak]{natbib}
\usepackage[stable]{footmisc}
\usepackage{amsmath}
\usepackage{amssymb}
\usepackage{txfonts}
\usepackage{placeins}
\bibpunct{(}{)}{;}{a}{}{,} %
\usepackage{graphicx}
\usepackage{epstopdf}
\usepackage{ifthen}
\usepackage[breaklinks, colorlinks, citecolor=blue]{hyperref}
\usepackage{overpic}
\usepackage{rotating}
\usepackage[table,usenames,dvipsnames]{xcolor}

\usepackage{longtable,lscape}
\usepackage{units}
\usepackage{mathrsfs}
\usepackage{mathbbol}

\def\setsymbol#1#2{\expandafter\def\csname #1\endcsname{#2}}
\def\getsymbol#1{\csname #1\endcsname}

\def\Planck{\textit{Planck}}

\newbox\tablebox    \newdimen\tablewidth
\def\leaderfil{\leaders\hbox to 5pt{\hss.\hss}\hfil}
\def\endPlancktablewide{\tablewidth=\textwidth 
    $$\hss\copy\tablebox\hss$$
    \vskip-\lastskip\vskip -2pt}
\def\tablenote#1 #2\par{\begingroup \parindent=0.8em
    \abovedisplayshortskip=0pt\belowdisplayshortskip=0pt
    \noindent
    $$\hss\vbox{\hsize\tablewidth \hangindent=\parindent \hangafter=1 \noindent
    \hbox to \parindent{$^#1$\hss}\strut#2\strut\par}\hss$$
    \endgroup}
\def\doubleline{\vskip 3pt\hrule \vskip 1.5pt \hrule \vskip 5pt}

\def\L2{\ifmmode L_2\else $L_2$\fi}

\def\DeltaT{\ifmmode \Delta T\else $\Delta T$\fi}
\def\deltat{\ifmmode \Delta t\else $\Delta t$\fi}
\def\fknee{\ifmmode f_{\rm knee}\else $f_{\rm knee}$\fi}
\def\Fmax{\ifmmode F_{\rm max}\else $F_{\rm max}$\fi}
\def\solar{\ifmmode{\rm M}_{\mathord\odot}\else${\rm M}_{\mathord\odot}$\fi}
\def\Msolar{\ifmmode{\rm M}_{\mathord\odot}\else${\rm M}_{\mathord\odot}$\fi}
\def\Lsolar{\ifmmode{\rm L}_{\mathord\odot}\else${\rm L}_{\mathord\odot}$\fi}
\def\inv{\ifmmode^{-1}\else$^{-1}$\fi}
\def\mo{\ifmmode^{-1}\else$^{-1}$\fi}
\def\sup#1{\ifmmode ^{\rm #1}\else $^{\rm #1}$\fi}
\def\expo#1{\ifmmode \times 10^{#1}\else $\times 10^{#1}$\fi}
\def\,{\thinspace}
\def\lsim{\mathrel{\raise .4ex\hbox{\rlap{$<$}\lower 1.2ex\hbox{$\sim$}}}}
\def\gsim{\mathrel{\raise .4ex\hbox{\rlap{$>$}\lower 1.2ex\hbox{$\sim$}}}}

\def\simprop{\mathrel{\raise .4ex\hbox{\rlap{$\propto$}\lower 1.2ex\hbox{$\sim$}}}}
\def\deg{\ifmmode^\circ\else$^\circ$\fi}
\def\pdeg{\ifmmode $\setbox0=\hbox{$^{\circ}$}\rlap{\hskip.11\wd0 .}$^{\circ}
          \else \setbox0=\hbox{$^{\circ}$}\rlap{\hskip.11\wd0 .}$^{\circ}$\fi}
\def\arcs{\ifmmode {^{\scriptstyle\prime\prime}}
          \else $^{\scriptstyle\prime\prime}$\fi}
\def\arcm{\ifmmode {^{\scriptstyle\prime}}
          \else $^{\scriptstyle\prime}$\fi}
\newdimen\sa  \newdimen\sb
\def\parcs{\sa=.07em \sb=.03em
     \ifmmode \hbox{\rlap{.}}^{\scriptstyle\prime\kern -\sb\prime}\hbox{\kern -\sa}
     \else \rlap{.}$^{\scriptstyle\prime\kern -\sb\prime}$\kern -\sa\fi}
\def\parcm{\sa=.08em \sb=.03em
     \ifmmode \hbox{\rlap{.}\kern\sa}^{\scriptstyle\prime}\hbox{\kern-\sb}
     \else \rlap{.}\kern\sa$^{\scriptstyle\prime}$\kern-\sb\fi}
\def\ra[#1 #2 #3.#4]{#1\sup{h}#2\sup{m}#3\sup{s}\llap.#4}
\def\dec[#1 #2 #3.#4]{#1\deg#2\arcm#3\arcs\llap.#4}
\def\deco[#1 #2 #3]{#1\deg#2\arcm#3\arcs}
\def\rra[#1 #2]{#1\sup{h}#2\sup{m}}

\def\dots{\relax\ifmmode \ldots\else $\ldots$\fi}
\def\WHzsr{\ifmmode $W\,Hz\mo\,sr\mo$\else W\,Hz\mo\,sr\mo\fi}
\def\mHz{\ifmmode $\,mHz$\else \,mHz\fi}
\def\GHz{\ifmmode $\,GHz$\else \,GHz\fi}
\def\mKs{\ifmmode $\,mK\,s$^{1/2}\else \,mK\,s$^{1/2}$\fi}
\def\muKs{\ifmmode \,\mu$K\,s$^{1/2}\else \,$\mu$K\,s$^{1/2}$\fi}
\def\muKRJs{\ifmmode \,\mu$K$_{\rm RJ}$\,s$^{1/2}\else \,$\mu$K$_{\rm RJ}$\,s$^{1/2}$\fi}
\def\muKHz{\ifmmode \,\mu$K\,Hz$^{-1/2}\else \,$\mu$K\,Hz$^{-1/2}$\fi}
\def\MJysr{\ifmmode \,$MJy\,sr\mo$\else \,MJy\,sr\mo\fi}
\def\MJysrmK{\ifmmode \,$MJy\,sr\mo$\,mK$_{\rm CMB}\mo\else \,MJy\,sr\mo\,mK$_{\rm CMB}\mo$\fi}
\def\microns{\ifmmode \,\mu$m$\else \,$\mu$m\fi}

\def\muK{\ifmmode \,\mu$K$\else \,$\mu$\hbox{K}\fi}
\def\microK{\ifmmode \,\mu$K$\else \,$\mu$\hbox{K}\fi}
\def\muW{\ifmmode \,\mu$W$\else \,$\mu$\hbox{W}\fi}
\def\kms{\ifmmode $\,km\,s$^{-1}\else \,km\,s$^{-1}$\fi}
\def\kmsMpc{\ifmmode $\,\kms\,Mpc\mo$\else \,\kms\,Mpc\mo\fi}
\providecommand{\sorthelp}[1]{}

\graphicspath{{./Figures/}}

\newcommand{\MUV}{M_{\rm UV}}
\newcommand{\WMAP}{WMAP}
\newcommand{\planckTT}{PlanckTT}
\newcommand{\lollipop}{{\tt lollipop}}
\newcommand{\VHL}{VHL}
\newcommand{\lowl}{\mbox{low-$\ell$}}
\newcommand{\highl}{\mbox{high-$\ell$}}
\newcommand{\lcdm}{$\Lambda$CDM}
\newcommand{\zre}{\ensuremath{z_\mathrm{re}}}

\newbox\tablebox    \newdimen\tablewidth
\def\leaderfil{\leaders\hbox to 5pt{\hss.\hss}\hfil}
\def\endPlancktablewide{\tablewidth=\textwidth 
    $$\hss\copy\tablebox\hss$$
    \vskip-\lastskip\vskip -2pt}
\def\tablenote#1 #2\par{\begingroup \parindent=0.8em
    \abovedisplayshortskip=0pt\belowdisplayshortskip=0pt
    \noindent
    $$\hss\vbox{\hsize\tablewidth \hangindent=\parindent \hangafter=1 \noindent
    \hbox to \parindent{$^#1$\hss}\strut#2\strut\par}\hss$$
    \endgroup}
\def\doubleline{\vskip 3pt\hrule \vskip 1.5pt \hrule \vskip 5pt}

\def\thetau {0.058 \pm 0.012}  

\begin{document}

\title{\textit{Planck} intermediate results. XLVII.\\ Planck constraints on reionization history}

\author{
\author{\small
Planck Collaboration: R.~Adam\inst{67}
\and
N.~Aghanim\inst{53}
\and
M.~Ashdown\inst{63, 7}
\and
J.~Aumont\inst{53}
\and
C.~Baccigalupi\inst{75}
\and
M.~Ballardini\inst{29, 45, 48}
\and
A.~J.~Banday\inst{85, 10}
\and
R.~B.~Barreiro\inst{58}
\and
N.~Bartolo\inst{28, 59}
\and
S.~Basak\inst{75}
% ...
}
\institute{\small
APC, AstroParticule et Cosmologie, Universit\'{e} Paris Diderot, CNRS/IN2P3, CEA/lrfu, Observatoire de Paris, Sorbonne Paris Cit\'{e}, 10, rue Alice Domon et L\'{e}onie Duquet, 75205 Paris Cedex 13, France\goodbreak
\and
Aalto University Mets\"{a}hovi Radio Observatory and Dept of Radio Science and Engineering, P.O. Box 13000, FI-00076 AALTO, Finland\goodbreak
\and
African Institute for Mathematical Sciences, 6-8 Melrose Road, Muizenberg, Cape Town, South Africa\goodbreak
\and
Agenzia Spaziale Italiana Science Data Center, Via del Politecnico snc, 00133, Roma, Italy\goodbreak
\and
Aix Marseille Universit\'{e}, CNRS, LAM (Laboratoire d'Astrophysique de Marseille) UMR 7326, 13388, Marseille, France\goodbreak
\and
Aix Marseille Universit\'{e}, Centre de Physique Th\'{e}orique, 163 Avenue de Luminy, 13288, Marseille, France\goodbreak
\and
Astrophysics Group, Cavendish Laboratory, University of Cambridge, J J Thomson Avenue, Cambridge CB3 0HE, U.K.\goodbreak
\and
Astrophysics \& Cosmology Research Unit, School of Mathematics, Statistics \& Computer Science, University of KwaZulu-Natal, Westville Campus, Private Bag X54001, Durban 4000, South Africa\goodbreak
\and
CITA, University of Toronto, 60 St. George St., Toronto, ON M5S 3H8, Canada\goodbreak
\and
CNRS, IRAP, 9 Av. colonel Roche, BP 44346, F-31028 Toulouse cedex 4, France\goodbreak
\and
California Institute of Technology, Pasadena, California, U.S.A.\goodbreak
\and
Computational Cosmology Center, Lawrence Berkeley National Laboratory, Berkeley, California, U.S.A.\goodbreak
\and
DTU Space, National Space Institute, Technical University of Denmark, Elektrovej 327, DK-2800 Kgs. Lyngby, Denmark\goodbreak
\and
D\'{e}partement de Physique Th\'{e}orique, Universit\'{e} de Gen\`{e}ve, 24, Quai E. Ansermet,1211 Gen\`{e}ve 4, Switzerland\goodbreak
\and
Departamento de Astrof\'{i}sica, Universidad de La Laguna (ULL), E-38206 La Laguna, Tenerife, Spain\goodbreak
\and
Departamento de F\'{\i}sica, Universidad de Oviedo, Avda. Calvo Sotelo s/n, Oviedo, Spain\goodbreak
\and
Department of Astrophysics/IMAPP, Radboud University Nijmegen, P.O. Box 9010, 6500 GL Nijmegen, The Netherlands\goodbreak
\and
Department of Physics \& Astronomy, University of British Columbia, 6224 Agricultural Road, Vancouver, British Columbia, Canada\goodbreak
\and
Department of Physics and Astronomy, Dana and David Dornsife College of Letter, Arts and Sciences, University of Southern California, Los Angeles, CA 90089, U.S.A.\goodbreak
\and
Department of Physics and Astronomy, University College London, London WC1E 6BT, U.K.\goodbreak
\and
Department of Physics and Astronomy, University of Sussex, Brighton BN1 9QH, U.K.\goodbreak
\and
Department of Physics, Gustaf H\"{a}llstr\"{o}min katu 2a, University of Helsinki, Helsinki, Finland\goodbreak
\and
Department of Physics, Princeton University, Princeton, New Jersey, U.S.A.\goodbreak
\and
Department of Physics, University of California, Berkeley, California, U.S.A.\goodbreak
\and
Department of Physics, University of California, One Shields Avenue, Davis, California, U.S.A.\goodbreak
\and
Department of Physics, University of California, Santa Barbara, California, U.S.A.\goodbreak
\and
Department of Physics, University of Illinois at Urbana-Champaign, 1110 West Green Street, Urbana, Illinois, U.S.A.\goodbreak
\and
Dipartimento di Fisica e Astronomia G. Galilei, Universit\`{a} degli Studi di Padova, via Marzolo 8, 35131 Padova, Italy\goodbreak
\and
Dipartimento di Fisica e Astronomia, Alma Mater Studiorum, Universit\`{a} degli Studi di Bologna, Viale Berti Pichat 6/2, I-40127, Bologna, Italy\goodbreak
\and
Dipartimento di Fisica e Scienze della Terra, Universit\`{a} di Ferrara, Via Saragat 1, 44122 Ferrara, Italy\goodbreak
\and
Dipartimento di Fisica, Universit\`{a} La Sapienza, P. le A. Moro 2, Roma, Italy\goodbreak
\and
Dipartimento di Fisica, Universit\`{a} degli Studi di Milano, Via Celoria, 16, Milano, Italy\goodbreak
\and
Dipartimento di Fisica, Universit\`{a} di Roma Tor Vergata, Via della Ricerca Scientifica, 1, Roma, Italy\goodbreak
\and
Dipartimento di Matematica, Universit\`{a} di Roma Tor Vergata, Via della Ricerca Scientifica, 1, Roma, Italy\goodbreak
\and
Discovery Center, Niels Bohr Institute, Copenhagen University, Blegdamsvej 17, Copenhagen, Denmark\goodbreak
\and
European Space Agency, ESAC, Planck Science Office, Camino bajo del Castillo, s/n, Urbanizaci\'{o}n Villafranca del Castillo, Villanueva de la Ca\~{n}ada, Madrid, Spain\goodbreak
\and
European Space Agency, ESTEC, Keplerlaan 1, 2201 AZ Noordwijk, The Netherlands\goodbreak
\and
Gran Sasso Science Institute, INFN, viale F. Crispi 7, 67100 L'Aquila, Italy\goodbreak
\and
HGSFP and University of Heidelberg, Theoretical Physics Department, Philosophenweg 16, 69120, Heidelberg, Germany\goodbreak
\and
Haverford College Astronomy Department, 370 Lancaster Avenue, Haverford, Pennsylvania, U.S.A.\goodbreak
\and
Helsinki Institute of Physics, Gustaf H\"{a}llstr\"{o}min katu 2, University of Helsinki, Helsinki, Finland\goodbreak
\and
INAF - Osservatorio Astronomico di Padova, Vicolo dell'Osservatorio 5, Padova, Italy\goodbreak
\and
INAF - Osservatorio Astronomico di Roma, via di Frascati 33, Monte Porzio Catone, Italy\goodbreak
\and
INAF - Osservatorio Astronomico di Trieste, Via G.B. Tiepolo 11, Trieste, Italy\goodbreak
\and
INAF/IASF Bologna, Via Gobetti 101, Bologna, Italy\goodbreak
\and
INAF/IASF Milano, Via E. Bassini 15, Milano, Italy\goodbreak
\and
INFN - CNAF, viale Berti Pichat 6/2, 40127 Bologna, Italy\goodbreak
\and
INFN, Sezione di Bologna, viale Berti Pichat 6/2, 40127 Bologna, Italy\goodbreak
\and
INFN, Sezione di Ferrara, Via Saragat 1, 44122 Ferrara, Italy\goodbreak
\and
INFN, Sezione di Roma 1, Universit\`{a} di Roma Sapienza, Piazzale Aldo Moro 2, 00185, Roma, Italy\goodbreak
\and
INFN, Sezione di Roma 2, Universit\`{a} di Roma Tor Vergata, Via della Ricerca Scientifica, 1, Roma, Italy\goodbreak
\and
Imperial College London, Astrophysics group, Blackett Laboratory, Prince Consort Road, London, SW7 2AZ, U.K.\goodbreak
\and
Institut d'Astrophysique Spatiale, CNRS, Univ. Paris-Sud, Universit\'{e} Paris-Saclay, B\^{a}t. 121, 91405 Orsay cedex, France\goodbreak
\and
Institut d'Astrophysique de Paris, CNRS (UMR7095), 98 bis Boulevard Arago, F-75014, Paris, France\goodbreak
\and
Institute of Astronomy, University of Cambridge, Madingley Road, Cambridge CB3 0HA, U.K.\goodbreak
\and
Institute of Theoretical Astrophysics, University of Oslo, Blindern, Oslo, Norway\goodbreak
\and
Instituto de Astrof\'{\i}sica de Canarias, C/V\'{\i}a L\'{a}ctea s/n, La Laguna, Tenerife, Spain\goodbreak
\and
Instituto de F\'{\i}sica de Cantabria (CSIC-Universidad de Cantabria), Avda. de los Castros s/n, Santander, Spain\goodbreak
\and
Istituto Nazionale di Fisica Nucleare, Sezione di Padova, via Marzolo 8, I-35131 Padova, Italy\goodbreak
\and
Jet Propulsion Laboratory, California Institute of Technology, 4800 Oak Grove Drive, Pasadena, California, U.S.A.\goodbreak
\and
Jodrell Bank Centre for Astrophysics, Alan Turing Building, School of Physics and Astronomy, The University of Manchester, Oxford Road, Manchester, M13 9PL, U.K.\goodbreak
\and
Kavli Institute for Cosmological Physics, University of Chicago, Chicago, IL 60637, USA\goodbreak
\and
Kavli Institute for Cosmology Cambridge, Madingley Road, Cambridge, CB3 0HA, U.K.\goodbreak
\and
LAL, Universit\'{e} Paris-Sud, CNRS/IN2P3, Orsay, France\goodbreak
\and
LERMA, CNRS, Observatoire de Paris, 61 Avenue de l'Observatoire, Paris, France\goodbreak
\and
Laboratoire Traitement et Communication de l'Information, CNRS (UMR 5141) and T\'{e}l\'{e}com ParisTech, 46 rue Barrault F-75634 Paris Cedex 13, France\goodbreak
\and
Laboratoire de Physique Subatomique et Cosmologie, Universit\'{e} Grenoble-Alpes, CNRS/IN2P3, 53, rue des Martyrs, 38026 Grenoble Cedex, France\goodbreak
\and
Laboratoire de Physique Th\'{e}orique, Universit\'{e} Paris-Sud 11 \& CNRS, B\^{a}timent 210, 91405 Orsay, France\goodbreak
\and
Lawrence Berkeley National Laboratory, Berkeley, California, U.S.A.\goodbreak
\and
Max-Planck-Institut f\"{u}r Astrophysik, Karl-Schwarzschild-Str. 1, 85741 Garching, Germany\goodbreak
\and
Mullard Space Science Laboratory, University College London, Surrey RH5 6NT, U.K.\goodbreak
\and
Nicolaus Copernicus Astronomical Center, Bartycka 18, 00-716 Warsaw, Poland\goodbreak
\and
Niels Bohr Institute, Copenhagen University, Blegdamsvej 17, Copenhagen, Denmark\goodbreak
\and
Nordita (Nordic Institute for Theoretical Physics), Roslagstullsbacken 23, SE-106 91 Stockholm, Sweden\goodbreak
\and
SISSA, Astrophysics Sector, via Bonomea 265, 34136, Trieste, Italy\goodbreak
\and
School of Chemistry and Physics, University of KwaZulu-Natal, Westville Campus, Private Bag X54001, Durban, 4000, South Africa\goodbreak
\and
School of Physics and Astronomy, University of Nottingham, Nottingham NG7 2RD, U.K.\goodbreak
\and
Simon Fraser University, Department of Physics, 8888 University Drive, Burnaby BC, Canada\goodbreak
\and
Sorbonne Universit\'{e}-UPMC, UMR7095, Institut d'Astrophysique de Paris, 98 bis Boulevard Arago, F-75014, Paris, France\goodbreak
\and
Space Research Institute (IKI), Russian Academy of Sciences, Profsoyuznaya Str, 84/32, Moscow, 117997, Russia\goodbreak
\and
Space Sciences Laboratory, University of California, Berkeley, California, U.S.A.\goodbreak
\and
Sub-Department of Astrophysics, University of Oxford, Keble Road, Oxford OX1 3RH, U.K.\goodbreak
\and
The Oskar Klein Centre for Cosmoparticle Physics, Department of Physics,Stockholm University, AlbaNova, SE-106 91 Stockholm, Sweden\goodbreak
\and
UPMC Univ Paris 06, UMR7095, 98 bis Boulevard Arago, F-75014, Paris, France\goodbreak
\and
Universit\'{e} de Toulouse, UPS-OMP, IRAP, F-31028 Toulouse cedex 4, France\goodbreak
% ...
}
}

\authorrunning{Planck Collaboration}

\date{Accepted 23 July 2016}

\abstract{ We investigate constraints on cosmic reionization extracted from the \Planck\ cosmic microwave background (CMB) data. We combine the \Planck\ CMB anisotropy data in temperature with the low-multipole polarization data to fit \lcdm\ models with various parameterizations of the reionization history. We obtain a Thomson optical depth $\tau=\thetau$ for the commonly adopted instantaneous reionization model. This confirms, with data solely from CMB anisotropies, the low value suggested by combining \Planck\ 2015 results with other data sets, and also reduces the uncertainties. We reconstruct the history of the ionization fraction using either a symmetric or an asymmetric model for the transition between the neutral and ionized phases. To determine better constraints on the duration of the reionization process, we also make use of measurements of the amplitude of the kinetic Sunyaev-Zeldovich (kSZ) effect using additional information from the high-resolution Atacama Cosmology Telescope and South Pole Telescope experiments. The average redshift at which reionization occurs is found to lie between $z=7.8$ and 8.8, depending on the model of reionization adopted. Using kSZ constraints and a redshift-symmetric reionization model, we find an upper limit to the width of the reionization period of $\Delta z < 2.8$. In all cases, we find that the Universe is ionized at less than the 10\,\% level at redshifts above $z\simeq10$. This suggests that an early onset of reionization is strongly disfavoured by the \Planck\ data. We show that this result also reduces the tension between CMB-based analyses and constraints from other astrophysical sources.}


\titlerunning{Planck constraints on reionization history}

\maketitle

\section{Introduction}

The process of cosmological recombination happened around redshift $z\simeq1100$, after which the ionized fraction fell precipitously \citep{Peebles1968,ZKS1969,SSS2000} and the Universe became mostly neutral. However, observations of the Gunn-Peterson effect
\citep{gunn1965} in quasar spectra \citep{becker01,fan06b,venemans13,becker15b} indicate that intergalactic gas had become almost fully reionized by redshift $z\simeq6$.  Reionization is thus the second major change in the ionization state of hydrogen in the Universe.  Details of the transition from the neutral to ionized Universe are still the subject of intense investigations \citep[for a recent review, see the book by][]{RevMes2016}. In the currently conventional picture, early galaxies reionize hydrogen progressively throughout the entire Universe between $z\simeq12$ and $z\simeq6$, while quasars take over to reionize helium from $z\simeq6$ to $\simeq2$. But many questions remain. When did the epoch of reionization (EoR) start, and how long did it last? Are early galaxies enough to reionize the entire Universe or is another source required?  We try to shed light on these questions using the traces left by the EoR in the cosmic microwave background (CMB) anisotropies.

The CMB is affected by the total column density of free electrons along each line of sight, parameterized by its Thomson scattering optical depth $\tau$. This is one of the six parameters of the baseline \lcdm\ cosmological model and is the key measurement for constraining reionization. Large-scale anisotropies in polarization are particularly sensitive to the value of $\tau$. The \WMAP\ mission was the first to extract a $\tau$ measurement through the correlation between the temperature field and the $E$-mode polarization (i.e., the $TE$ power spectrum) over a large fraction of the sky. This measurement is very demanding, since the expected level of the $E$-mode polarization power spectrum at low multipoles ($\ell<10$) is only a few times $10^{-2}\,\mu{\rm K}^2$, lower by more than two orders of magnitude than the level of the temperature anisotropy power spectrum. For such weak signals the difficulty is not only to have enough detector sensitivity, but also to reduce and control both instrumental systematic effects and foreground residuals to a very low level.  This difficulty is illustrated by the improvements over time in the \WMAP-derived $\tau$ estimates. The 1-year results gave a value of $\tau=0.17\pm0.04$, based on the temperature-polarization $TE$ cross-power spectrum \citep{kogut2003}. In the 3-year release, this was revised down to $0.10\pm0.03$ using $E$-modes alone, whereas the combined $TT$, $TE$, and $EE$ power spectra gave $0.09\pm0.03$ \citep{page2007}.  Error bars improved in further \WMAP\ analyses, ending up with $0.089\pm0.014$ after the 9-year release \citep[see][]{dunkley2009,komatsu2010,hinshaw2012}. In 2013, the first \Planck\ satellite\footnote{\Planck\ (\url{http://www.esa.int/Planck}) is a project of the European Space Agency  (ESA) with instruments provided by two scientific consortia funded by ESA member states and led by Principal Investigators from France and Italy, telescope reflectors provided through a collaboration between ESA and a scientific consortium led and funded by Denmark, and additional contributions from NASA (USA).} cosmological results were based on \Planck\ temperature power spectra combined with the polarized \WMAP\ data and gave the same value $\tau=0.089\pm0.014$ \citep{planck2013-p11}. However, using a preliminary version of the \Planck\ 353\GHz\ polarization maps to clean the dust emission (in place of the \WMAP\ dust model), the optical depth was reduced by approximately $1\,\sigma$ to $\tau=0.075\pm0.013$ \citep{planck2013-p08}.

In the 2015 Planck analysis \citep{planck2014-a15}, the Low Frequency Instrument (LFI) low-resolution maps polarization at 70\GHz\ were used. Foreground cleaning was performed using the LFI 30\GHz\ and High Frequency Instrument (HFI) 353\GHz\ maps, operating effectively as polarized synchrotron and dust templates, respectively. The optical depth was found to be $\tau=0.078\pm 0.019$, and this decreased to $0.066\pm0.016$ when adding CMB lensing data. This value is also in agreement with the constraints from the combination ``PlanckTT+lensing+BAO,'' yielding $\tau=0.067\pm 0.016$, which uses no information from \lowl\ polarization.

In this paper and its companion \citep{planck2014-a10}, we derive the first estimate of $\tau$ from the \Planck-HFI polarization data at large scales. For the astrophysical interpretation, the power spectra are estimated using a PCL estimate which is more conservative. Indeed, it gives a slightly larger distribution on $\tau$ than the QML estimator used in \citet{planck2014-a10} but is less sensitive to the limited number of simulations available for the analysis. Using only $E$-mode polarization, the \Planck\ \lollipop\ likelihood gives $\tau = 0.053^{+0.014}_{-0.016}$ for a standard instantaneous reionization model, when all other \lcdm\ parameters are fixed to their \Planck-2015 best-fit values. We show that in combination with the \Planck\ temperature data the error bars are improved and we find $\tau = \thetau$.

In the \lcdm\ model, improved accuracy on the reionization optical
depth helps to reduce the degeneracies with other parameters. In
particular, the measurement of $\tau$ reduces the correlation with the
normalization of the initial power spectrum $A_{\rm s}$ and its
spectral index $n_{\rm s}$.  In addition to this $\tau$ is a
particularly important source of information for constraining the
history of reionization, which is the main subject of this paper.
When combined with direct probes at low redshift, a better knowledge
of the value of the CMB optical depth parameter may help to
characterize the duration of the EoR, and thus tell us when it
started.

In addition to the effect of reionization on the polarized large-scale
CMB anisotropies, reionization generates CMB temperature anisotropies
through the kinetic Sunyaev-Zeldovich (kSZ) effect~\citep{SZ80},
caused by the Doppler shift of photons scattering off electrons moving
with bulk velocities. Simulations have shown that early homogeneous
and patchy reionization scenarious differently affect the shape of the
kSZ power spectrum, allowing us to place constraints on the
reionization history \citep[e.g.,][]{mcquinn05,aghanim08}.
\citet{zahn12} derived the first constraints on the epoch of
reionization from the combination of kSZ and low-$\ell$ CMB
polarization, specifically using the \lowl\ polarization power
spectrum from WMAP and the very high multipoles of the temperature
angular power spectrum from the South Pole Telescope
\citep[SPT,][]{reichardt12}.
However one should keep in mind that kSZ signal is complicated to predict and depends on detailed astrophysics which makes the constraints on reionization difficult to interpret~\citep{mesinger2012}.

In this paper, we investigate constraints on the epoch of reionization
coming from \Planck. Section~\ref{sec:data_lik} first briefly
describes the pre-2016 data and likelihood used in this paper, which are
presented in detail in \citet{planck2014-a10}.  In
Sect.~\ref{sec:model} we then present the parameterizations we adopt
for the ionization fraction, describing the reionization history as a
function of redshift.  In Sect.~\ref{sec:observables}, we show the
results obtained from the CMB observables (i.e., the optical depth
$\tau$ and the amplitude of the kSZ effect) in the case of
``instantaneous'' reionization. Section~\ref{sec:cmbresults} presents
results based on the CMB measurements by considering different models for the ionization history.  In particular, we derive limits
on the reionization redshift and duration.  Finally, in
Sect.~\ref{sec:xe}, we derive the ionization fraction as a function of
redshift and discuss how our results relate to other astrophysical
constraints.


\section{Data and likelihood}
\label{sec:data_lik}

\subsection{Data \label{sec:data}}
The present analysis is based on the pre-2016 full mission intensity and polarization \Planck-HFI maps at 100 and 143\GHz. The data processing and the beam description are the same as in the \Planck\ 2015 release and have been detailed in~\cite{planck2014-a08}. \Planck-HFI polarization maps are constructed from the combination of polarized detectors that have fixed polarization direction. The \Planck\ scanning strategy produces a relatively low level of polarization angle measurement redundancy on the sky, resulting in a high level of $I$-$Q$-$U$ mixing, as shown in~\cite{planck2014-a09}. As a consequence, any instrumental mismatch between detectors from the same frequency channel produces leakage from intensity to polarization. This temperature-to-polarization leakage was at one point the main systematic effect present in the \Planck-HFI data, and prevented robust low-$\ell$ polarization measurements from being included in the previous \Planck\ data releases.

The maps that we use here differ in some respects from those data released in 2015. The updated mapmaking procedure, presented in \citet{planck2014-a10}, now allows for a significant reduction of the systematic effects in the maps. In particular, the relative calibration within a channel is now accurate to better than 0.001\,\%, which ensures a very low level of gain-mismatch between detectors. The major systematic effect that remains in the pre-2016 maps is due to
imperfections of the correction for nonlinearity in the analogue-to-digital converters (ADCs) but produces very low level of residuals in the maps. In addition to the 100\GHz\ and 143\GHz\ maps, we also make use of 30\GHz\ LFI data \citep{planck2014-a03} and 353\GHz\ HFI data to remove polarized foregrounds.

\begin{figure}[htbp!]
\centering
\includegraphics[width=\columnwidth]{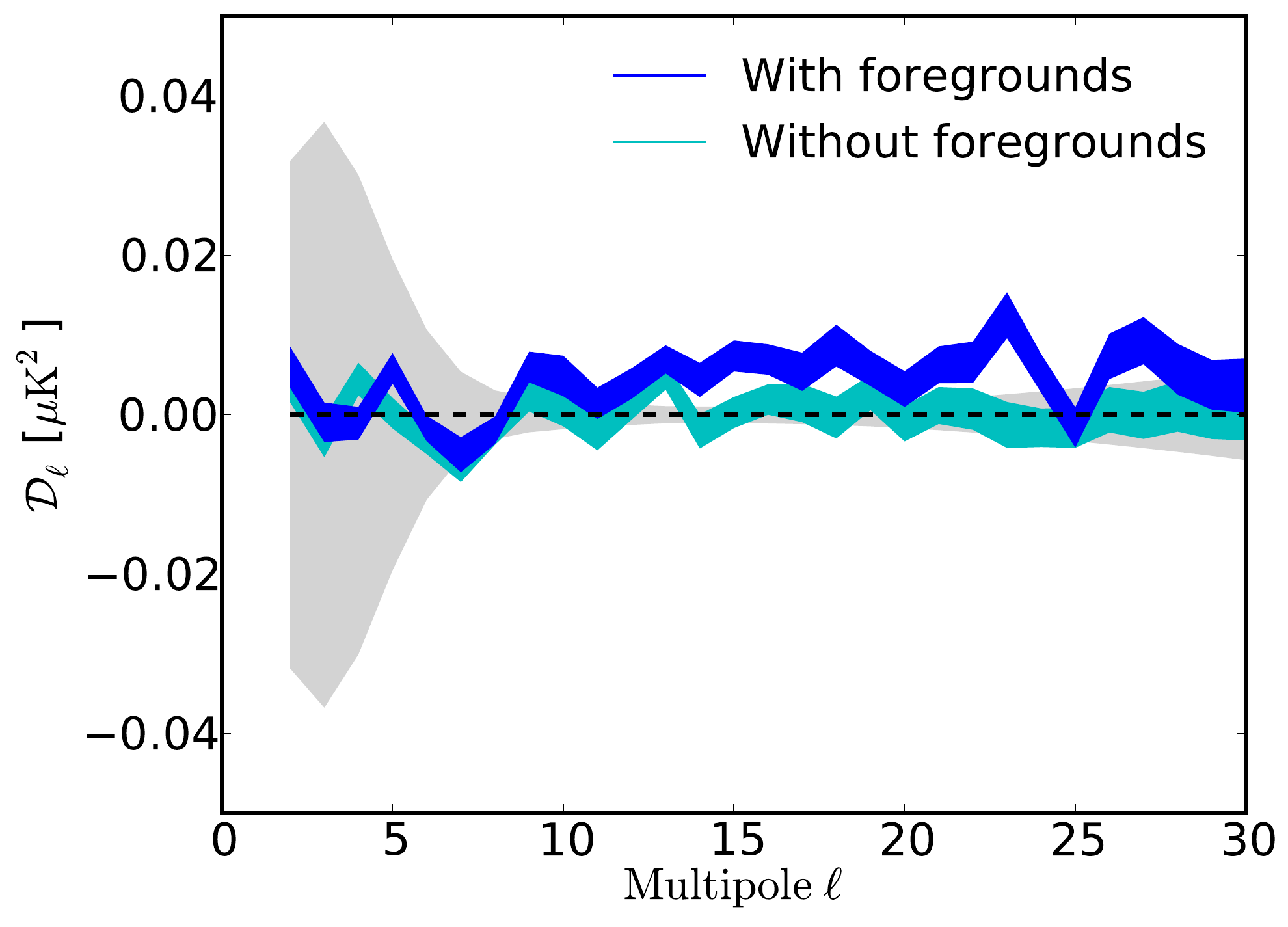}
\caption{Bias in the $100\times143$ cross-power spectrum computed from
  simulations, including instrumental noise and systematic effects,
  with or without foregrounds (dark blue and light blue), compared to
  the cosmic variance level (in grey).}
\label{fig:cl_bias}
\end{figure}

Using the pre-2016 end-to-end simulations, we show that the power spectrum bias induced by the remaining nonlinearities is very small and properly accounted for in the likelihood. Figure~\ref{fig:cl_bias} shows the bias (in the quantity ${\cal D}_\ell\equiv\ell(\ell+1)C_\ell/2\pi$, where $C_\ell$ is the conventional power spectrum) computed as the mean of the $EE$ cross-power spectra from simulated maps, including realistic noise and systematic effects without and with Galactic foregrounds. In the latter case, the foregrounds are removed for each simulation using the 30\GHz\ and 353\GHz\ maps as templates for synchrotron and dust,
respectively. The resulting bias in the $EE$ $100\times143$ cross-power spectrum can be used to correct the measured cross-spectrum, but in fact has very little impact on the likelihood.

\begin{figure}[htbp!]
\centering
\includegraphics[width=\columnwidth]{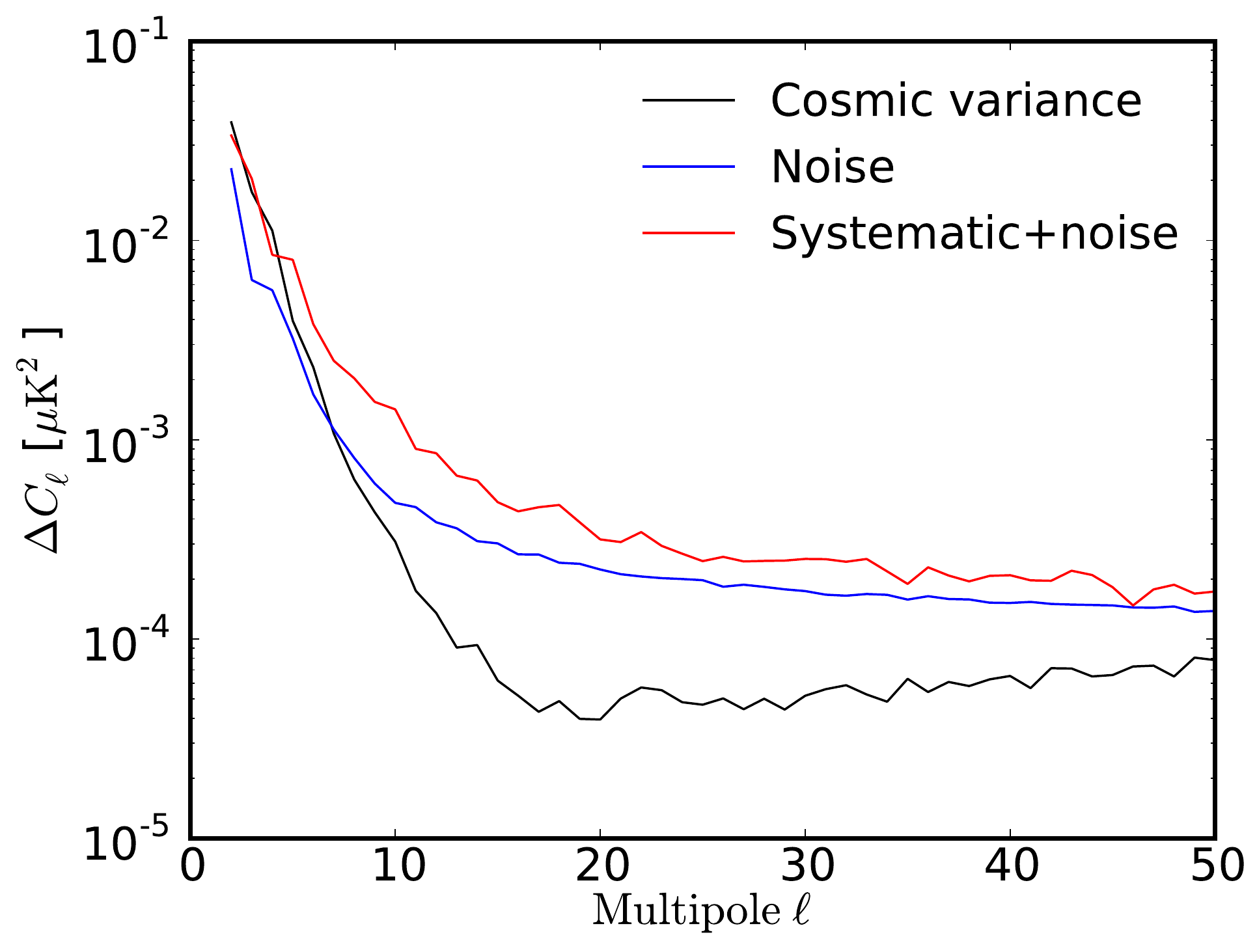}
\caption{Variance of the $100\times143$ $EE$ cross-power spectrum for
  simulations, including instrumental noise and noise plus systematic
  effects, compared to cosmic variance.}
\label{fig:cl_var}
\end{figure}
Furthermore, we use end-to-end simulations to propagate the systematic uncertainties to the cross-power spectra and all the way to the cosmological parameters. Figure~\ref{fig:cl_var} shows the impact on the variance due to the inclusion of the main ADC nonlinearity systematic effect, compared to realistic noise and cosmic variance. The resulting $C_\ell$ covariance matrix is estimated from these Monte Carlos. In the presence of such systematic effects, the variance of
the $C_\ell$ is shown to be higher by roughly a factor of 2 compared to the pure noise case.

Polarized foregrounds at \Planck-HFI frequencies are essentially dominated by Galactic dust emission, but also include a small contribution from synchrotron emission. We use the 353\GHz\ and 30\GHz\ \Planck\ maps as templates to subtract dust and synchrotron, respectively, using a single coefficient for each component over 95\,\% of the sky \citep[see][]{planck2014-a11,planck2014-a12}. However, foreground residuals in the maps are still dominant over the
CMB polarized signal near the Galactic plane. We therefore apply a very conservative mask, based on the amplitude of the polarized dust emission, which retains 50\,\% of the sky for the cosmological analysis. Outside this mask, the foreground residuals are found to be lower than 0.3 and 0.4\muK\ in $Q$ and $U$ Stokes polarization maps at 100 and 143\GHz, respectively. We have checked that our results are very stable when using a larger sky fraction of 60\,\%.

In this paper, we also make use of the constraints derived from the observation of the Gunn-Peterson effect on high-redshift quasars. As suggested by \citet{fan06a}, these measurements show that the Universe was almost fully reionized at redshift $z \simeq6$. We later discuss the results obtained with and without imposing a prior on the redshift of the end of reionization.

\begin{figure*}[htbp!]
  \centering
  \includegraphics[height=185px]{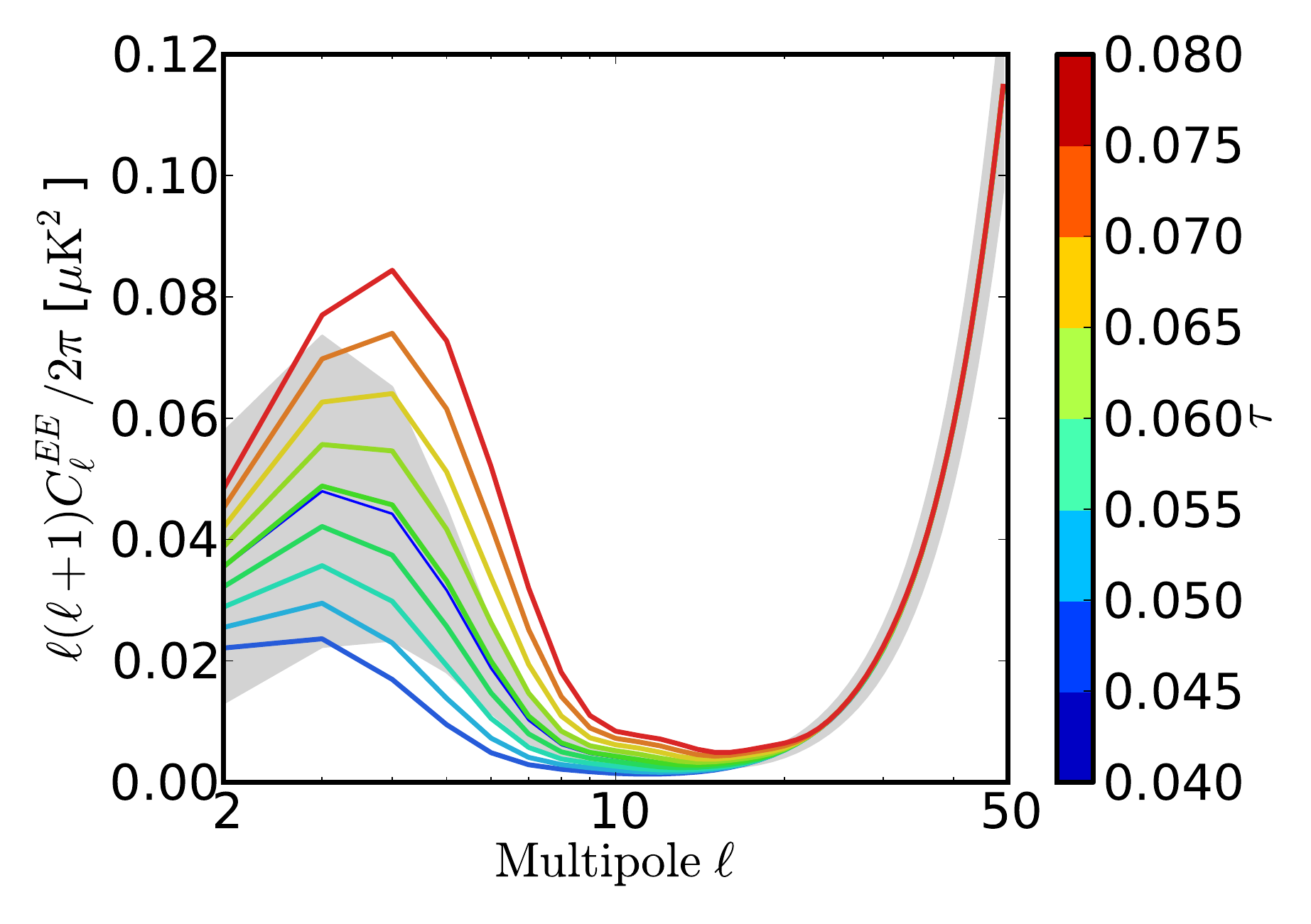}
  \includegraphics[height=185px]{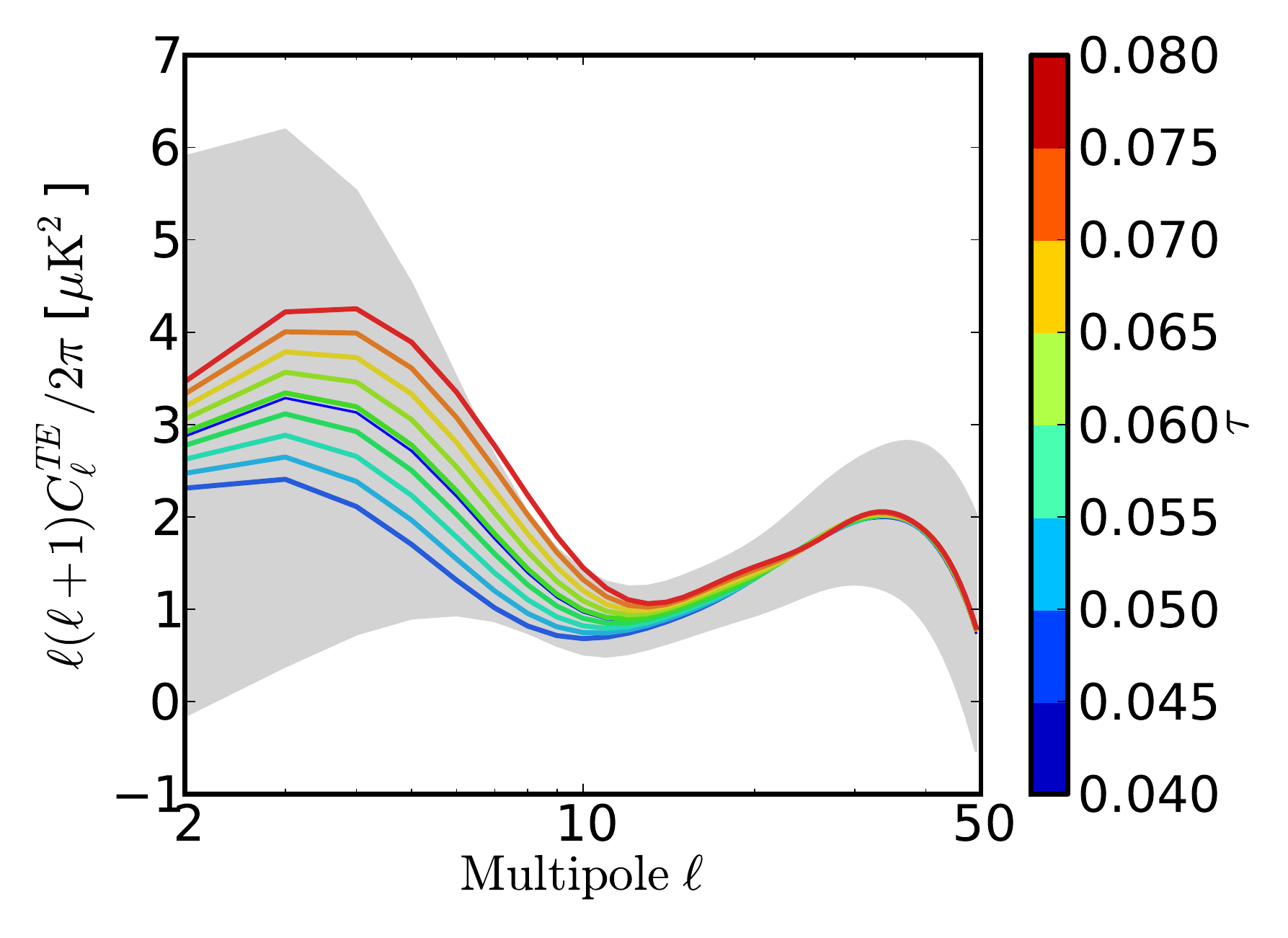}
  \caption{$EE$ and $TE$ power spectra for various $\tau$ values
    ranging from 0.04 to 0.08. The ionization fraction is modelled
    using a redshift-symmetric tanh function with $\delta z=0.5$. Grey
    bands represent the cosmic variance (full-sky) associated with the
    $\tau=0.06$ model.}
  \label{fig:clee_models}
\end{figure*}

\subsection{Likelihood \label{sec:likelihood}}
For temperature anisotropies, we use the combined \Planck\ likelihood
(hereafter ``\planckTT''), which includes the $TT$ power spectrum
likelihood at multipoles $\ell>30$ (using the {\tt Plik} code) and the
low-$\ell$ temperature-only likelihood based on the CMB map recovered
from the component-separation procedure (specifically {\tt Commander})
described in detail in \citet{planck2014-a13}.

For polarization, we use the \Planck\ \lowl\ $EE$ polarization
likelihood (hereafter \lollipop), a cross-spectra-based likelihood
approach described in detail in \citet{mangilli15} and applied to
\Planck\ data as discussed here in Appendix~\ref{sec:lollipop}. 
The multipole range used is $\ell=4$--20.
Cross-spectra are estimated using the pseudo-$C_\ell$ estimator {\tt
  Xpol} \citep[a generalization to polarization of the algorithm
presented in][]{tristram2005}. For a full-sky analysis, the statistics
of the reconstructed $C_\ell$ are given by a $\chi^2$ distribution
that is uncorrelated between multipoles. For a cut-sky analysis, the
distribution is more complex and includes $\ell$-to-$\ell$
correlations. \citet{hamimeche08} proposed an approximation of the
likelihood for cut-sky auto-power spectra that was adapted by
\cite{mangilli15} to be suitable for cross-spectra. Cross-spectra
between independent data sets show common sky signal, but are not
biased by the noise because this should be uncorrelated. This
approximation assumes that any systematic residuals are {\it not\/}
correlated between the different data sets; We have shown using realistic simulations (including \Planck-HFI noise
characteristics and systematic effect residuals), that the bias in the
cross-spectra is very small and can be corrected for at the
power-spectrum level. Nevertheless, we choose to remove the first two
multipoles ($\ell=2$ and $\ell=3$), since they may still be partially
contaminated by systematics.  Using those simulations, we derive the
$C_\ell$ covariance matrix used in the likelihood, which propagates
both the noise and the systematic uncertainties.
For the astrophysical interpretation, the power-spectra are estimated with a PCL estimate which is more conservative. Indeed, it gives a slightly larger distribution on $\tau$ than a QML estimator but is less sensitive to the limited number of simulations available for the analysis.

With \Planck\ sensitivity in polarization, the results from the \lowl\
$EE$ power spectrum dominate the constraints compared to the $TE$
power spectrum, as can be seen in Fig.~\ref{fig:clee_models}.  This is
because of the relatively larger cosmic variance for $TE$ (arising
from the temperature term) and the intrinsically weaker dependence on
$\tau$ ($\propto\tau$ compared with $\tau^2$ for $EE$), as well as the
fact that there is only partial correlation between $T$ and $E$. As a
consequence, we do not consider the $TE$ data in this analysis.
Furthermore, we do not make use of the high-$\ell$ likelihoods in $EE$
and $TE$ from \Planck, since they do not carry additional information
on reionization parameters.

\Planck\ temperature observations are complemented at smaller angular
scales by measurements from the ground-based Atacama Cosmology
Telescope (ACT) and South Pole Telescope (SPT). As explained in
\citet{planck2014-a13}, the high-$\ell$ likelihood (hereafter \VHL)
includes ACT power spectra at 148 and 218\GHz\ \citep{das14}, with a
revised binning \citep[described in][]{calabrese13} and final beam
estimates \citep{hasselfield13}, together with SPT measurements in the
range $2000<\ell<13\,000$ from the $2540\,{\rm deg}^2$ SPT-SZ survey
at 95, 150, and 220\GHz\ \citep{george15}.  To assess the consistency
between these data sets, we extend the \Planck\ foreground models up
to $\ell=13\,000$, with additional nuisance parameters for ACT and SPT
\citep[as described in][]{planck2014-a15}. We use the same models for
cosmic infrared background (CIB) fluctuations, the thermal SZ (tSZ)
effect, kSZ effect, and ${\rm CIB}\times{\rm tSZ}$ components.  The
kSZ template used in the \Planck\ 2015 results assumed homogeneous
reionization. In order to investigate inhomogeneous reionization, we
have modified the kSZ template when necessary, as discussed in
Sect.~\ref{sec:ksz_model}.

We use the CMB lensing likelihood \citep{planck2014-a17} in addition
to the CMB anisotropy likelihood. The lensing information can be used
to break the degeneracy between the normalization of the initial power
spectrum $A_{\rm s}$ and $\tau$ \citep[as discussed
in][]{planck2014-a15}. Despite this potential for improvement, we show in Sect.~\ref{sec:cmb_model} that \Planck's \lowl\
polarization signal-to-noise ratio is sufficiently high that the
lensing does not bring much additional information for the
reionization constraints.

The \Planck\ reference cosmology used in this paper corresponds to the
\planckTT+lowP+lensing best fit, as described in table 4, column 2 of
\citet{planck2014-a15}, namely $\Omega_{\rm b}h^2=0.02226$,
$\Omega_{\rm c}h^2 = 0.1197$, $\Omega_{\rm m}=0.308$, $n_{\rm
  s}=0.9677$, $H_0=67.81\,{\rm km}\,{\rm s}^{-1}\,{\rm Mpc}^{-1}$, for
which $Y_{\rm P}=0.2453$. This best-fit model comes from the
combination of three \Planck\ likelihoods: the temperature power
spectrum likelihood at high $\ell$; the ``lowP''
temperature+polarization likelihood, based on the foreground-cleaned
LFI 70\GHz\ polarization maps, together with the temperature map from
the {\tt Commander} component-separation algorithm; and the power
spectrum of the lensing potential as measured by \Planck.

\section{Parametrization of reionization history \label{sec:model}}
The epoch of reionization (EoR) is the period during which the cosmic gas
transformed from a neutral to ionized state at the onset of the first sources.
Details of the transition are thus strongly connected to many fundamental
questions in cosmology, such as what were the properties of the first galaxies
and the first (mini-)quasars, how did the formation of very metal-poor stars
proceed, etc. We certainly know that, at some point, luminous sources started
emitting ultraviolet radiation that reionized the neutral regions around them.
After a sufficient number of ionizing sources had formed, the average ionized
fraction of the gas in the Universe rapidly increased until hydrogen became
fully ionized. Empirical, analytic, and numerical models of the reionization
process have highlighted many pieces of the essential physics that led to the
birth to the ionized intergalactic medium (IGM) at late times
\citep{Couchman86,Miralda90,Meiksin93,aghanim96,gru98,Madau99,
gne00,Barkana01,cia03,Furlanetto04,PLW10,Pan11,Mitra11,Iliev14}.
Such studies provide predictions on the various reionization observables,
including those associated with the CMB.

The most common physical quantity used to characterize reionization is the
Thomson scattering optical depth defined as
\begin{equation}\label{eq:tau}
  \tau(z) = \int_{t(z)}^{t_0} n_{\rm e} \sigma_{\rm T}\,c\mathrm{d}t' ,
\end{equation}
where $n_{\rm e}$ is the number density of free electrons at time $t'$,
$\sigma_{\rm T}$ is the Thomson scattering cross-section, $t_0$ is the time
today, $t(z)$ is the time at redshift $z$, and we can use the Friedmann
equation to convert ${\rm d}t$ to ${\rm d}z$.
The reionization history is conveniently expressed in terms of the ionized
fraction $x_{\rm e}(z) \equiv n_{\rm e}(z)/n_{\rm H}(z)$ where $n_{\rm H}(z)$
is the hydrogen number density.
In practice, the CMB is sensitive to the average over all sky directions of $x_{\rm e} (1+\delta_b)$ (where $\delta_b$ denotes the baryon overdensity). The IGM is likely to be very inhomogeneous during reionization process, with ionized bubbles embedded in neutral surroundings, which would impact the relation between the optical depth and the reionisation parameters~\citep[see][]{liu2015} at a level which is neglected in this paper.

In this study, we define the redshift of reionization, $\zre \equiv
z_{50\,\%}$, as the redshift at which $x_{\rm e}=0.5 \times f$.  Here
the normalization, $f = 1+f_{\rm He} = 1 + n_{\rm He}/n_{\rm H}$,
takes into account electrons injected into the IGM by the first
ionization of helium (corresponding to 25\,eV), which is assumed to
happen roughly at the same time as hydrogen reionization. We define
the beginning and the end of the EoR by the redshifts $z_{\rm beg}
\equiv z_{10\,\%}$ and $z_{\rm end} \equiv z_{99\,\%}$ at which
$x_{\rm e} = 0.1 \times f$ and $0.99 \times f$, respectively. The
duration of the EoR is then defined as $\Delta
z=z_{10\,\%}-z_{99\,\%}$.\footnote{The reason this is not defined
  symmetrically is that in practice we have tighter constraints on the
  end of reionization than on the beginning.}  Moreover, to ensure
that the Universe is fully reionized at low redshift, we impose the
condition that the EoR is completed before the second helium
reionization phase (corresponding to 54\,eV), noting that it is
commonly assumed that quasars are necessary to produce the hard
photons needed to ionize helium.  To be explicit about how we treat
the lowest redshifts we assume that the full reionization of helium
happens fairly sharply at $z_{\rm He} = 3.5$ \citep{becker11},
following a transition of hyperbolic tangent shape with width $\delta
z=0.5$.  While there is still some debate on whether helium
reionization could be inhomogeneous and extended \citep[and thus have
an early start,][]{worseck14}, we have checked that varying the helium
reionization redshift between 2.5 and 4.5 changes the total optical
depth by less than 1\,\%.

The simplest and most widely-used parameterizations describes the EoR as a
step-like transition between an essentially vanishing ionized
fraction\footnote{The ionized fraction is actually matched to the relic
free electron density from recombination, calculated using {\tt recfast}
\cite{SSS2000}.} $x_{\rm e}$ at early times, to a value of unity at low
redshifts. When calculating the effect on anisotropies it is necessary to
give a non-zero width to the transition, and it can be modelled using a
$\tanh$ function \citep{lewis08}:
\begin{equation}
  x_{\rm e}(z) = \frac{f}{2} \left[ 1 +
    \tanh\left(\frac{y-y_{\rm re}}{\delta y}\right) \right] \, ,
\end{equation}
where $y = (1+z)^{3/2}$ and $\delta y = \frac{3}{2} (1+z)^{1/2} \delta
z$.  The key parameters are thus $\zre$, which measures the redshift
at which the ionized fraction reaches half its maximum and a width
$\delta z$.  The $\tanh$ parameterization of the EoR transition allows
us to compute the optical depth of Eq.~(\ref{eq:tau}) for a one-stage
almost redshift-symmetric\footnote{For convenience, we refer to
  this parameterization as ``redshift symmetric'' in the rest of the
  paper, even although it is actually symmetric in $y$ rather than
  $z$. The asymmetry is maximum in the instantaneous case, but the
  difference in $x_{\rm e}$ values around, for example, $\zre = 8 \pm
  1$, is less than 1\,\%.}  reionization transition, where the
redshift interval between the onset of the reionization process and
its half completion is (by construction) equal to the interval between
half completion and full completion.  In this parameterization, the
optical depth is mainly determined by \zre\ and almost degenerate with
the width $\delta z$.  This is the model used in the \Planck\ 2013 and
2015 cosmological papers, for which we have fixed $\delta z=0.5$
(corresponding to $\Delta z = 1.73$). In this case, we usually talk
about ``instantaneous'' reionization.

\begin{figure*}[htbp!]
  \centering
  \includegraphics[height=190px]{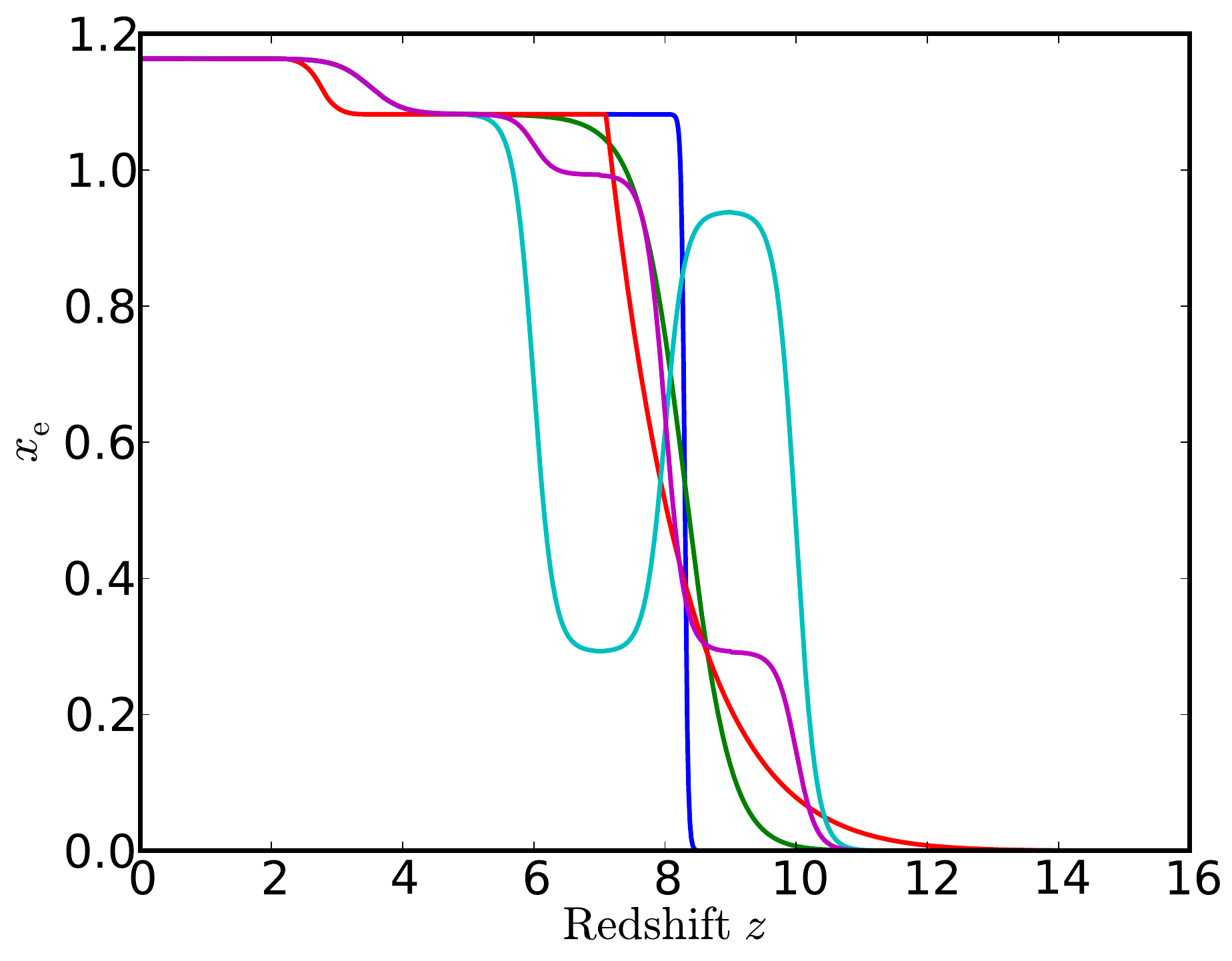}
  \includegraphics[height=190px]{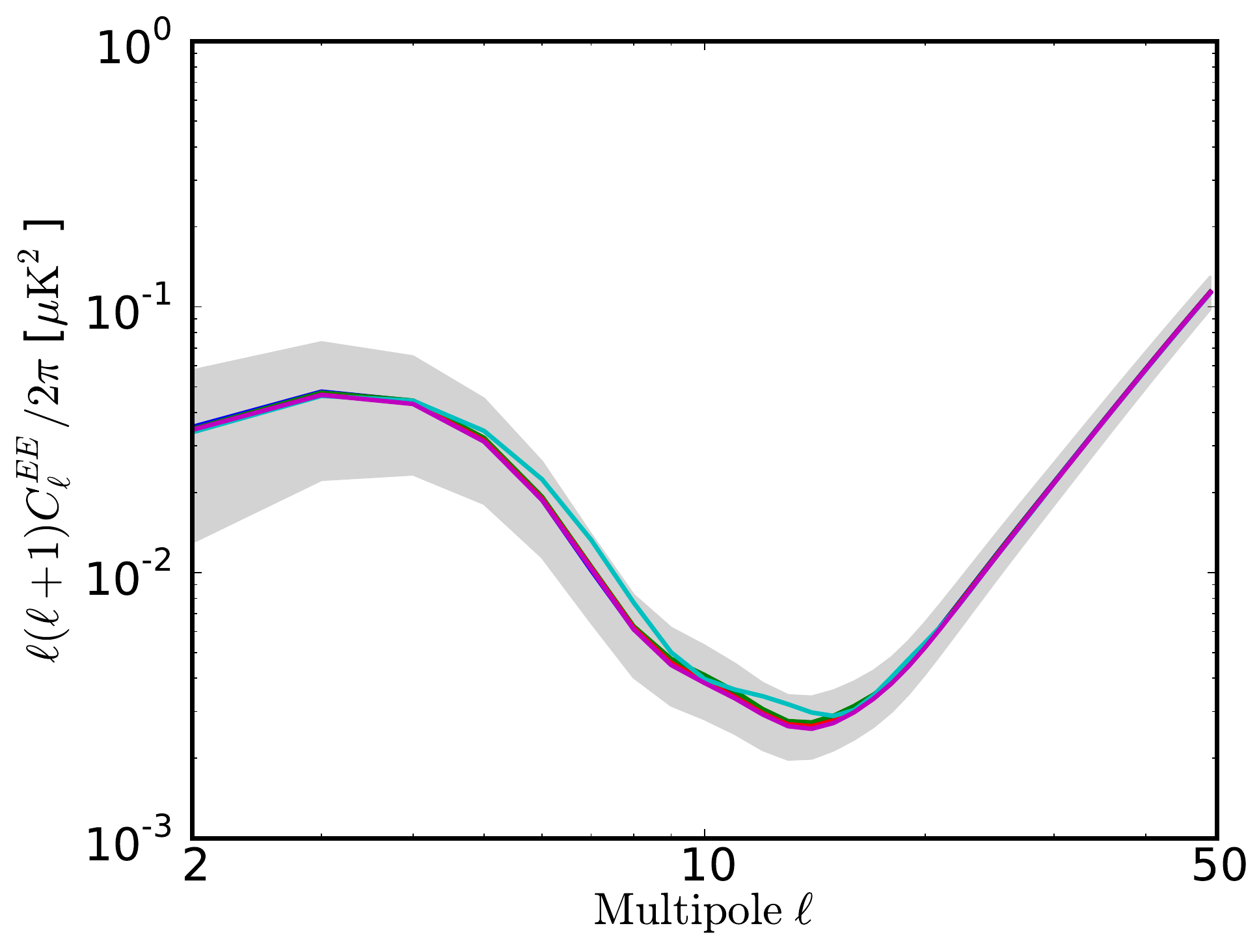}
  \caption{{\it Left}: Evolution of the ionization fraction for
    several functions, all having the same optical depth, $\tau=0.06$:
    green and blue are for redshift-symmetric instantaneous ($\delta
    z=0.05$) and extended reionization ($\delta z=0.7$), respectively;
    red is an example of a redshift-asymmetric parameterization; and
    light blue and magenta are examples of an ionization fraction
    defined in redshift bins, with two bins inverted between these two
    examples.  {\it Right}: corresponding $EE$ power spectra with
    cosmic variance in grey.  All models have the same optical depth
    $\tau=0.06$ and are essentially indistinguishable at the
    reionization bump scale.}
  \label{fig:xe_models}
\end{figure*}

A redshift-asymmetric parameterization is a better, more flexible
description of numerical simulations of the reionization process
\citep[e.g.,][]{ahn12,park13,douspis15}.  A function with this
behaviour is also suggested by the constraints from ionizing
background measurements of star-forming galaxies and from low-redshift
line-of-sight probes such as quasars, Lyman-$\alpha$ emitters, or
$\gamma$-ray bursts \citep{Faisst14,Chornock14,Ishi15,Rob15,Bou15}.
The two simplest choices of redshift-asymmetric parameterizations are
polynomial or exponential functions of redshift~\citep{douspis15}.
These two parameterizations are in fact very similar, and we adopt
here a power law defined by two parameters: the redshift at which
reionization ends ($z_{\rm end}$); and the exponent $\alpha$.
Specifically we have
\begin{equation}
  x_{\rm e}(z) =
  \begin{cases}
    f & \mbox{for } z<z_{\rm end}, \\
    f \left(\frac{z_{\rm early}-z}{z_{\rm early}-z_{\rm end}}\right)^\alpha
      & \mbox{for }z>z_{\rm end}.
  \end{cases}
\end{equation}
In the following, we fix $z_{\rm early}=20$, the redshift around which
the first emitting sources form, and at which we smoothly match
$x_{\rm e}(z)$ to the ionized fraction left over from recombination.
We checked that our results are not sensitive to the precise value of
$z_{\rm early}$, as long as it is not dramatically different.

Non-parametric reconstructions of the ionization fraction have also
been proposed to probe the reionization history. Such methods are
based on exploring reionization parameters in bins of redshift
\citep{lewis06}. They should be particularly useful for investigating
exotic reionization histories, e.g., double reionization \citep{Cen03}.
However, the CMB large-scale ($\ell \la 10$) polarization anisotropies
are mainly sensitive to the overall value of the optical depth, which
determines the amplitude of the reionization bump in the $EE$ power
spectrum (see Fig.~\ref{fig:clee_models}). We have estimated the
impact on $C_\ell^{EE}$ for the two different models (tanh and power
law) having the same $\tau=0.06$ and found differences of less than
4\,\% for $\ell < 10$. Even for a double reionization model,
Fig.~\ref{fig:xe_models} shows that the impact on $C_\ell^{EE}$ is
quite weak, given the actual measured value of $\tau$, and cannot be
distinguished relative to the cosmic variance spread (i.e., even for a
full-sky experiment).  We also checked that \Planck\ data do not allow
for model-independent reconstruction of $x_{\rm e}$ in redshift bins.
Principal component analysis has been proposed as an explicit approach
to try to capture the details of the reionization history in a small
set of parameters \citep{hu03,mortonson08}. Although these methods are
generally considered to be non-parametric, they are in fact based on a
description of $x_{\rm e}(z)$ in bins of redshift, expanded around a
given fiducial model for $C_\ell^{EE}$.  Moreover, the potential bias
on the $\tau$ measurement when analysing a more complex reionization
history using a simple sharp transition model
\citep{holder03,colombo09} is considerably reduced for the (lower)
$\tau$ values as suggested by the \Planck\ results.  Consequently, we
do not consider the non-parametric approach further.

\section{Measuring reionization observables}\label{sec:observables}
Reionization leaves imprints in the CMB power spectra, both in
polarization at very large scales and in intensity via the suppression
of $TT$ power at higher $\ell$. Reionization also affects the kSZ
effect, due to the re-scattering of photons off newly liberated
electrons.
We sample from the space of possible parameters with MCMC exploration using CAMEL\footnote{available at \url{camel.in2p3.fr}}. This uses an adaptative-Metropolis algorithm to generate chains of samples for a set of parameters.

\subsection{Large-scale CMB polarization} \label{sec:cmb_model}
Thomson scattering between the CMB photons and free electrons
generates linear polarization from the quadrupole moment of the CMB
radiation field at the scattering epoch. This occurs at recombination
and also during the epoch of reionization. Re-scattering of the CMB
photons at reionization generates an additional polarization
anisotropy at large angular scales, because the horizon size at this
epoch subtends a much larger angular size. The multipole location of
this additional anisotropy (essentially a bump) in the $EE$ and $TE$
angular power spectra relates to the horizon size at the new
``last-rescattering surface'' and thus depends on the redshift of
reionization. The height of the bump is a function of the optical
depth or, in other words, of the history of the reionization process.
Such a signature (i.e., a polarization bump at large scales) was first
observed by WMAP, initially in the $TE$ angular power spectrum
\citep{kogut2003}, and later in combination with all power spectra
\citep{hinshaw2012}.

In Fig.~\ref{fig:clee_models} we show for the ``instantaneous''
reionization case (specifically the redshift-symmetric
parameterization with $\delta z = 0.5$) power spectra for the $E$-mode
polarization power spectrum $C_\ell^{EE}$ and the
temperature-polarization cross-power spectrum $C_\ell^{TE}$.  The
curves are computed with the {\tt CLASS} Boltzmann solver
\citep{lesgourgues11} using $\tau$ values ranging from 0.04 to 0.08.
For the range of optical depth considered here and given the amount of
cosmic variance, the $TE$ spectrum has only a marginal sensitivity to
$\tau$, while in $EE$ the ability to distinguish different values of
$\tau$ is considerably stronger.

In Fig.~\ref{fig:xe_models} (left panel), the evolution of the ionized
fraction $x_{\rm e}$ during the EoR is shown for five different
parameterizations of the reionization history, all yielding the same
optical depth $\tau=0.06$. Despite the differences in the evolution of
the ionization fraction, the associated $C_\ell^{EE}$ curves
(Fig.~\ref{fig:xe_models}, right panel) are almost indistinguishable.
This illustrates that while CMB large-scale anisotropies in
polarization are only weakly sensitive to the details of the
reionization history, they can nevertheless be used to measure the
reionization optical depth, which is directly related to the amplitude
of the \lowl\ bump in the $E$-mode power spectrum.

\begin{figure}[htbp!]
  \centering
  \includegraphics[width=\columnwidth]{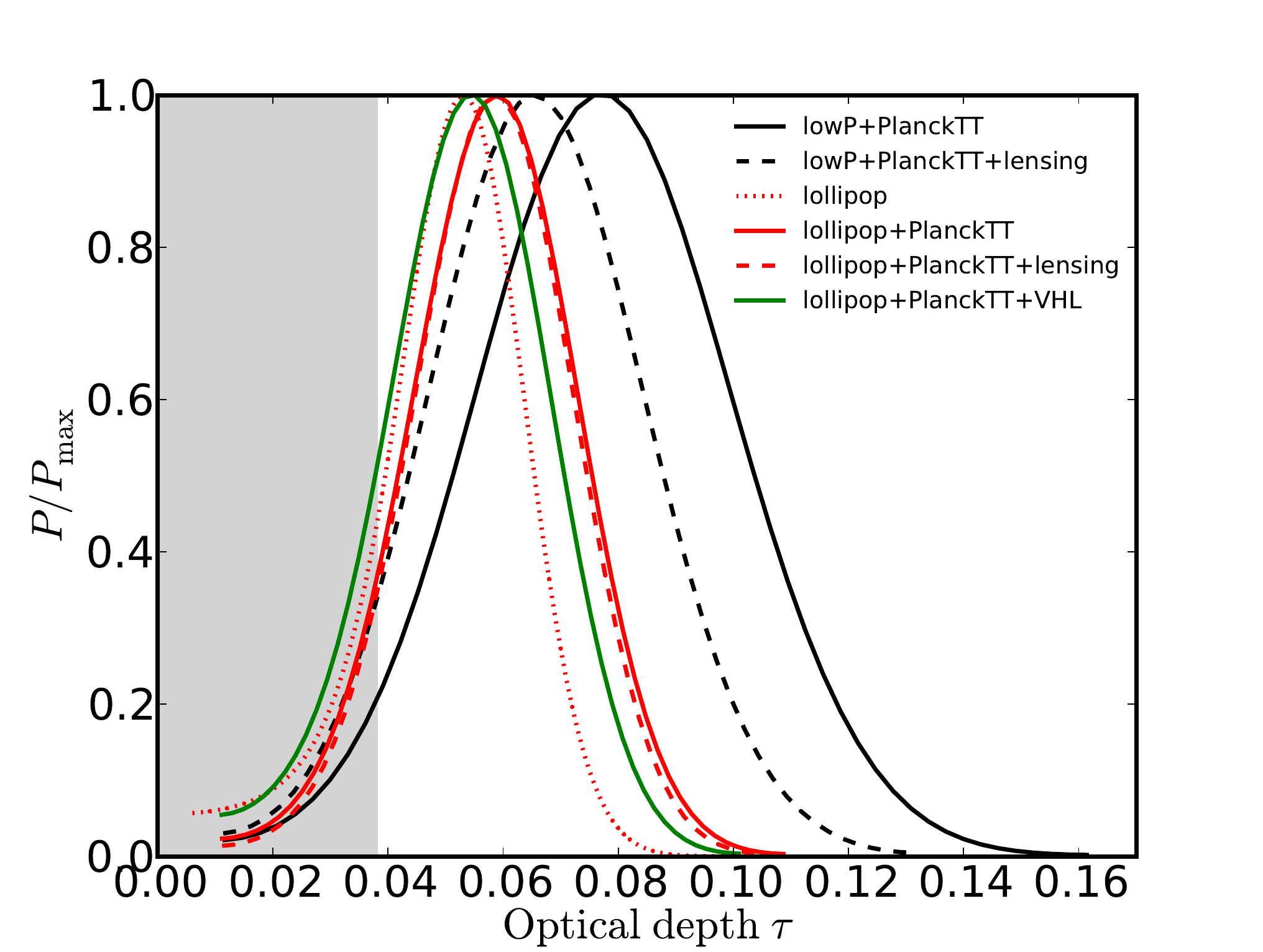}
  \caption{Posterior distribution for $\tau$ from the various
    combinations of \Planck\ data. The grey band shows the lower limit
    on $\tau$ from the Gunn-Peterson effect.}
  \label{fig:tau_datasets}
\end{figure}

We use the \Planck\ data to provide constraints on the Thomson
scattering optical depth for ``instantaneous'' reionization.
Figure~\ref{fig:tau_datasets} shows the posterior distributions for
$\tau$ obtained with the different data sets described in
Sect.~\ref{sec:data_lik} and compared to the 2015 \planckTT+lowP
results \citep{planck2014-a15}.  We show the posterior distribution
for the \lowl\ \Planck\ polarized likelihood (\lollipop) and in
combination with the high-$\ell$ \Planck\ likelihood in temperature
(\planckTT). We also consider the effect of adding the SPT and ACT
likelihoods (\VHL) and the \Planck\ lensing likelihood, as described
in \citet{planck2014-a17}.

\begin{figure}[htbp!]
  \centering
  \includegraphics[width=\columnwidth]{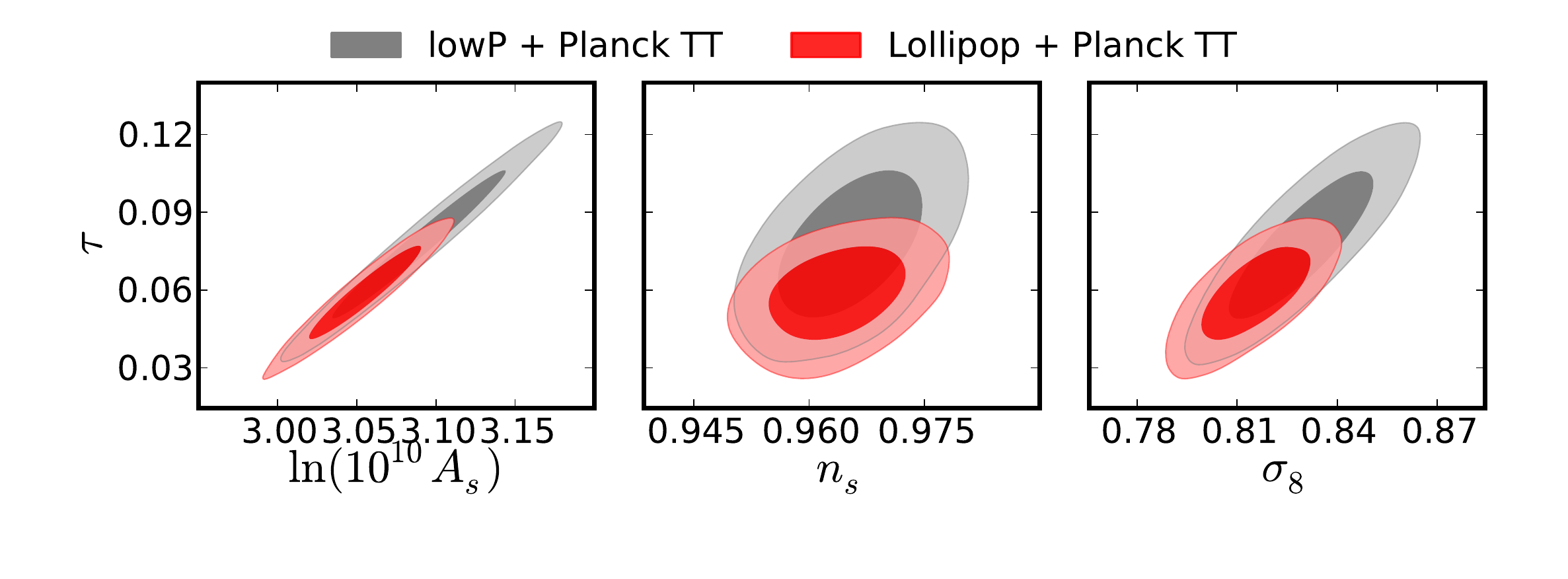}
  \includegraphics[width=\columnwidth]{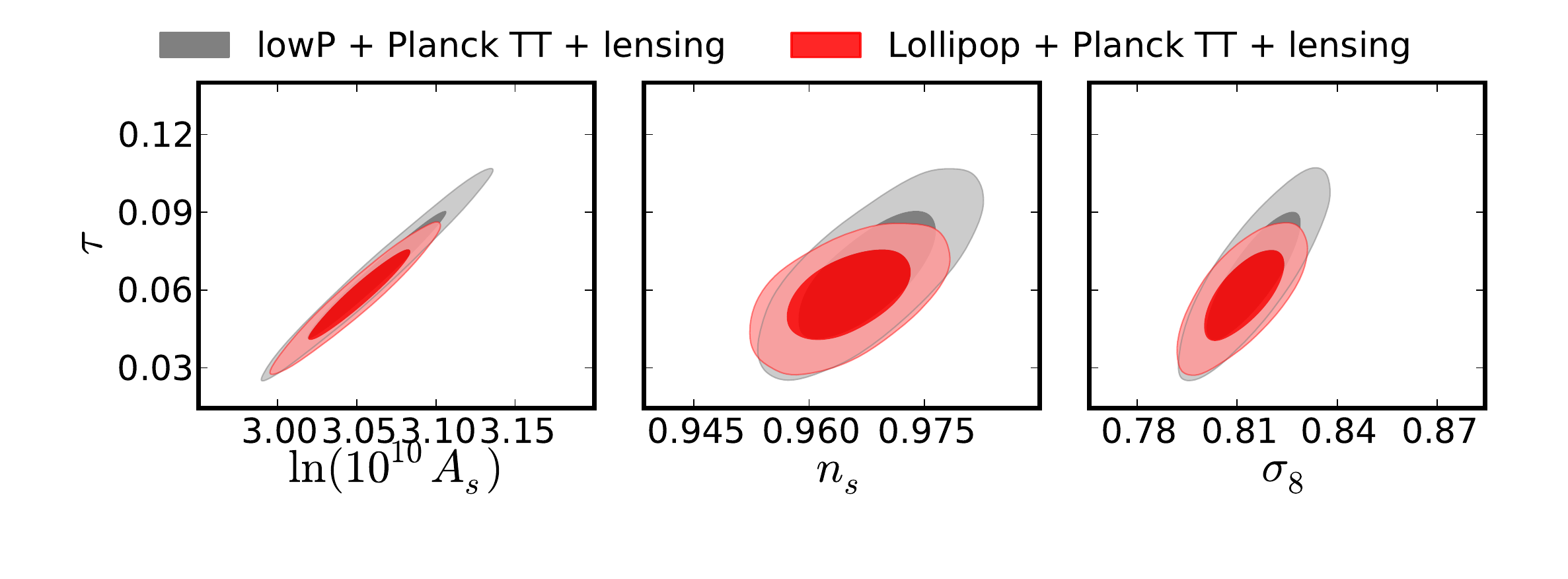}
  \caption{Constraints on $\tau$, $A_{\rm s}$, $n_{\rm s}$, and
    $\sigma_8$ for the \lcdm\ cosmology from \planckTT, showing the
    impact of replacing the lowP likelihood from \Planck\ 2015 release
    with the new \lollipop\ likelihood.  The top panels show results
    without lensing, while the bottom panels are with lensing.}
  \label{fig:As_tau}
\end{figure}

The different data sets show compatible constraints on the optical
depth $\tau$.  The comparison between posteriors indicates that the
optical depth measurement is driven by the \lowl\ likelihood in
polarization (i.e., \lollipop).  The \Planck\ constraints on $\tau$
for a \lcdm\ model when considering the standard ``instantaneous''
reionization assumption (symmetric model with fixed $\delta z=0.5$),
for the various data combinations are:
\begin{align}
  \tau =\, &0.053^{+0.014}_{-0.016}\,, & \mbox{\lollipop \footnotemark}\,;		\label{tau:lol}\\
  \tau =\, &0.058_{-0.012}^{+0.012}\,, & \mbox{\lollipop+\planckTT}\,;			\label{tau:plikTT_Comm_lol} \\
  \tau =\, &0.058_{-0.012}^{+0.011}\,, & \mbox{\lollipop+\planckTT+lensing}\,;	\label{tau:plikTT_Comm_lol_lens} \\
  \tau =\, &0.054_{-0.013}^{+0.012}\,, & \mbox{\lollipop+\planckTT+\VHL}\,.	\label{tau:plikTT_Comm_lol_vhl}
\end{align}
\footnotetext{In this case only, other \lcdm\ parameters are held
  fixed, including $A_{\rm s} \exp{(-2\tau)}$.}  We can see an
improvement of the posterior width when adding temperature anisotropy
data to the \lollipop\ likelihood. This comes from the fact that the
temperature anisotropies help to fix other \lcdm\ parameters, in
particular the normalization of the initial power spectrum $A_{\rm
  s}$, and its spectral index, $n_{\rm s}$.  CMB lensing also helps to
reduce the degeneracy with $A_{\rm s}$, while getting rid of the
tension with the phenomenological lensing parameter $A_{\rm L}$ when
using \planckTT\ only \citep[see][]{planck2014-a15}, even if the
impact on the error bars is small.  Comparing the posteriors in
Fig.~\ref{fig:As_tau} with the constraints from \planckTT\ alone
\citep[see figure~45 in][]{planck2014-a13} shows that indeed, the
polarization likelihood is sufficiently powerful that it breaks the
degeneracy between $n_{\rm s}$ and $\tau$.  The impact on other \lcdm\
parameters is small, typically below $0.3\,\sigma$ (as shown more
explicitly in Appendix~\ref{app:lcdm_impact}).  The largest changes
are for $\tau$ and $A_{\rm s}$, where the \lollipop\ likelihood
dominates the constraint. The parameter $\sigma_8$ shifts towards
slightly smaller values by about $1\,\sigma$. This is in the right
direction to help resolve some of the tension with cluster abundances
and weak galaxy lensing measurements, discussed in \citet{planck2013-p15} and
\citet{planck2014-a15}; however, some tension still remains.

Combining with \VHL\ data gives compatible results, with consistent
error bars.  The slight shift toward lower $\tau$ value (by
$0.3\,\sigma$) is related to the fact that the \planckTT\ likelihood
alone pushes towards higher $\tau$ values
\citep[see][]{planck2014-a15}, while the addition of \VHL\ data helps
to some extent in reducing the tension on $\tau$ between \highl\ and
\lowl\ polarization.

As mentioned earlier, astrophysics constraints from measurements of the Gunn-Peterson effect provide strong evidence that the IGM was highly ionized by a redshift of $z \simeq 6$. This places a lower limit on the optical depth (using Eq.~\ref{eq:tau}), which in the case of instantaneous reionization in the standard $\Lambda$CDM cosmology corresponds to $\tau=0.038$.

\subsection{Kinetic Sunyaev-Zeldovich effect}
\label{sec:ksz_model}
The Thomson scattering of CMB photons off ionized electrons induces secondary anisotropies at different stages of the reionization process. In particular, we are interested here in the effect of photons scattering off electrons moving with bulk velocity, which is called the ``kinetic Sunyaev Zeldovich'' or kSZ effect.  It is common to distinguish between the ``homogeneous'' kSZ effect, arising when the reionization is complete \citep[e.g.,][]{ostriker86}, and ``patchy'' (or inhomogeneous) reionization \citep[e.g.,][]{aghanim96}, which arises during the process of reionization, from the proper motion of ionized bubbles around emitting sources. These two components can be described by their power spectra, which can be computed analytically or derived from numerical simulations. In \citet{planck2014-a13}, we used a kSZ template based on homogeneous simulations, as described in \citet{trac11}.

In the following, we assume that the kSZ power spectrum is given by
\begin{equation}
\label{eq:clksz}
	{\cal D}_\ell^{\rm kSZ}={\cal D}_\ell^{\rm h-kSZ}+{\cal D}_\ell^{\rm p-kSZ},
\end{equation}
where ${\cal D}_\ell = \ell(\ell+1)C_\ell / 2\pi$ and the superscripts ``h-kSZ'' and ``p-kSZ'' stand for ``homogeneous'' and ``patchy'' reionization, respectively. For the homogeneous reionization, we use the kSZ template power spectrum given by \citet{Shaw12} calibrated with a simulation that includes the effects of cooling and star-formation (which we label ``CSF'').  For the patchy reionization kSZ effect we use the fiducial model of \citet{Battaglia13}.

\begin{figure}[htbp!]
  \centering
  \includegraphics[width=\columnwidth]{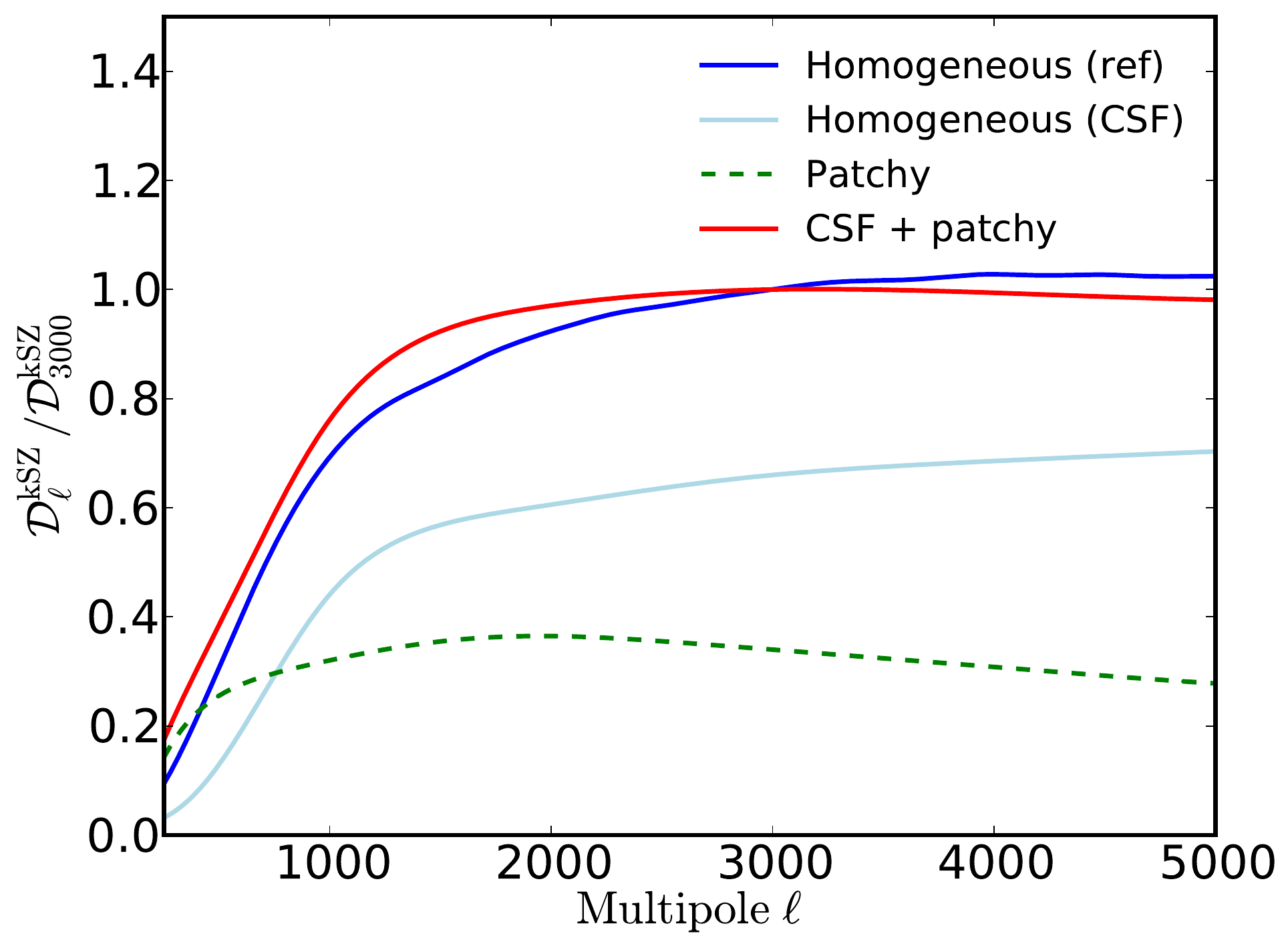}
  \caption{Power spectrum templates for the kSZ effect.  The different
    lines correspond to: homogeneous reionization as used in
    \citet{planck2014-a13} (dark blue), based on \citet{trac11};
    ``CSF'' (light blue), which is a homogeneous reionization model
    from \citet{Shaw12}; Patchy (green dashed) based on patchy
    reionization model from \citet{Battaglia13}; and the sum of CSF
    and patchy (red).}
  \label{fig:Cl_ksz}
\end{figure}

In the range $\ell=1000$--7000, the shape of the kSZ power spectrum is relatively flat and does not vary much with the detailed reionization history. The relative contributions (specifically ``CSF'' and ``patchy'') to the kSZ power spectrum are shown in Fig~\ref{fig:Cl_ksz} and compared to the ``homogeneous'' template used
in \cite{planck2014-a13}, rescaled to unity at $\ell=3000$.

The kSZ power spectrum amplitude does depend on the cosmological parameters \citep{Shaw12,zahn12}. To deal with this, we adopt the scalings from \citet{Shaw12}, which gives the amplitude at $\ell=3000$, $A_{\rm kSZ} \equiv \mathcal{D}_{\ell=3000}^{\rm kSZ}$:
\begin{align}
  A_{\rm kSZ} \propto
  \left(\frac{h}{0.7}\right)^{1.7}
  \left(\frac{\sigma_8}{0.8}\right)^{4.5}
  \left(\frac{\Omega_{\rm b}}{0.045}\right)^{2.1}
  \left(\frac{0.27}{\Omega_{\rm m}}\right)^{0.44}
  \left(\frac{0.96}{n_{\rm s}}\right)^{0.19}.
  \label{eq:ksz_scaling}
\end{align}

The amplitude of the kSZ power spectrum at $\ell=3000$ for the fiducial cosmology, $A_{\rm kSZ}$ is another observable of the reionization history that can be probed by CMB data. Its scalings with the reionization redshift and the duration of the EoR can be extracted from simulations. We assume for the patchy and homogeneous kSZ effect, the scalings of \cite{Battaglia13} and \cite{Shaw12}, respectively. For the \Planck\ base $\Lambda$CDM cosmology given in Sect.~\ref{sec:likelihood}, we find (in $\muK^2$):
\begin{align}
  A_{\rm kSZ}^{\rm h} =\,& 2.02 \times
  \left( \frac{\tau}{0.076} \right)^{0.44}; \label{eq:hksz_reio}\\
  A_{\rm kSZ}^{\rm p} =\,& 2.03 \times \left[ \left( \frac{1+{z_{\rm
            re}}}{11} \right) - 0.12 \right] \left(
    \frac{z_{25\,\%}-z_{75\,\%}}{1.05} \right)^{0.51} \, .
  \label{eq:pksz_reio}
\end{align}

For the measured value $\tau = \thetau$, Eqs.~(\ref{eq:hksz_reio}) and (\ref{eq:pksz_reio}) give amplitudes for the homogeneous and patchy reionization contributions of $A_{\rm kSZ}^{\rm h}=1.79\muK^2$ and $A_{\rm kSZ}^{\rm p}=1.01\muK^2$, respectively.

For the multipole range of \Planck, the amplitude of the kSZ spectrum is dominated by other foregrounds, including Galactic dust, point sources, CIB fluctuations, and the tSZ effect. Moreover, the \Planck\ signal-to-noise ratio decreases rapidly above $\ell=2000$, where the kSZ signal is maximal. This is why we cannot constrain the kSZ amplitude using \Planck\ data alone. Combining with additional data at higher multipoles helps to disentangle the different foregrounds. We explicitly use the band powers from SPT~\citep{george15} and ACT~\citep{das14}, covering the multipole range up to $\ell=13\,000$.

Despite our best efforts to account for the details, the kSZ amplitude is weak and there are large uncertainties in the models (both homogeneous and patchy). Moreover, there are correlations between the different foreground components, coming both from the astrophysics (we use the same halo model to derive the power spectra for the CIB and for CIB$\times$tSZ as the one used for the kSZ effect) and from the adjustments in the data. We carried out several tests to check the robustness of the constraints on $A_{\rm kSZ}$ with respect to the template used for the CIB, CIB$\times$tSZ, and kSZ contributions. In particular, the CIB$\times$tSZ power spectrum amplitude is strongly anti-correlated with the kSZ amplitude and poorly constrained by the CMB data. As a consequence, if we neglect the CIB$\times$tSZ contribution, the kSZ amplitude measured in CMB data is substantially reduced, leading to an upper limit much lower than the one derived when including the CIB$\times$tSZ correlation. In the following discussion we consider only the more realistic case (and thus more conservative in terms of constraints on $A_{\rm kSZ}$) where the CIB$\times$tSZ correlation contributes to the \highl\ signal.

\begin{figure}[htbp!]
  \centering
  \includegraphics[width=\columnwidth]{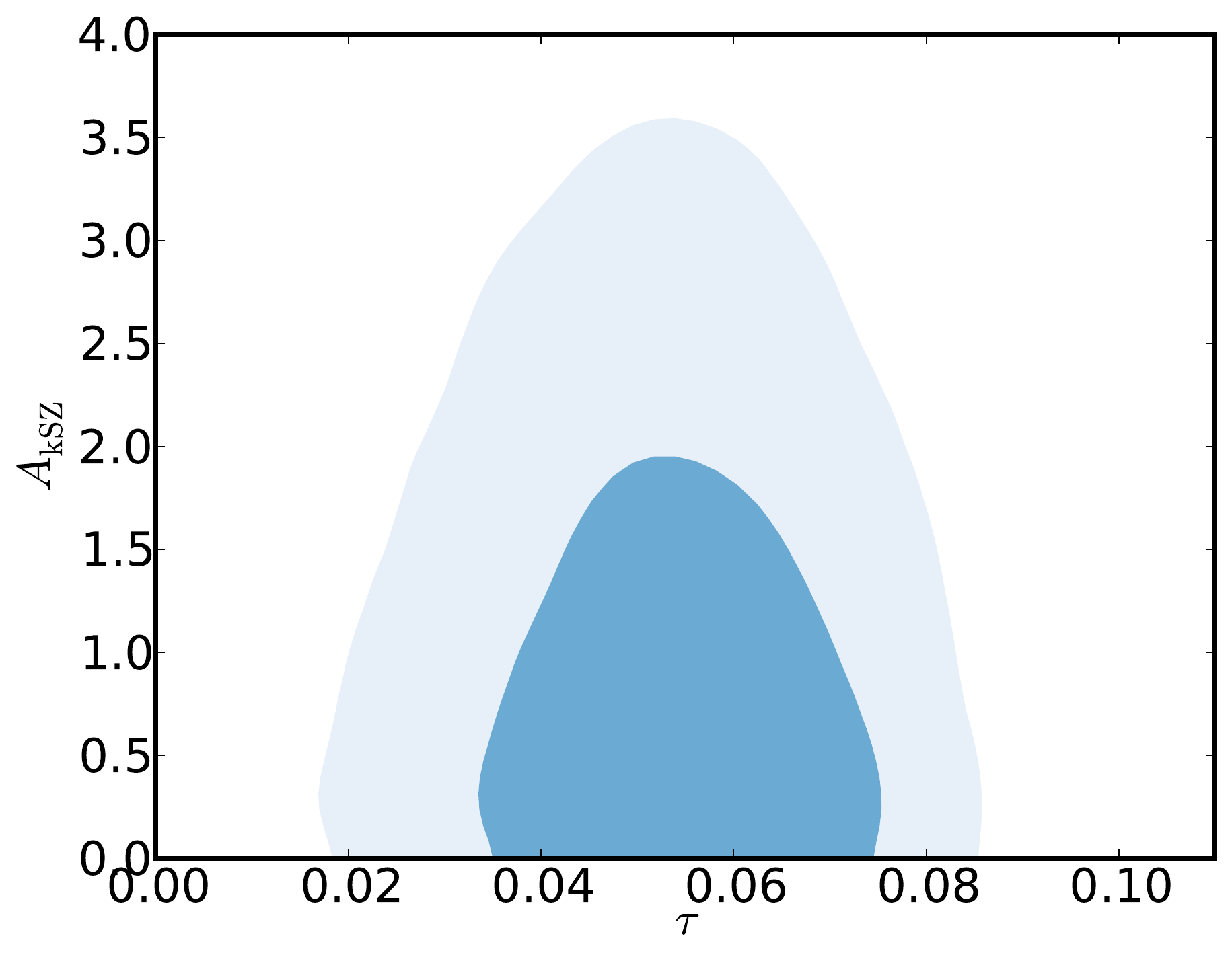}
  \caption{68\,\% and 95\,\% confidence intervals on the reionization
    optical depth, $\tau$, and the amplitude of the kinetic SZ effect,
    $A_{\rm kSZ}$, from the CMB (\lollipop+\planckTT+\VHL).}
  \label{fig:tau_Aksz}
\end{figure}

We combine the \Planck\ likelihoods in $TT$ (\planckTT) and from \lowl\ $EE$ polarization (\lollipop) with the very high-$\ell$ data from ACT and SPT (\VHL), assuming a redshift-symmetric parameterization of the reionization. Figure~\ref{fig:tau_Aksz} shows the 2D posterior distribution for $\tau$ and $A_{\rm kSZ}$ after marginalization over the other cosmological and nuisance parameters.

\begin{figure}[htbp!]
  \centering
  \includegraphics[width=\columnwidth]{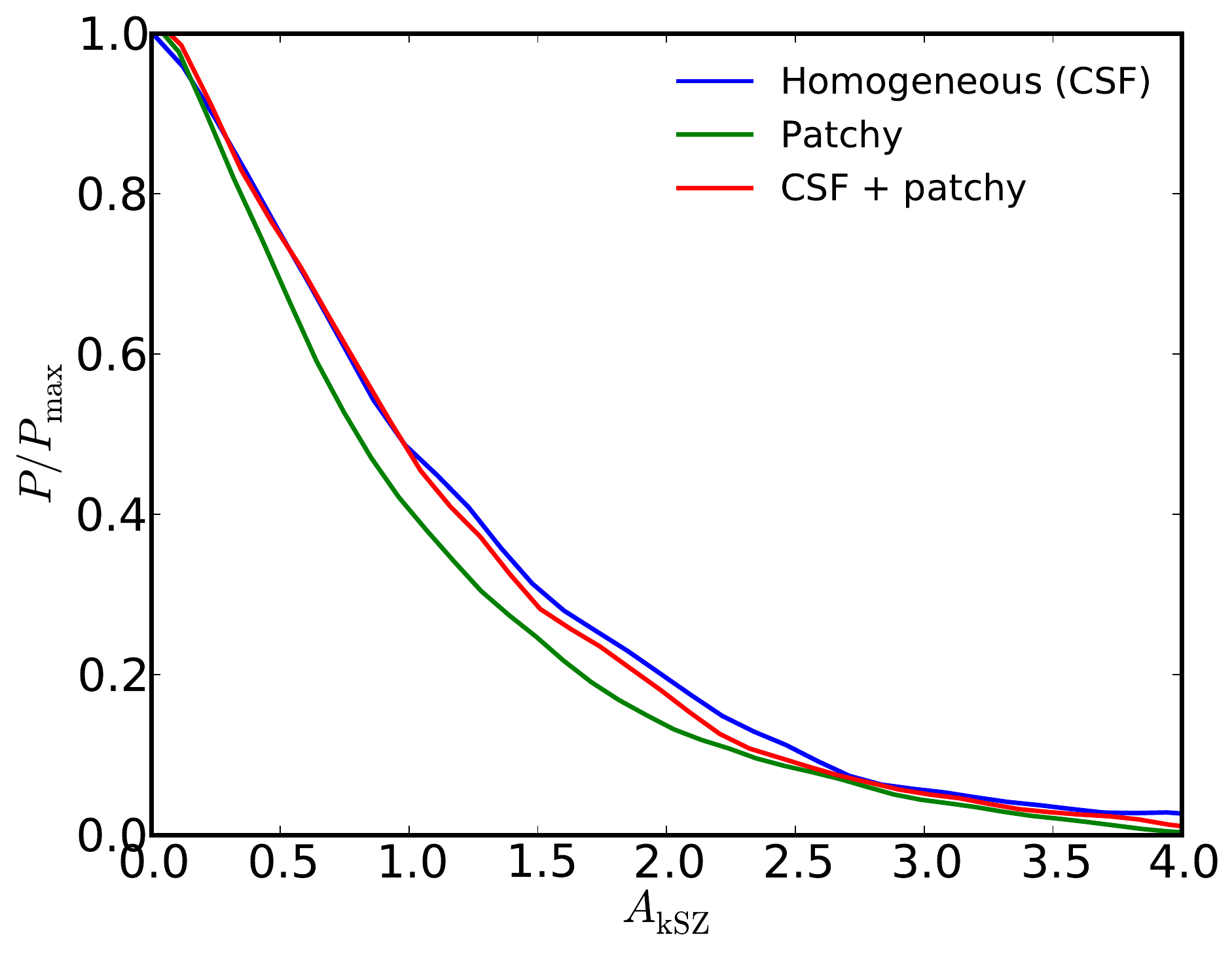}
  \caption{Constraints on the kSZ amplitude at $\ell=3000$ using
    \lollipop+\planckTT+\VHL\ likelihoods. The three cases correspond
    to different kSZ templates.}
  \label{fig:ksz_norm}
\end{figure}

Figure~\ref{fig:ksz_norm} compares the constraints on the kSZ power at $\ell=3000$, $A_{\rm kSZ}$, obtained for three different kSZ templates: the ``homogeneous'' reionization template from \citet{trac11}, which neglects contributions from inhomogeneous reionization; a more complex model ``CSF \& patchy,'' including both homogeneous and patchy contributions; and a pure ``patchy'' template from \citet{Battaglia13}. We find very similar upper limits on $A_{\rm kSZ}$, even in the case of pure patchy reionization.

Using the ``CSF \& patchy'' model, the upper limit is
\begin{equation}
  A_{\rm kSZ} < 2.6 \, \mu{\rm K}^2\quad \mbox{(95\,\% CL)} \, .
\end{equation}
Compared to \Planck~2013 results, the maximum likelihood value $A_{\rm kSZ} = 5.3^{+2.8}_{-1.9}\,\muK^2$
\citep[\planckTT+WP+highL,][]{planck2013-p11} is reduced to an upper
limit in this new analysis.  The data presented here provide the best
constraint to date on the kSZ power and is a factor of 2 lower than
the limit reported in \citet{george15}. Our limit is certainly not in
tension with the homogeneous kSZ template, which predicts $A_{\rm kSZ}
= 1.79\,\muK^2$.  However, it does not leave much room for any
additional kSZ power coming from patchy reionization.

Consistent with \citet{george15}, we find the total kSZ power to be
stable against varying tSZ and CIB templates. We also find very little
dependence on the choice of the kSZ template
(Fig.~\ref{fig:ksz_norm}). This confirms that there is only a modest
amount of information in the angular shape of the kSZ signal with the
current data.

\section{Constraints on the reionization history\label{sec:cmbresults}}

We now interpret our measurements of the reionization observables in terms of constraint on the reionization history. We mainly focus on the determination of the reionization redshift $\zre$ and its duration $\Delta z = z_{\rm beg}-z_{\rm end}$. We show only the results for $\Delta z$ greater than unity, which corresponds to approximatively 90\,Myr at redshift $z=8$. We first begin by looking at constraints on the EoR for symmetric and asymmetric models using \Planck\ data only (\lollipop+\planckTT). Then we introduce the \VHL\ data and discuss additional constraints from the kSZ amplitude. In each case, we also derive the constraints that follow from postulating that reionization should be completed at a redshift of 6 (see Sect.~\ref{sec:data}), i.e., when imposing the prior $z_{\rm end} > 6$.

\subsection{Redshift-symmetric parameterization}
\label{sec:sym_results}

\begin{figure}[htbp!]
  \centering
  \includegraphics[width=\columnwidth]{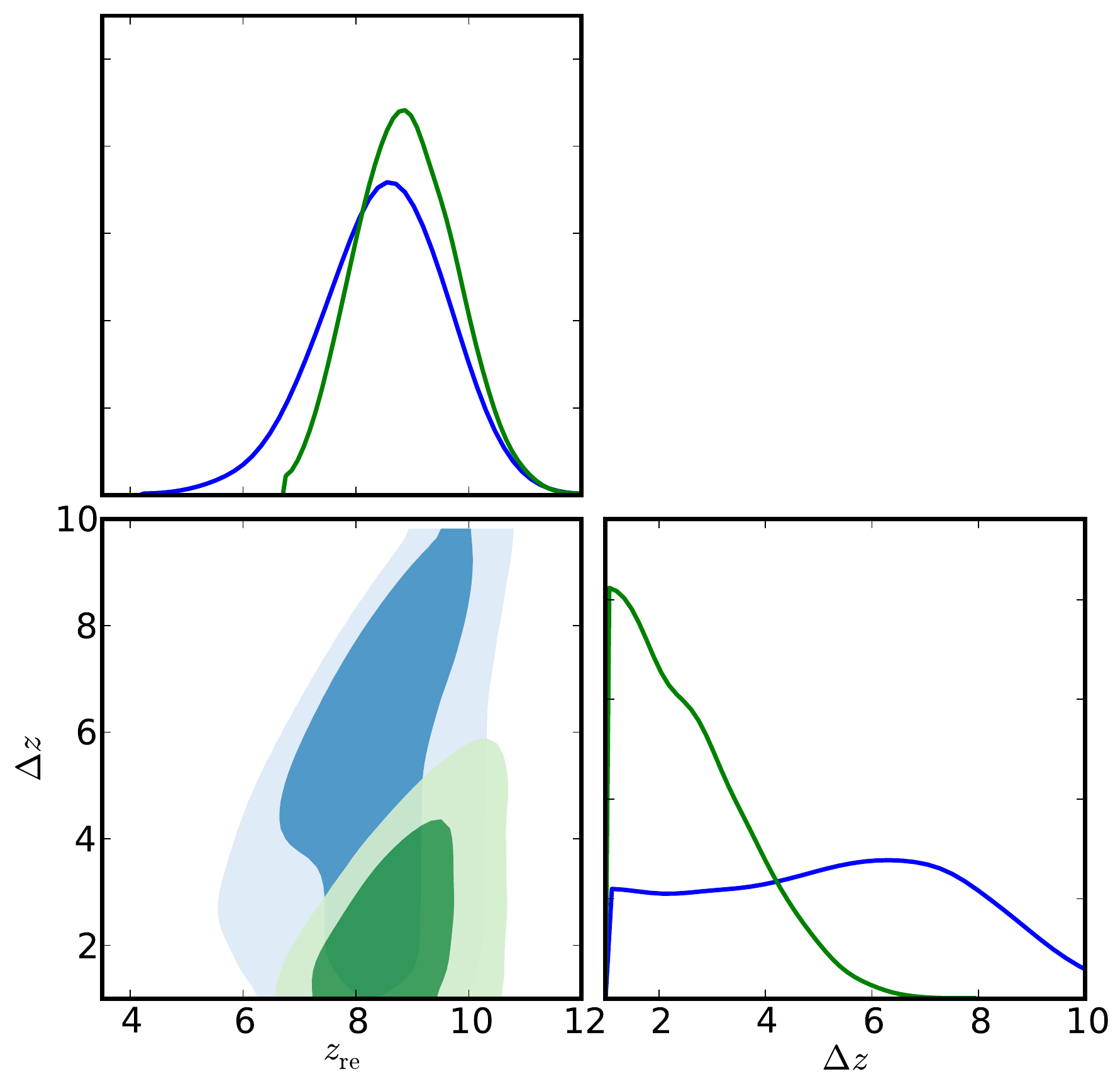}
  \caption{Posterior distributions (in blue) of $\zre$ and $\Delta z$
    for a redshift-symmetric parameterization using the CMB
    likelihoods in polarization and temperature (\lollipop+\planckTT).
    The green contours and lines show the distribution after imposing
    the additional prior $z_{\rm end} > 6$.}
  \label{fig:triangle_sym}
\end{figure}

We use the \Planck\ CMB likelihoods in temperature (\planckTT) and
polarization (\lollipop) to derive constraints on \lcdm\ parameters,
including the reionization redshift $z_{\rm re}$ and width $\Delta z$
for a redshift-symmetric parameterization.
Figure~\ref{fig:triangle_sym} shows (in blue) the posterior on $z_{\rm
  re}$ and $\Delta z$ after marginalization over the other
cosmological and nuisance parameters. As discussed in
Sect.~\ref{sec:model}, the large-scale polarized CMB anisotropies are
almost insensitive to the width $\delta z$ of the tanh function. We
thus recover the degeneracy in the direction of $\Delta z$.  Imposing
an additional Gunn-Peterson constraint on the ionization fraction at
very low redshift can break this degeneracy. This is illustrated in
Fig.~\ref{fig:triangle_sym}, where we show (in green) the results of
the same analysis with an additional prior $z_{\rm end} > 6$.  In this
case, we find $\delta z < 1.3$ at 95\,\% CL, which corresponds to a
reionization duration ($z_{\rm beg}-z_{\rm end}$) of
\begin{equation}
  \Delta z < 4.6\quad \mbox{(95\,\% CL).}
\end{equation}

The posterior distribution of $z_{\rm re}$ is shown in Fig.~\ref{fig:triangle_sym} after marginalizing over $\Delta z$, with and without the additional constraint $z_{\rm end} > 6$. This suggests that the reionization process occurred at redshift
\begin{align}
  \zre =&\, 8.5_{-1.1}^{+1.0}\quad \mbox{(uniform prior)\,,} \\
  \zre =&\, 8.8_{-0.9}^{+0.9}\quad \mbox{(prior $z_{\rm end}>6$)}\, .
  \label{eq:zre_sym_z6}
\end{align}
This redshift is lower than the values derived previously from WMAP-9 data, in combination with ACT and SPT \citep{hinshaw2012}, namely $\zre=10.3\pm1.1$.  It is also lower than the value $\zre=11.1\pm1.1$ derived in \citet{planck2013-p11}, based on \Planck~2013 data and the WMAP-9 polarization likelihood.

Although the uncertainty is now smaller, this new reionization redshift value is entirely consistent with the \Planck~2015 results \citep{planck2014-a15} for \planckTT+lowP alone, $\zre=9.9^{+1.8}_{-1.6}$ or in combination with other data sets, $\zre=8.8^{+1.3}_{-1.2}$ (specifically for \planckTT+lowP+lensing+BAO)
estimated with $\delta z$ fixed to 0.5. The constraint from \lollipop+\planckTT\ when fixing $\delta z$ to 0.5 is $\zre=8.2_{-1.2}^{+1.0}$. This slightly lower value (compared to the one obtained when letting the reionization width be free) is explained by the shape of the degeneracy surface. Allowing for larger duration
when keeping the same value of $\tau$ pushes towards higher reionization redshifts; marginalizing over $\Delta z$ thus shifts the posterior distribution to slightly larger $\zre$ values.

\begin{figure}[htbp!]
  \centering
  \includegraphics[width=\columnwidth]{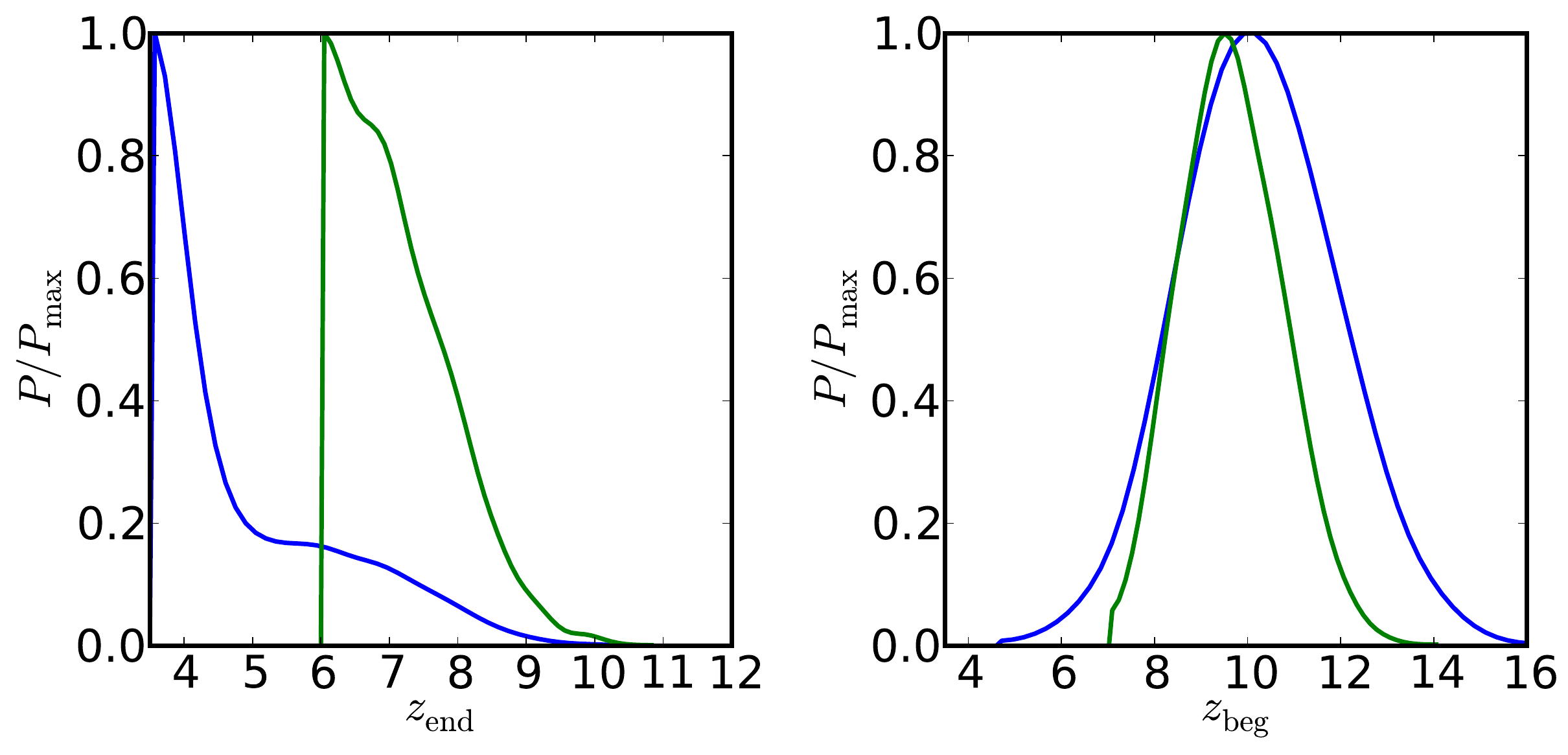}
  \caption{Posterior distributions on the end and beginning of
    reionization, i.e., $z_{\rm end}$ and $z_{\rm beg}$, using the
    redshift-symmetric parameterization without (blue) and with
    (green) the prior $z_{\rm end}>6$.}
  \label{fig:dist_sym}
\end{figure}

In addition to the posteriors for $\zre$ and $\delta z$ using the redshift-symmetric parameterization, the distributions of the end and beginning of reionization, $z_{\rm end}$ (i.e., $z_{99\,\%}$) and $z_{\rm beg}$ (i.e., $z_{10\,\%}$), are plotted in Fig.~\ref{fig:dist_sym}. In such a model, the end of reionization
strongly depends on the constraint at low redshift. On the other hand, the constraints on $z_{\rm beg}$ depend only slightly on the low-redshift prior. These results show that the Universe is ionized at less than the 10\,\% level above $z=9.4\pm1.2$.

\subsection{Redshift-asymmetric parameterization}
\label{sec:asym_results}
We now explore more complex reionization histories using the redshift-asymmetric parameterization of $x_{\rm e}(z)$ described in Sect.~\ref{sec:model}.  In the same manner as in Sect.~\ref{sec:sym_results}, also examine the effect of imposing the additional constraint from the Gunn-Peterson effect.

\begin{figure}[htbp!]
  \centering
  \includegraphics[width=\columnwidth]{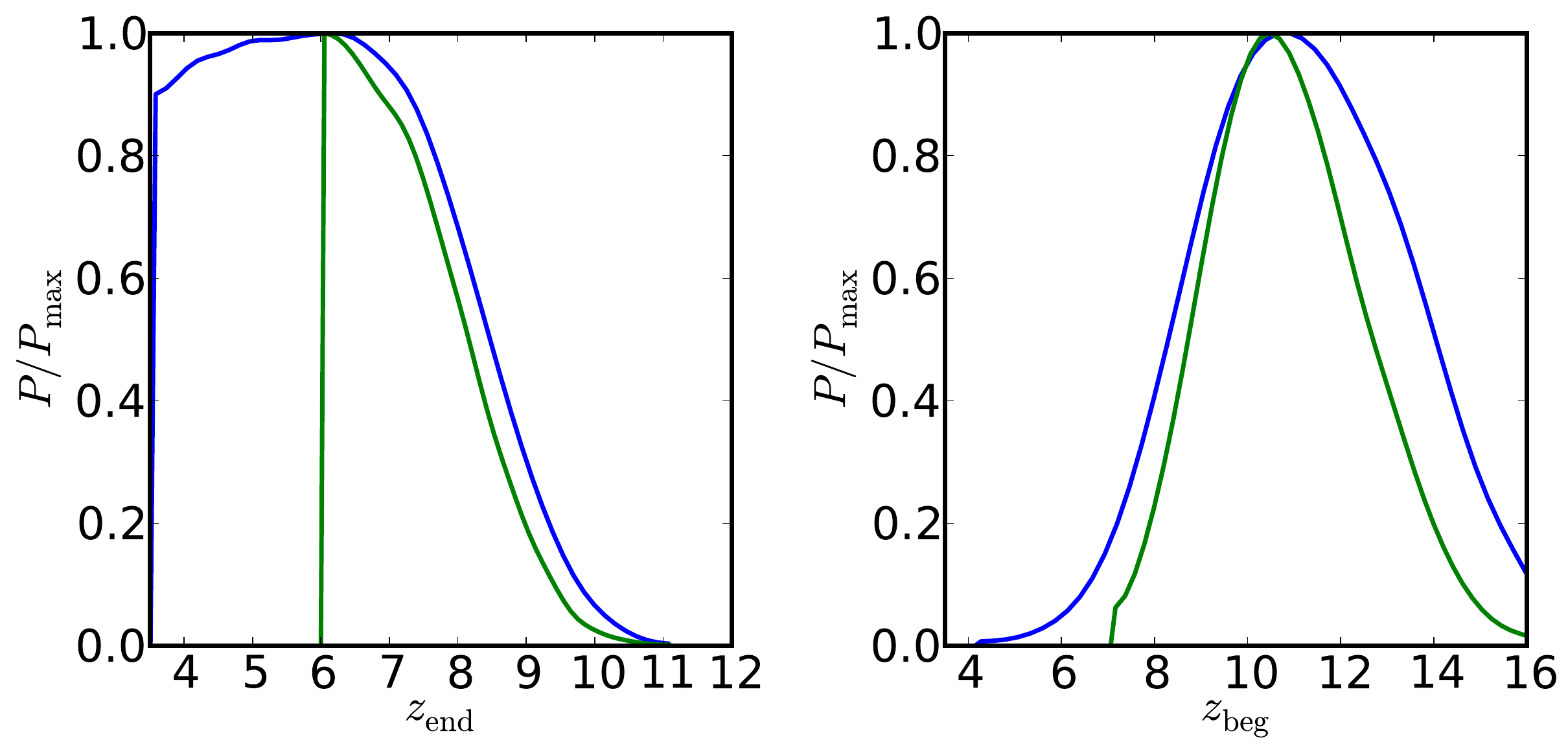}
  \caption{Posterior distributions of $z_{\rm end}$ and $z_{\rm beg}$
    using the redshift-asymmetric parameterization without (blue) and
    with (green) the prior $z_{\rm end}>6$.}
  \label{fig:dist_asym}
\end{figure}

The distributions of the two parameters, $z_{\rm end}$ and $z_{\rm beg}$, are plotted in Fig.~\ref{fig:dist_asym}. With the redshift-asymmetric parameterization, we obtain $z_{\rm beg} = 10.4_{-1.6}^{+1.9}$ (imposing the prior on $z_{\rm end}$), which disfavours any major contribution to the ionized fraction from sources
that could form as early as $z\ga15$.

\begin{figure}[htbp!]
  \centering
  \includegraphics[width=\columnwidth]{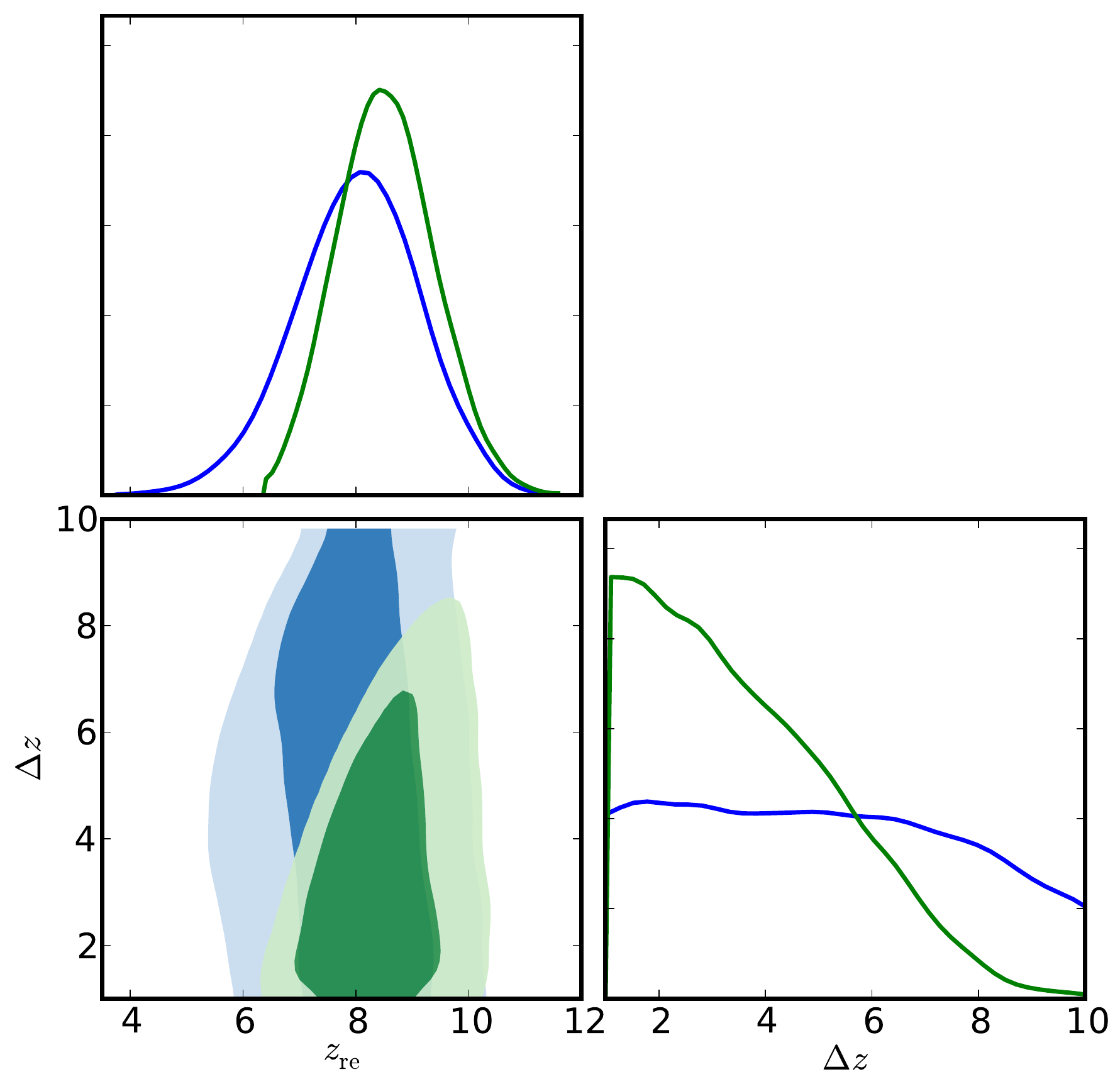}
  \caption{Posterior distributions for $\zre$ and $\Delta z$ using the
    redshift-asymmetric parameterization without (blue) and with
    (green) the prior $z_{\rm end}>6$.}
  \label{fig:triangle_asym}
\end{figure}

In Fig.~\ref{fig:triangle_asym}, we interpret the results in terms of reionization redshift and duration of the EoR, finding
\begin{align}
  \zre =\,& 8.0_{-1.1}^{+0.9}\quad \mbox{(uniform prior)}\, , \\
  \zre =\,& 8.5_{-0.9}^{+0.9}\quad \mbox{(prior $z_{\rm end}>6$)}\, .
\end{align}
These values are within $0.4\,\sigma$ of the results for the redshift-symmetric model.  For the duration of the EoR, the upper limits on $\Delta z$ are
\begin{align}
  \Delta z <\,& 10.2\quad \mbox{(95\,\% CL, unform prior)} \, ,\\
  \Delta z <\,& \phantom{0}6.8\quad \mbox{(95\,\% CL, prior $z_{\rm
      end}>6$)}\, .
\end{align}

\subsection{Combination with the kSZ effect}
\label{sec:ksz_results}
In order to try to obtain better constraints on the reionization width, we now make use of the additional information coming from the amplitude of the kinetic SZ effect. Since \Planck\ alone is not able to provide accurate limits on the kSZ amplitude, we combine the \Planck\ likelihoods in temperature and polarization with the measurements of the CMB $TT$ power spectrum at high-resolution from the ACT and SPT experiments, ``\VHL.''

Using the redshift-symmetric model, when adding the \VHL\ data, we recover essentially the same results as in Sect.~\ref{sec:sym_results}. The reionization redshift is slightly lower, as suggested by the results on $\tau$ (see Eq.~\ref{tau:plikTT_Comm_lol_vhl} and the discussion in Sect.~\ref{sec:cmb_model}). We also see the same degeneracy along the $\Delta z$ direction.

With the addition of kSZ information, we are able to break the degeneracy with $\Delta z$. This might allow us to determine how much kSZ power originated during reionization (i.e., patchy kSZ) and how much at later times, when the Universe became fully ionized (i.e., homogeneous kSZ). We use the templates from \cite{Shaw12} and \cite{Battaglia13} for the homogeneous and patchy kSZ contributions, respectively, with the dependency on \lcdm\ cosmological parameters as
described in Sect.~\ref{sec:ksz_model}. Those specific relations rely on a redshift-symmetric model for the description of the EoR. Note, however, that the results presented here are derived from specific simulations of the reionization process, and so explicit scalings need to be assumed, as discussed by \citet{zahn12} and \citet{george15}.

\begin{figure}[htbp!]
  \centering
  \includegraphics[width=\columnwidth]{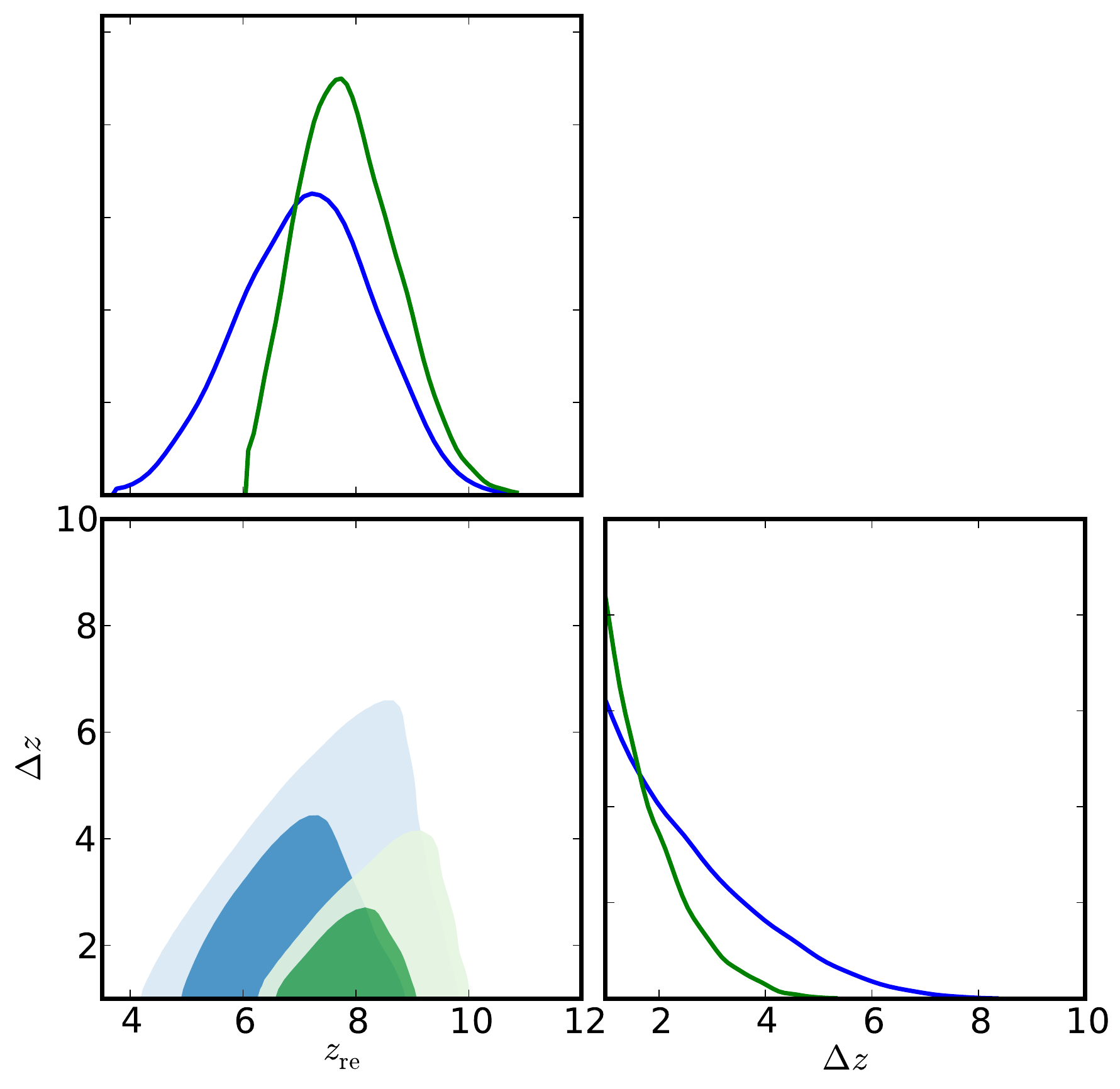}
  \caption{Posterior distributions on the duration $\Delta z$ and the redshift $\zre$ of reionization from the combination of CMB polarization and kSZ effect constraints using the redshift-symmetric parameterization without (blue) and with (green) the prior $z_{\rm end}>6$.}
  \label{fig:triangle_ksz}
\end{figure}

As described in Sect.~\ref{sec:ksz_model}, the amplitude of the kSZ power primarily depends on the duration of reionization, while the epoch is essentially constrained by the optical depth. Using the 2D distribution for $\tau$ and $A_{\rm kSZ}$, as measured by \Planck\ in combination with very high-$\ell$ temperature data (Fig.~\ref{fig:tau_Aksz}), we derive a 2D likelihood function for $z_{\rm re}$ and $\Delta z$.  We can then sample the reionization
parameters (the epoch $z_{\rm re}$ and duration $\Delta z$ of the EoR), compute the associated optical depth and kSZ power and derive constraints based on the 2D likelihood.  The allowed models in terms of $z_{\rm re}$ and $\Delta z$ are shown in Fig.~\ref{fig:triangle_ksz} (in blue).  We also plot (in green) the
same constraints with the additional prior $z_{\rm end}>6$.

As discussed in Sect.~\ref{sec:ksz_model}, the measurement of the total kSZ power constrains the amplitude of patchy reionization, resulting in an upper limit of
\begin{align}
  \Delta z <&\, 4.8\quad \mbox{(95\,\% CL, uniform prior)} \, ,\\
  \Delta z <&\, 2.8\quad \mbox{(95\,\% CL, prior $z_{\rm end}>6$)} \, .
\end{align}
This is compatible with the constraints from \citet{george15}, where an upper limit was quoted of $z_{20\,\%} - z_{99\,\%} < 5.4$ at 95\,\% CL. Our 95\,\% CL upper limits on this same quantity are $4.3$ and $2.5$ without and with the prior on $z_{\rm end}$, respectively.

For the reionization redshift, we find
\begin{align}
  \zre =\,& 7.2_{-1.2}^{+1.2}\quad \mbox{(uniform prior)} \, , \\
  \zre =\,& 7.8_{-0.8}^{+1.0}\quad \mbox{(prior $z_{\rm end}>6$)} \, ,
\end{align}
which is compatible within $1\,\sigma$ with the results from CMB
\Planck\ data alone without the kSZ constraint
(Sect.~\ref{sec:sym_results}).

\begin{figure}[htbp!]
  \centering
  \includegraphics[width=\columnwidth]{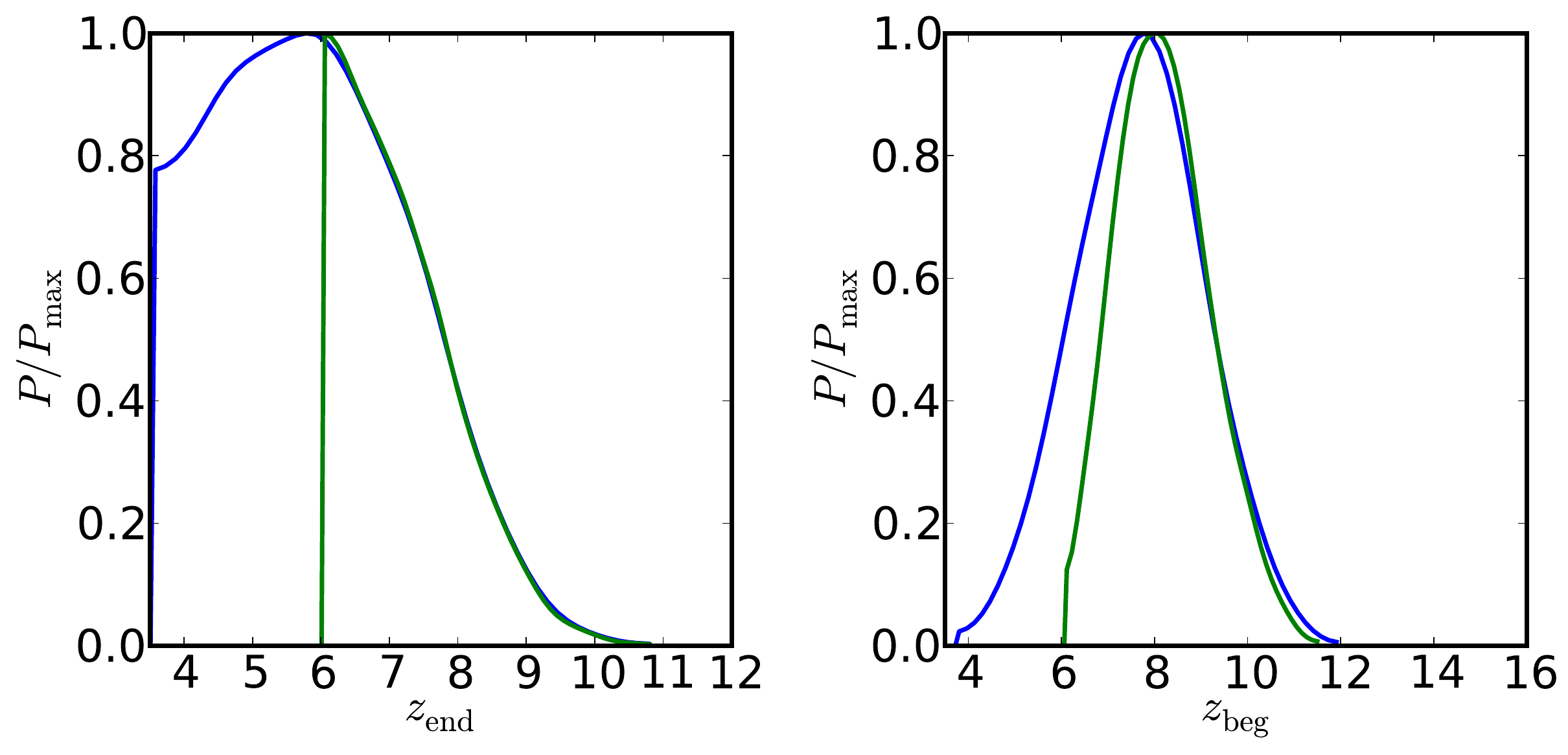}
  \caption{Posterior distributions of $z_{\rm end}$ and $z_{\rm beg}$
    using the redshift-symmetric parameterization, combining \Planck\
    and \VHL\ data, and using information from the kSZ amplitude,
    without (blue) and with (green) the prior $z_{\rm end}>6$.}
  \label{fig:dist_ksz}
\end{figure}

The distributions of $z_{\rm end}$ and $z_{\rm beg}$ are plotted in
Fig.~\ref{fig:dist_ksz}. Within the redshift-symmetric
parameterization, we obtain $z_{\rm beg} = 8.1_{-0.9}^{+1.1}$ (with
the prior on $z_{\rm end}$).

Adding information from the kSZ amplitude allows for somewhat tighter
constraints to be placed on the reionization duration $\Delta z$ and
the beginning of reionization (corresponding to the 10\,\% ionization
limit) $z_{\rm beg}$. However, as discussed in
Sect.~\ref{sec:ksz_model}, those results are very sensitive to details
of the simulations used to predict both the shape and the parameter
dependences of the kSZ template in the different reionization
scenarios (patchy or homogeneous).

\section{Discussion} \label{sec:xe}

The CMB has long held the promise of measuring the Thomson optical depth in
order to derive constraints on the reionization history of the Universe.
Despite its importance, this constraint is fundamentally limited by cosmic
variance in polarization and is further challenged by foregrounds and
systematic effects.  The first results, from \WMAP, gave $\tau=0.17\pm0.04$,
suggesting a reionization redshift between 11 and 30 \citep{kogut2003}.  This
was revised in the final 9-year \WMAP\ results to a central value of
$\tau=0.084$ \citep{hinshaw2012}, which, in the instantaneous reionization
model, implies $\zre = 10.4$. However, with the context of the same model, the
\Planck~2015 results \citep{planck2014-a15}, either alone
($\zre=9.9^{+1.8}_{-1.6}$) or in combination with other data sets
($\zre=8.8^{+1.3}_{-1.2}$), showed that the reionization redshift was smaller.
The main result we present here, $\zre= 8.2^{+1.0}_{-1.2}$, further confirms
that reionization occurred rather late, leaving little room for any significant
ionization at $z\ga15$.  This is consistent with what is suggested by other
reionization probes, which we now discuss \citep[for reviews, see
e.g.,][]{becker15b, mcquinn15}.

The transition from neutral to ionized gas is constrained by absorption
spectra of very distant quasars and gamma ray bursts (GRBs), revealing neutral
hydrogen in intergalactic clouds. They show, through the Gunn-Peterson effect,
that the diffuse gas in the Universe is mostly ionized up to a redshift of
about 6 \citep{fan06a}. Given the decline in their abundance beyond redshift
$z\simeq6$, quasars and other active galactic nuclei (AGN) cannot be major
contributors to the early stages of reionization \citep[e.g.,][but see
\citeauthor{madau15} \citeyear{madau15}; \citeauthor{khaire2016} \citeyear{khaire2016}, for alternative AGN-only
models]{willott2010,fontanot2012}. A faint AGN population can produce
significant photoionization rates at redshifts of 4--6.5, consistent with the
observed highly ionized IGM in the Ly-$\alpha$ forest of high-$z$ quasar
spectra \citep{gia15}. Star-forming galaxies at redshifts $z\ga6$ have
therefore been postulated to be the most likely sources of early reionization,
and their time-dependent abundance and spectral properties are crucial
ingredients for understanding how intergalactic hydrogen ceased to be neutral
\citep[for reviews, see][]{Barkana01,fan06a,robertson2010,mcquinn15}. The
luminosity function of early star-forming galaxies, in particular in the UV
domain, is thus an additional and powerful probe of the reionization history
\citep[e.g.,][]{kuh12,Robertson13,Rob15, Bou15}. Based on comparison of the
9-year WMAP results to optical depth values inferred from the UV luminosity
function of high-$z$ galaxies, it has been suggested that either the UV
luminosity density flattens, or physical parameters such as the escape fraction
and the clumping factor evolved significantly, or alternatively, additional,
undetected sources (such as X-ray binaries and faint AGN) must have existed
at $z \ga11$ \citep[e.g.,][]{kuh12,ellis2013,cai14,Ishi15}.

\begin{figure}[htbp!]
  \centering
  \includegraphics[width=\columnwidth]{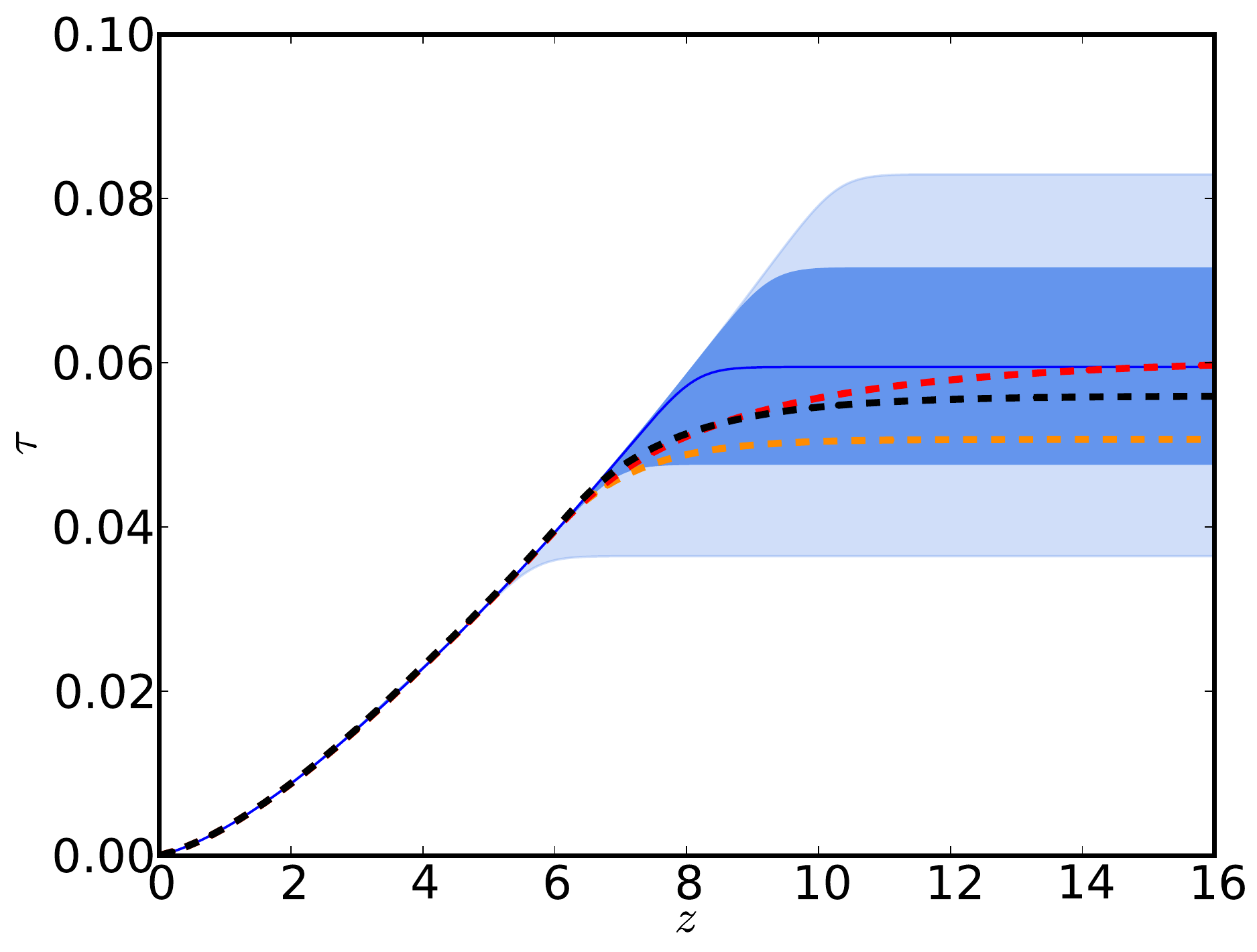}
  \caption{Evolution of the integrated optical depth for the tanh
    functional form (with $\delta z=0.5$, blue shaded area). The two
    envelopes mark the 68\,\% and 95\,\% confidence intervals. The
    red, black, and orange dashed lines are the models from
    \citet{Bou15}, \citet{Rob15}, and \citet{Ishi15}, respectively,
    using high-redshift galaxy UV and IR fluxes and/or direct
    measurements.}
  \label{fig:rob}
\end{figure}

The \Planck\ results, both from the 2015 data release and those presented here, strongly reduce the need for a significant contribution of Lyman continuum emission at early times. Indeed, as shown in Fig.~\ref{fig:rob}, the present CMB results on the Thomson optical depth, $\tau = \thetau$, are perfectly consistent with the best models of star-formation rate densities derived from the UV and IR luminosity functions, as directly estimated from observations of high-redshift galaxies \citep{Ishi15, Rob15, Bou15}. With the present value of $\tau$, if we maintain a UV-luminosity density at the maximum level allowed by the luminosity density constraints at redshifts $z<9$, then the currently observed galaxy population at $\MUV<-17$ seems to be sufficient to comply with all the observational constraints without the need for high-redshift ($z=10$--15) galaxies.

\begin{figure}[htbp!]
  \centering
  \includegraphics[width=\columnwidth]{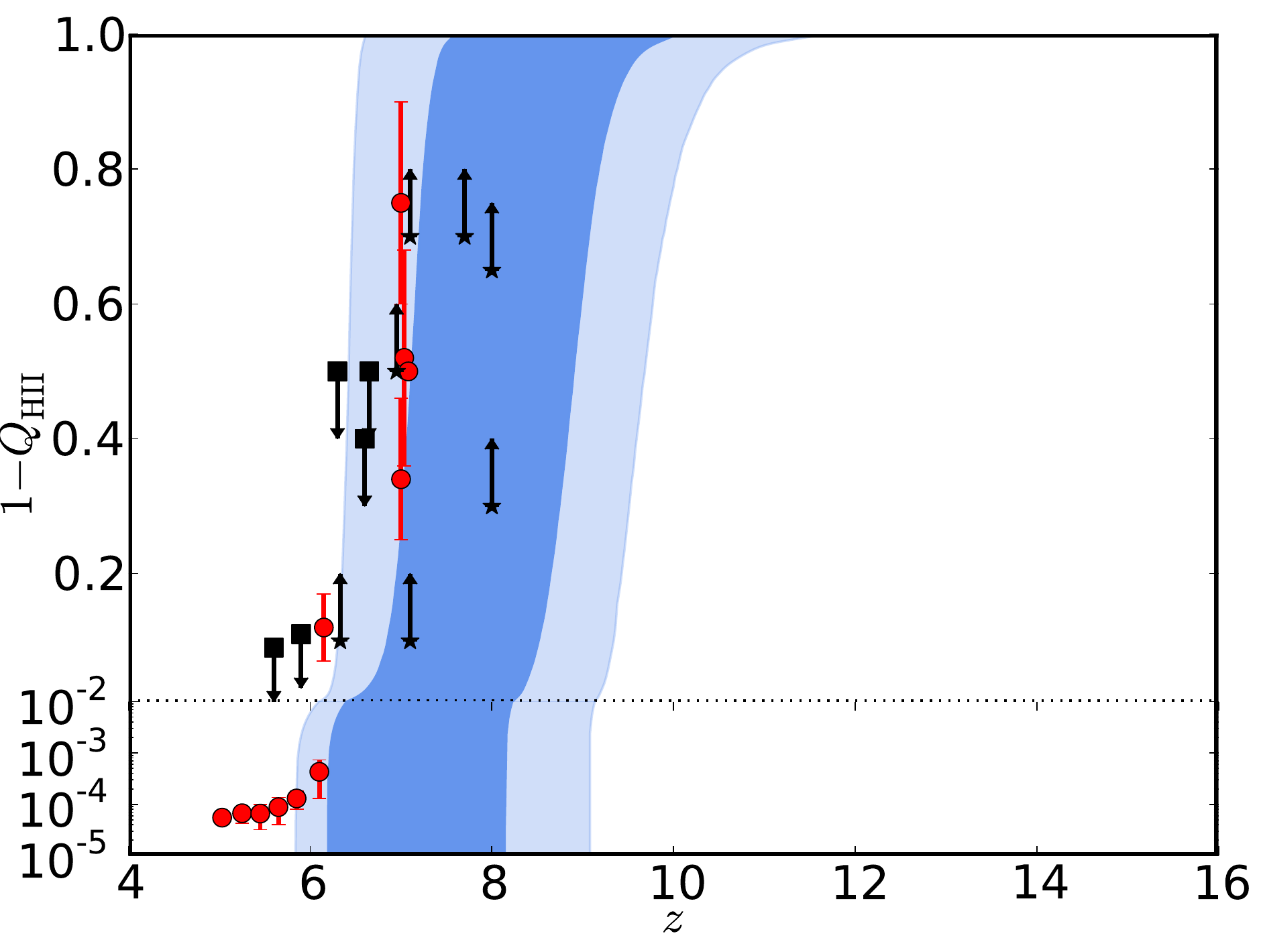}
  \caption{Reionization history for the redshift-symmetric
    parameterization compared with other observational constraints coming from quasars, Ly-$\alpha$ emitters, and the Ly-$\alpha$ forest
    \citep[compiled by][]{Bou15}. The red points are measurements of
    ionized fraction, while black arrows mark upper and lower limits.
    The dark and light blue shaded areas show the 68\,\% and 95\,\%
    allowed intervals, respectively.}
  \label{fig:qhii}
\end{figure}

The \Planck\ data are certainly consistent with a fully reionized
Universe at $z\simeq6$. Moreover, they seem to be in good agreement
with recent observational constraints on reionization in the direction
of particular objects. The \ion{H}{i} absorption along the line of
sight to a distant $\gamma$-ray burst, GRB-140515A \citep{Chornock14},
suggests a Universe containing about a 10\,\% fraction of neutral
hydrogen at $z=6$--6.3. At even higher redshifts $z\simeq 7$,
observation of Ly-$\alpha$ emitters suggests that at least 70\,\% of
the IGM is neutral \citep{til14, schenker14, Faisst14}.  Similarly,
quasar near-zone detection and analysis (including sizes, and
Ly-$\alpha$ and $\beta$ transmission properties) have been used to
place constraints on $z_{\rm end}$ from signatures of the ionization
state of the IGM around individual sources
\citep{wyithe04,mesinger2004,wyithe05,mesinger2007,
  carilli2010,mortlock11,schroeder2013}.  However, interpretation of
the observed evolution of the near-zone sizes may be complicated by
the opacity caused by absorption systems within the ionized IGM
\citep[e.g.,][]{bolton2011, bolton2013, becker15b}. Similarly, it is
difficult to completely exclude the possibility that damped
Ly-$\alpha$ systems contribute to the damping wings of quasar spectra
blueward of the Ly-$\alpha$ line
\citep[e.g.,][]{mesinger2008,schroeder2013}. Nevertheless, most such
studies, indicate that the IGM is significantly neutral at redshifts
between 6 and 7 \citep[see also][]{keating2015}, in agreement with the
current \Planck\ results, as shown in Fig.~\ref{fig:qhii}.

\begin{figure*}[!ht]
  \centering
  \includegraphics[width=\textwidth]{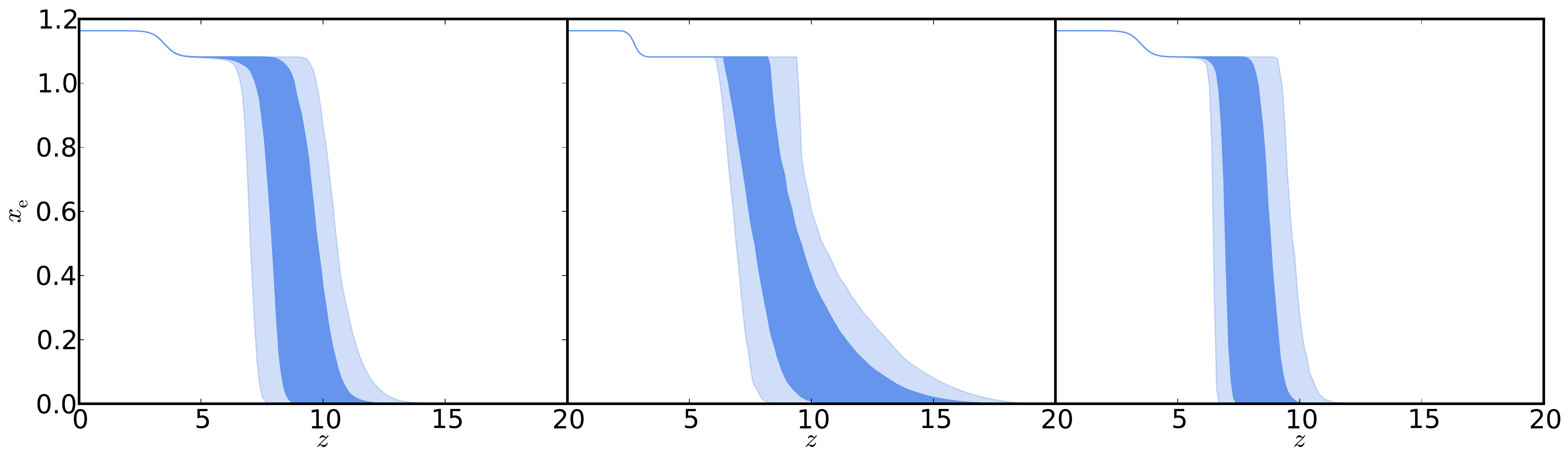}
  \caption{Constraints on ionization fraction during reionization. The
    allowed models, in terms of $\zre$ and $\Delta z$, translate into
    an allowed region in $x_{\rm e}(z)$ (68\,\% and 95\,\% in dark
    blue and light blue, respectively), including the $z_{\rm end}>6$
    prior here. {\it Left}: Constraints from CMB data using a
    redshift-symmetric function ($x_{\rm e}(z)$ as a hyperbolic
    tangent with $\delta z=0.5$). {\it Centre}: Constraints from CMB
    data using a redshift-asymmetric parameterization ($x_{\rm e}(z)$
    as a power law).  {\it Right}: Constraints from CMB data using a
    redshift-symmetric parameterization with additional constraints
    from the kSZ effect.}
  \label{fig:xe_constraints}
\end{figure*}

Although there are already all the constraints described above,
understanding the formation of the first luminous sources in the
Universe is still very much a work in progress.  Our new (and lower)
value of the optical depth leads to better agreement between the CMB
and other astrophysical probes of reionization; however, the
fundamental questions remain regarding how reionization actually
proceeded.


\section{Conclusions}
We have derived constraints on cosmic reionization using \Planck\ data. The CMB \Planck\ power spectra, combining the $EE$ polarization at \lowl\ with the temperature data, give, for a so-called ``instantaneous'' reionization history (a redshift-symmetric tanh function $x_{\rm e}(z)$ with $\delta z = 0.5$), a measurement of the Thomson optical depth
\begin{equation}
  \tau = \thetau\quad \mbox{(\lollipop+\planckTT),}
\end{equation}
which is significantly more accurate than previous measurements.
Thanks to the relatively high signal-to-noise ratio of the \lowl\
polarization signal, the combination with lensing or data from high-resolution CMB anisotropy experiments (ACT and SPT) does not bring
much additional constraining power.  The impact on other \lcdm\
parameters is only significant for the amplitude of the initial scalar
power spectrum $A_{\rm s}$ and (to a lesser extent) on its tilt
$n_{\rm s}$.  Other parameters are very stable compared to the
\Planck\ 2015 results.

Using \Planck\ data, we have derived constraints on two models for the
reionization history $x_{\rm e}(z)$ that are commonly used in the
literature: a redshift-symmetric form using a hyperbolic tangent
transition function; and a redshift-asymmetric form parameterized by a
power law.  We have also investigated the effect of imposing the
condition that the reionization is completed by $z=6$.

Allowing the ionization fraction shape and duration to vary, we have
found very compatible best-fit estimates for the optical depth (0.059
and 0.060 for the symmetric and asymmetric model, respectively),
showing that the CMB is indeed more sensitive to the value of the
optical depth than to the exact shape of the reionization history.
However, the value of the reionization redshift does slightly depend
on the model considered.  In the case of a symmetric parameterization,
we have found slightly larger estimates of $\zre$ than in the case of
instantaneous reionization. This can be understood through the shape
of the degeneracy surface between the reionization parameters.  For an
asymmetric parameterization, $\zre$ is smaller, due to the fact that
$x_{\rm e}(z)$ changes more rapidly at the end of reionization than
the beginning.  We specifically find:
\begin{align}
  \zre = 8.8\pm 0.9 & \quad \mbox{(redshift-symmetric)}\,, \\
  \zre = 8.5\pm 0.9 & \quad \mbox{(redshift-asymmetric)\, .}
\end{align}

Assuming two different parameterizations of the reionization history shows how much results on effective parameters (like the redshift of reionization or its duration) are sensitive to the assumption of the reionization history shape. The best models of symmetric and asymmetric parameterization give similar values for $\tau$, and provide reionization redshifts which differ by less than $0.4\,\sigma$. Constraints on the limits of possible early reionization are similar, leading to 10\,\% reionization levels at around $z=10$.

To derive constraints on the duration of the reionization epoch, we
combined CMB data with measurements of the amplitude of the kSZ
effect. In the case of a redshift-symmetric model, we found
\begin{equation}
  \Delta z < 2.8\quad \mbox{(95\,\% CL)},
\end{equation}
using the additional constraint that the Universe is entirely
reionized at redshift 6 (i.e., $z_{\rm end}>6$).

\begin{table}[!b]
\begingroup
\newdimen\tblskip \tblskip=5pt
\caption{Constraints on reionization parameters for the different models presented in this paper when including the $z_{\rm end}>6$ prior. We show 68\% limit for $\zre$ and $z_{\rm beg}$, while we quote 95\% upper limit for $\Delta z$ and $z_{\rm end}$.}
\label{tab:constraints}
\nointerlineskip
\vskip -3mm
\footnotesize
\setbox\tablebox=\vbox{
   \newdimen\digitwidth 
   \setbox0=\hbox{\rm 0} 
   \digitwidth=\wd0 
   \catcode`*=\active 
   \def*{\kern\digitwidth}
   \newdimen\signwidth 
   \setbox0=\hbox{-} 
   \signwidth=\wd0 
   \catcode`!=\active 
   \def!{\kern\signwidth}
%
\halign{\hbox to 3cm{#\leaderfil}\tabskip 1em&\hfil# \tabskip 1em\hfil&\hfil# \tabskip 1em&\hfil# \tabskip 1em&\hfil# \tabskip 0pt\cr
\noalign{\doubleline}
	\omit\hfil model	\hfil	& \omit\hfil  $\zre$ \hfil	& \omit\hfil  $\Delta z$ \hfil & \omit\hfil  $z_{\rm end}$ \hfil & \omit\hfil  $z_{\rm beg}$ \hfil \cr
	\noalign{\vskip 3pt\hrule\vskip 5pt}
	redshift-symmetric 	& $8.8 \pm 0.9$	& $<4.6$	& $<8.6$	& $~9.4 \pm 1.2$	\cr
	redshift-asymmetric 	& $8.5 \pm 0.9$ 	& $<6.8$ 	& $<8.9$ 	& $10.4 \pm 1.8$ 	\cr
	\omit redshift-symetrical & & \cr
	with kSZ 			& $7.8 \pm 0.9$ 	& $<2.8$ 	& $<8.8$ 	& $8.1 \pm 1.0$ \cr
\noalign{\vskip 5pt\hrule\vskip 3pt}
}
}
\endPlancktablewide                 
\endgroup
\end{table}

Our final constraints on the reionization history are summarized on Table~\ref{tab:constraints} and plotted in
Fig.~\ref{fig:xe_constraints} for each of the aforementioned cases,
i.e., the redshift-symmetric and redshift-asymmetric models, using
only the CMB, and the redshift-symmetric case using CMB+kSZ (all with
prior $z_{\rm end}>6$).  Plotted this way, the constraints are not
very tight and are still fairly model dependent. Given the low value
of $\tau$ as measured now by \Planck, the CMB is not able to give
tight constraints on details of the reionization history.  However,
the \Planck\ data suggest that an early onset of reionization is
disfavoured.  In particular, in all cases, we found that the Universe
was less than 10\,\% ionized for redshift $z>10$.  Furthermore,
comparisons with other tracers of the ionization history show that our
new result on the optical depth eliminates most of the tension between
CMB-based analyses and constraints from other astrophysical data.
Additional sources of reionization, non-standard early galaxies, or
significantly evolving escape fractions or clumping factors, are thus
not needed.

Ongoing and future experiments like LOFAR, MWA, and SKA, aimed at
measuring the redshifted 21-cm signal from neutral hydrogen during the
EoR, should be able to probe reionization directly and measure its
redshift and duration to high accuracy.  Moreover, since reionization
appears to happen at redshifts below 10, experiments measuring the
global emission of the 21-m line over the sky (e.g., EDGES,
\citealt{Bowman10}, LEDA, \citealt{LEDA}, DARE, \citealt{DARE}),
NenuFAR, \citealt{NenuFAR}, SARAS, \citealt{SARAS}, SCI-HI,
\citealt{SCI-HI}, ZEBRA, \citealt{ZEBRA}, and BIGHORNS,
\citealt{BIGHORNS}) will also be able to derive very competitive
constraints on the models \citep[e.g.,][]{liu2015,fialkov2016}.

\begin{acknowledgements}
The Planck Collaboration acknowledges the support of: ESA; CNES, and
CNRS/INSU-IN2P3-INP (France); ASI, CNR, and INAF (Italy); NASA and DoE
(USA); STFC and UKSA (UK); CSIC, MINECO, JA, and RES (Spain); Tekes, AoF,
and CSC (Finland); DLR and MPG (Germany); CSA (Canada); DTU Space
(Denmark); SER/SSO (Switzerland); RCN (Norway); SFI (Ireland);
FCT/MCTES (Portugal); ERC and PRACE (EU). A description of the Planck
Collaboration and a list of its members, indicating which technical
or scientific activities they have been involved in, can be found at
\href{http://www.cosmos.esa.int/web/planck/planck-collaboration}{http://www.cosmos.esa.int/web/planck/planck-collaboration}.
\end{acknowledgements}

\bibliographystyle{aat}
\bibliography{Planck_bib,reio}

\def\eprinttmppp@#1arXiv:@{#1}
\providecommand{\arxivlink[1]}{\href{http://arxiv.org/abs/#1}{arXiv:#1}}
\def\eprinttmp@#1arXiv:#2 [#3]#4@{\ifthenelse{\equal{#3}{x}}{\ifthenelse{
\equal{#1}{}}{\arxivlink{\eprinttmppp@#2@}}{\arxivlink{#1}}}{\arxivlink{#2}
  [#3]}}
\providecommand{\eprintlink}[1]{\eprinttmp@#1arXiv: [x]@}
\providecommand{\eprint}[1]{\eprintlink{#1}}
\providecommand{\adsurl}[1]{\href{#1}{ADS}}
\begin{thebibliography}{110}
\expandafter\ifx\csname natexlab\endcsname\relax\def\natexlab#1{#1}\fi

\bibitem[{{Aghanim} {et~al.}(1996){Aghanim}, {Desert}, {Puget}, \&
  {Gispert}}]{aghanim96}
{Aghanim}, N., {Desert}, F.~X., {Puget}, J.~L., \& {Gispert}, R., {Ionization
  by early quasars and cosmic microwave background anisotropies.} 1996, \aap,
  311, 1, \eprint{astro-ph/9604083}

\bibitem[{{Aghanim} {et~al.}(2008){Aghanim}, {Majumdar}, \& {Silk}}]{aghanim08}
{Aghanim}, N., {Majumdar}, S., \& {Silk}, J., {Secondary anisotropies of the
  CMB}. 2008, Reports on Progress in Physics, 71, 066902, \eprint{0711.0518}

\bibitem[{{Ahn} {et~al.}(2012){Ahn}, {Iliev}, {Shapiro}, {Mellema}, {Koda}, \&
  {Mao}}]{ahn12}
{Ahn}, K., {Iliev}, I.~T., {Shapiro}, P.~R., {et~al.}, {Detecting the Rise and
  Fall of the First Stars by Their Impact on Cosmic Reionization}. 2012, \apjl,
  756, L16, \eprint{1206.5007}

\bibitem[{{Barkana} \& {Loeb}(2001)}]{Barkana01}
{Barkana}, R. \& {Loeb}, A., {In the beginning: the first sources of light and
  the reionization of the universe}. 2001, \physrep, 349, 125,
  \eprint{astro-ph/0010468}

\bibitem[{{Battaglia} {et~al.}(2013){Battaglia}, {Natarajan}, {Trac}, {Cen}, \&
  {Loeb}}]{Battaglia13}
{Battaglia}, N., {Natarajan}, A., {Trac}, H., {Cen}, R., \& {Loeb}, A.,
  {Reionization on Large Scales. III. Predictions for Low-l Cosmic Microwave
  Background Polarization and High-l Kinetic Sunyaev-Zel'dovich Observables}.
  2013, \apj, 776, 83, \eprint{1211.2832}

\bibitem[{{Becker} {et~al.}(2011){Becker}, {Bolton}, {Haehnelt}, \&
  {Sargent}}]{becker11}
{Becker}, G.~D., {Bolton}, J.~S., {Haehnelt}, M.~G., \& {Sargent}, W.~L.~W.,
  {Detection of extended He II reionization in the temperature evolution of the
  intergalactic medium}. 2011, \mnras, 410, 1096, \eprint{1008.2622}

\bibitem[{{Becker} {et~al.}(2015){Becker}, {Bolton}, \& {Lidz}}]{becker15b}
{Becker}, G.~D., {Bolton}, J.~S., \& {Lidz}, A., {Reionisation and
  High-Redshift Galaxies: The View from Quasar Absorption Lines}. 2015, \pasa,
  32, e045, \eprint{1510.03368}

\bibitem[{{Becker} {et~al.}(2001){Becker}, {Fan}, {White}, {Strauss},
  {Narayanan}, {Lupton}, {Gunn}, {Annis}, {Bahcall}, {Brinkmann}, {Connolly},
  {Csabai}, {Czarapata}, {Doi}, {Heckman}, {Hennessy}, {Ivezi{\'c}}, {Knapp},
  {Lamb}, {McKay}, {Munn}, {Nash}, {Nichol}, {Pier}, {Richards}, {Schneider},
  {Stoughton}, {Szalay}, {Thakar}, \& {York}}]{becker01}
{Becker}, R.~H., {Fan}, X., {White}, R.~L., {et~al.}, {Evidence for
  Reionization at z\~{}6: Detection of a Gunn-Peterson Trough in a z=6.28
  Quasar}. 2001, \aj, 122, 2850, \eprint{astro-ph/0108097}

\bibitem[{{Bolton} \& {Haehnelt}(2013)}]{bolton2013}
{Bolton}, J.~S. \& {Haehnelt}, M.~G., {On the rapid demise of Ly {$\alpha$}
  emitters at redshift $z \gtrsim 7$ due to the increasing incidence of
  optically thick absorption systems}. 2013, \mnras, 429, 1695,
  \eprint{1208.4417}

\bibitem[{{Bolton} {et~al.}(2011){Bolton}, {Haehnelt}, {Warren}, {Hewett},
  {Mortlock}, {Venemans}, {McMahon}, \& {Simpson}}]{bolton2011}
{Bolton}, J.~S., {Haehnelt}, M.~G., {Warren}, S.~J., {et~al.}, {How neutral is
  the intergalactic medium surrounding the redshift z = 7.085 quasar ULAS
  J1120+0641?} 2011, \mnras, 416, L70, \eprint{1106.6089}

\bibitem[{{Bouwens} {et~al.}(2015){Bouwens}, {Illingworth}, {Oesch}, {Caruana},
  {Holwerda}, {Smit}, \& {Wilkins}}]{Bou15}
{Bouwens}, R.~J., {Illingworth}, G.~D., {Oesch}, P.~A., {et~al.}, {Reionization
  After Planck: The Derived Growth of the Cosmic Ionizing Emissivity Now
  Matches the Growth of the Galaxy UV Luminosity Density}. 2015, \apj, 811,
  140, \eprint{1503.08228}

\bibitem[{{Bowman} \& {Rogers}(2010)}]{Bowman10}
{Bowman}, J.~D. \& {Rogers}, A.~E.~E., {A lower limit of {$\Delta z > 0.06$}
  for the duration of the reionization epoch}. 2010, \nat, 468, 796

\bibitem[{{Burns} {et~al.}(2012){Burns}, {Lazio}, {Bale}, {Bowman}, {Bradley},
  {Carilli}, {Furlanetto}, {Harker}, {Loeb}, \& {Pritchard}}]{DARE}
{Burns}, J.~O., {Lazio}, J., {Bale}, S., {et~al.}, {Probing the first stars and
  black holes in the early Universe with the Dark Ages Radio Explorer (DARE)}.
  2012, Advances in Space Research, 49, 433, \eprint{1106.5194}

\bibitem[{{Cai} {et~al.}(2014){Cai}, {Lapi}, {Bressan}, {De Zotti}, {Negrello},
  \& {Danese}}]{cai14}
{Cai}, Z.-Y., {Lapi}, A., {Bressan}, A., {et~al.}, {A Physical Model for the
  Evolving Ultraviolet Luminosity Function of High Redshift Galaxies and their
  Contribution to the Cosmic Reionization}. 2014, \apj, 785, 65,
  \eprint{1403.0055}

\bibitem[{{Calabrese} {et~al.}(2013){Calabrese}, {Hlozek}, {Battaglia},
  {Battistelli}, {Bond}, {Chluba}, {Crichton}, {Das}, {Devlin}, {Dunkley},
  {D{\"u}nner}, {Farhang}, {Gralla}, {Hajian}, {Halpern}, {Hasselfield},
  {Hincks}, {Irwin}, {Kosowsky}, {Louis}, {Marriage}, {Moodley}, {Newburgh},
  {Niemack}, {Nolta}, {Page}, {Sehgal}, {Sherwin}, {Sievers}, {Sif{\'o}n},
  {Spergel}, {Staggs}, {Switzer}, \& {Wollack}}]{calabrese13}
{Calabrese}, E., {Hlozek}, R.~A., {Battaglia}, N., {et~al.}, {Cosmological
  parameters from pre-planck cosmic microwave background measurements}. 2013,
  \prd, 87, 103012, \eprint{1302.1841}

\bibitem[{{Carilli} {et~al.}(2010){Carilli}, {Wang}, {Fan}, {Walter}, {Kurk},
  {Riechers}, {Wagg}, {Hennawi}, {Jiang}, {Menten}, {Bertoldi}, {Strauss}, \&
  {Cox}}]{carilli2010}
{Carilli}, C.~L., {Wang}, R., {Fan}, X., {et~al.}, {Ionization Near Zones
  Associated with Quasars at z \~{} 6}. 2010, \apj, 714, 834,
  \eprint{1003.0016}

\bibitem[{{Cen}(2003)}]{Cen03}
{Cen}, R., {The Universe Was Reionized Twice}. 2003, \apj, 591, 12,
  \eprint{astro-ph/0210473}

\bibitem[{{Chornock} {et~al.}(2014){Chornock}, {Berger}, {Fox}, {Fong},
  {Laskar}, \& {Roth}}]{Chornock14}
{Chornock}, R., {Berger}, E., {Fox}, D.~B., {et~al.}, {GRB 140515A at z=6.33:
  Constraints on the End of Reionization From a Gamma-ray Burst in a Low
  Hydrogen Column Density Environment}. 2014, ArXiv e-prints,
  \eprint{1405.7400}

\bibitem[{{Ciardi} {et~al.}(2003){Ciardi}, {Ferrara}, \& {White}}]{cia03}
{Ciardi}, B., {Ferrara}, A., \& {White}, S.~D.~M., {Early reionization by the
  first galaxies}. 2003, \mnras, 344, L7, \eprint{astro-ph/0302451}

\bibitem[{{Colombo} \& {Pierpaoli}(2009)}]{colombo09}
{Colombo}, L.~P.~L. \& {Pierpaoli}, E., {Model independent approaches to
  reionization in the analysis of upcoming CMB data}. 2009, \na, 14, 269,
  \eprint{0804.0278}

\bibitem[{{Couchman} \& {Rees}(1986)}]{Couchman86}
{Couchman}, H.~M.~P. \& {Rees}, M.~J., {Pregalactic evolution in cosmologies
  with cold dark matter}. 1986, \mnras, 221, 53

\bibitem[{{Das} {et~al.}(2014){Das}, {Louis}, {Nolta}, {Addison},
  {Battistelli}, {Bond}, {Calabrese}, {Crichton}, {Devlin}, {Dicker},
  {Dunkley}, {D{\"u}nner}, {Fowler}, {Gralla}, {Hajian}, {Halpern},
  {Hasselfield}, {Hilton}, {Hincks}, {Hlozek}, {Huffenberger}, {Hughes},
  {Irwin}, {Kosowsky}, {Lupton}, {Marriage}, {Marsden}, {Menanteau}, {Moodley},
  {Niemack}, {Page}, {Partridge}, {Reese}, {Schmitt}, {Sehgal}, {Sherwin},
  {Sievers}, {Spergel}, {Staggs}, {Swetz}, {Switzer}, {Thornton}, {Trac}, \&
  {Wollack}}]{das14}
{Das}, S., {Louis}, T., {Nolta}, M.~R., {et~al.}, {The Atacama Cosmology
  Telescope: temperature and gravitational lensing power spectrum measurements
  from three seasons of data}. 2014, \jcap, 4, 14, \eprint{1301.1037}

\bibitem[{{Douspis} {et~al.}(2015){Douspis}, {Aghanim}, {Ili{\'c}}, \&
  {Langer}}]{douspis15}
{Douspis}, M., {Aghanim}, N., {Ili{\'c}}, S., \& {Langer}, M., {A new
  parameterization of the reionisation history}. 2015, \aap, 580, L4,
  \eprint{1509.02785}

\bibitem[{{Dunkley} {et~al.}(2009){Dunkley}, {Komatsu}, {Nolta}, {Spergel},
  {Larson}, {Hinshaw}, {Page}, {Bennett}, {Gold}, {Jarosik}, {Weiland},
  {Halpern}, {Hill}, {Kogut}, {Limon}, {Meyer}, {Tucker}, {Wollack}, \&
  {Wright}}]{dunkley2009}
{Dunkley}, J., {Komatsu}, E., {Nolta}, M.~R., {et~al.}, {Five-Year Wilkinson
  Microwave Anisotropy Probe (WMAP) Observations: Likelihoods and Parameters
  from the WMAP data}. 2009, \apjs, 180, 306, \eprint{0803.0586}

\bibitem[{{Ellis} {et~al.}(2013){Ellis}, {McLure}, {Dunlop}, {Robertson},
  {Ono}, {Schenker}, {Koekemoer}, {Bowler}, {Ouchi}, {Rogers}, {Curtis-Lake},
  {Schneider}, {Charlot}, {Stark}, {Furlanetto}, \& {Cirasuolo}}]{ellis2013}
{Ellis}, R.~S., {McLure}, R.~J., {Dunlop}, J.~S., {et~al.}, {The Abundance of
  Star-forming Galaxies in the Redshift Range 8.5-12: New Results from the 2012
  Hubble Ultra Deep Field Campaign}. 2013, \apjl, 763, L7, \eprint{1211.6804}

\bibitem[{{Faisst} {et~al.}(2014){Faisst}, {Capak}, {Carollo}, {Scarlata}, \&
  {Scoville}}]{Faisst14}
{Faisst}, A.~L., {Capak}, P., {Carollo}, C.~M., {Scarlata}, C., \& {Scoville},
  N., {Spectroscopic Observation of Ly{$\alpha$} Emitters at z \~{} 7.7 and
  Implications on Re-ionization}. 2014, \apj, 788, 87

\bibitem[{{Fan} {et~al.}(2006{\natexlab{a}}){Fan}, {Strauss}, {Becker},
  {White}, {Gunn}, {Knapp}, {Richards}, {Schneider}, {Brinkmann}, \&
  {Fukugita}}]{fan06a}
{Fan}, X., {Strauss}, M.~A., {Becker}, R.~H., {et~al.}, {Constraining the
  Evolution of the Ionizing Background and the Epoch of Reionization with
  z\~{}6 Quasars. II. A Sample of 19 Quasars}. 2006{\natexlab{a}}, \aj, 132,
  117, \eprint{astro-ph/0512082}

\bibitem[{{Fan} {et~al.}(2006{\natexlab{b}}){Fan}, {Strauss}, {Richards},
  {Hennawi}, {Becker}, {White}, {Diamond-Stanic}, {Donley}, {Jiang}, {Kim},
  {Vestergaard}, {Young}, {Gunn}, {Lupton}, {Knapp}, {Schneider}, {Brandt},
  {Bahcall}, {Barentine}, {Brinkmann}, {Brewington}, {Fukugita}, {Harvanek},
  {Kleinman}, {Krzesinski}, {Long}, {Neilsen}, {Nitta}, {Snedden}, \&
  {Voges}}]{fan06b}
{Fan}, X., {Strauss}, M.~A., {Richards}, G.~T., {et~al.}, {A Survey of
  z\textgreater 5.7 Quasars in the Sloan Digital Sky Survey. IV. Discovery of
  Seven Additional Quasars}. 2006{\natexlab{b}}, \aj, 131, 1203,
  \eprint{astro-ph/0512080}

\bibitem[{{Fialkov} \& {Loeb}(2016)}]{fialkov2016}
{Fialkov}, A. \& {Loeb}, A., {Precise Measurement of the Reionization Optical
  Depth from the Global 21 cm Signal Accounting for Cosmic Heating}. 2016,
  \apj, 821, 59, \eprint{1601.03058}

\bibitem[{{Fontanot} {et~al.}(2012){Fontanot}, {Cristiani}, \&
  {Vanzella}}]{fontanot2012}
{Fontanot}, F., {Cristiani}, S., \& {Vanzella}, E., {On the relative
  contribution of high-redshift galaxies and active galactic nuclei to
  reionization}. 2012, \mnras, 425, 1413, \eprint{1206.5810}

\bibitem[{{Furlanetto} {et~al.}(2004){Furlanetto}, {Zaldarriaga}, \&
  {Hernquist}}]{Furlanetto04}
{Furlanetto}, S.~R., {Zaldarriaga}, M., \& {Hernquist}, L., {The Growth of H II
  Regions During Reionization}. 2004, \apj, 613, 1, \eprint{astro-ph/0403697}

\bibitem[{{George} {et~al.}(2015){George}, {Reichardt}, {Aird}, {Benson},
  {Bleem}, {Carlstrom}, {Chang}, {Cho}, {Crawford}, {Crites}, {de Haan},
  {Dobbs}, {Dudley}, {Halverson}, {Harrington}, {Holder}, {Holzapfel}, {Hou},
  {Hrubes}, {Keisler}, {Knox}, {Lee}, {Leitch}, {Lueker}, {Luong-Van},
  {McMahon}, {Mehl}, {Meyer}, {Millea}, {Mocanu}, {Mohr}, {Montroy}, {Padin},
  {Plagge}, {Pryke}, {Ruhl}, {Schaffer}, {Shaw}, {Shirokoff}, {Spieler},
  {Staniszewski}, {Stark}, {Story}, {van Engelen}, {Vanderlinde}, {Vieira},
  {Williamson}, \& {Zahn}}]{george15}
{George}, E.~M., {Reichardt}, C.~L., {Aird}, K.~A., {et~al.}, {A Measurement of
  Secondary Cosmic Microwave Background Anisotropies from the 2500
  Square-degree SPT-SZ Survey}. 2015, \apj, 799, 177, \eprint{1408.3161}

\bibitem[{{Giallongo} {et~al.}(2015){Giallongo}, {Grazian}, {Fiore}, {Fontana},
  {Pentericci}, {Vanzella}, {Dickinson}, {Kocevski}, {Castellano}, {Cristiani},
  {Ferguson}, {Finkelstein}, {Grogin}, {Hathi}, {Koekemoer}, {Newman}, \&
  {Salvato}}]{gia15}
{Giallongo}, E., {Grazian}, A., {Fiore}, F., {et~al.}, {Faint AGNs at z gt 4 in
  the CANDELS GOODS-S field: looking for contributors to the reionization of
  the Universe}. 2015, \aap, 578, A83, \eprint{1502.02562}

\bibitem[{{Gnedin}(2000)}]{gne00}
{Gnedin}, N.~Y., {Cosmological Reionization by Stellar Sources}. 2000, \apj,
  535, 530, \eprint{astro-ph/9909383}

\bibitem[{{Greenhill} \& {Bernardi}(2012)}]{LEDA}
{Greenhill}, L.~J. \& {Bernardi}, G., {HI Epoch of Reionization Arrays}. 2012,
  ArXiv e-prints, \eprint{1201.1700}

\bibitem[{{Gruzinov} \& {Hu}(1998)}]{gru98}
{Gruzinov}, A. \& {Hu}, W., {Secondary Cosmic Microwave Background Anisotropies
  in a Universe Reionized in Patches}. 1998, \apj, 508, 435,
  \eprint{astro-ph/9803188}

\bibitem[{{Gunn} \& {Peterson}(1965)}]{gunn1965}
{Gunn}, J.~E. \& {Peterson}, B.~A., {On the Density of Neutral Hydrogen in
  Intergalactic Space.} 1965, \apj, 142, 1633

\bibitem[{{Hamimeche} \& {Lewis}(2008)}]{hamimeche08}
{Hamimeche}, S. \& {Lewis}, A., {Likelihood analysis of CMB temperature and
  polarization power spectra}. 2008, \prd, 77, 103013, \eprint{0801.0554}

\bibitem[{{Hasselfield} {et~al.}(2013){Hasselfield}, {Moodley}, {Bond}, {Das},
  {Devlin}, {Dunkley}, {D{\"u}nner}, {Fowler}, {Gallardo}, {Gralla}, {Hajian},
  {Halpern}, {Hincks}, {Marriage}, {Marsden}, {Niemack}, {Nolta}, {Page},
  {Partridge}, {Schmitt}, {Sehgal}, {Sievers}, {Staggs}, {Swetz}, {Switzer}, \&
  {Wollack}}]{hasselfield13}
{Hasselfield}, M., {Moodley}, K., {Bond}, J.~R., {et~al.}, {The Atacama
  Cosmology Telescope: Beam Measurements and the Microwave Brightness
  Temperatures of Uranus and Saturn}. 2013, \apjs, 209, 17, \eprint{1303.4714}

\bibitem[{{Hinshaw} {et~al.}(2013){Hinshaw}, {Larson}, {Komatsu}, {Spergel},
  {Bennett}, {Dunkley}, {Nolta}, {Halpern}, {Hill}, {Odegard}, {Page}, {Smith},
  {Weiland}, {Gold}, {Jarosik}, {Kogut}, {Limon}, {Meyer}, {Tucker}, {Wollack},
  \& {Wright}}]{hinshaw2012}
{Hinshaw}, G., {Larson}, D., {Komatsu}, E., {et~al.}, {Nine-year Wilkinson
  Microwave Anisotropy Probe (WMAP) Observations: Cosmological Parameter
  Results}. 2013, \apjs, 208, 19, \eprint{1212.5226}

\bibitem[{{Holder} {et~al.}(2003){Holder}, {Haiman}, {Kaplinghat}, \&
  {Knox}}]{holder03}
{Holder}, G.~P., {Haiman}, Z., {Kaplinghat}, M., \& {Knox}, L., {The
  Reionization History at High Redshifts. II. Estimating the Optical Depth to
  Thomson Scattering from Cosmic Microwave Background Polarization}. 2003,
  \apj, 595, 13, \eprint{astro-ph/0302404}

\bibitem[{Hu \& Holder(2003)}]{hu03}
Hu, W. \& Holder, G.~P., Model-independent reionization observables in the CMB.
  2003, Phys. Rev. D, 68, 023001

\bibitem[{{Iliev} {et~al.}(2014){Iliev}, {Mellema}, {Ahn}, {Shapiro}, {Mao}, \&
  {Pen}}]{Iliev14}
{Iliev}, I.~T., {Mellema}, G., {Ahn}, K., {et~al.}, {Simulating cosmic
  reionization: how large a volume is large enough?} 2014, \mnras, 439, 725,
  \eprint{1310.7463}

\bibitem[{{Ishigaki} {et~al.}(2015){Ishigaki}, {Kawamata}, {Ouchi}, {Oguri},
  {Shimasaku}, \& {Ono}}]{Ishi15}
{Ishigaki}, M., {Kawamata}, R., {Ouchi}, M., {et~al.}, {Hubble Frontier Fields
  First Complete Cluster Data: Faint Galaxies at z \~{} 5-10 for UV Luminosity
  Functions and Cosmic Reionization}. 2015, \apj, 799, 12, \eprint{1408.6903}

\bibitem[{{Keating} {et~al.}(2015){Keating}, {Haehnelt}, {Cantalupo}, \&
  {Puchwein}}]{keating2015}
{Keating}, L.~C., {Haehnelt}, M.~G., {Cantalupo}, S., \& {Puchwein}, E.,
  {Probing the end of reionization with the near zones of $z \gtrsim 6$ QSOs}.
  2015, \mnras, 454, 681, \eprint{1506.03396}

\bibitem[{{Khaire} {et~al.}(2016){Khaire}, {Srianand}, {Choudhury}, \&
  {Gaikwad}}]{khaire2016}
{Khaire}, V., {Srianand}, R., {Choudhury}, T.~R., \& {Gaikwad}, P., {The
  redshift evolution of escape fraction of hydrogen ionizing photons from
  galaxies}. 2016, \mnras, 457, 4051, \eprint{1510.04700}

\bibitem[{{Kogut} {et~al.}(2003){Kogut}, {Spergel}, {Barnes}, {Bennett},
  {Halpern}, {Hinshaw}, {Jarosik}, {Limon}, {Meyer}, {Page}, {Tucker},
  {Wollack}, \& {Wright}}]{kogut2003}
{Kogut}, A., {Spergel}, D.~N., {Barnes}, C., {et~al.}, {First-Year Wilkinson
  Microwave Anisotropy Probe (WMAP) Observations: Temperature-Polarization
  Correlation}. 2003, \apjs, 148, 161, \eprint{astro-ph/0302213}

\bibitem[{{Komatsu} {et~al.}(2011){Komatsu}, {Smith}, {Dunkley}, {Bennett},
  {Gold}, {Hinshaw}, {Jarosik}, {Larson}, {Nolta}, {Page}, {Spergel},
  {Halpern}, {Hill}, {Kogut}, {Limon}, {Meyer}, {Odegard}, {Tucker}, {Weiland},
  {Wollack}, \& {Wright}}]{komatsu2010}
{Komatsu}, E., {Smith}, K.~M., {Dunkley}, J., {et~al.}, {Seven-year Wilkinson
  Microwave Anisotropy Probe (WMAP) Observations: Cosmological Interpretation}.
  2011, \apjs, 192, 18, \eprint{1001.4538}

\bibitem[{{Kuhlen} \& {Faucher-Gigu{\`e}re}(2012)}]{kuh12}
{Kuhlen}, M. \& {Faucher-Gigu{\`e}re}, C.-A., {Concordance models of
  reionization: implications for faint galaxies and escape fraction evolution}.
  2012, \mnras, 423, 862, \eprint{1201.0757}

\bibitem[{{Lesgourgues}(2011)}]{lesgourgues11}
{Lesgourgues}, J., {The Cosmic Linear Anisotropy Solving System (CLASS) I:
  Overview}. 2011, ArXiv e-prints, \eprint{1104.2932}

\bibitem[{{Lewis}(2008)}]{lewis08}
{Lewis}, A., {Cosmological parameters from WMAP 5-year temperature maps}. 2008,
  \prd, 78, 023002, \eprint{0804.3865}

\bibitem[{{Lewis} {et~al.}(2006){Lewis}, {Weller}, \& {Battye}}]{lewis06}
{Lewis}, A., {Weller}, J., \& {Battye}, R., {The cosmic microwave background
  and the ionization history of the Universe}. 2006, \mnras, 373, 561,
  \eprint{astro-ph/0606552}

\bibitem[{{Liu} {et~al.}(2016){Liu}, {Pritchard}, {Allison}, {Parsons},
  {Seljak}, \& {Sherwin}}]{liu2015}
{Liu}, A., {Pritchard}, J.~R., {Allison}, R., {et~al.}, {Eliminating the
  optical depth nuisance from the CMB with 21 cm cosmology}. 2016, \prd, 93,
  043013, \eprint{1509.08463}

\bibitem[{{Madau} \& {Haardt}(2015)}]{madau15}
{Madau}, P. \& {Haardt}, F., {Cosmic Reionization after Planck: Could Quasars
  Do It All?} 2015, \apjl, 813, L8, \eprint{1507.07678}

\bibitem[{{Madau} {et~al.}(1999){Madau}, {Haardt}, \& {Rees}}]{Madau99}
{Madau}, P., {Haardt}, F., \& {Rees}, M.~J., {Radiative Transfer in a Clumpy
  Universe. III. The Nature of Cosmological Ionizing Sources}. 1999, \apj, 514,
  648, \eprint{astro-ph/9809058}

\bibitem[{{Mahesh} {et~al.}(2014){Mahesh}, {Subrahmanyan}, {Udaya Shankar}, \&
  {Raghunathan}}]{ZEBRA}
{Mahesh}, N., {Subrahmanyan}, R., {Udaya Shankar}, N., \& {Raghunathan}, A., {A
  Resistive Wideband Space Beam Splitter}. 2014, ArXiv e-prints,
  \eprint{1406.2585}

\bibitem[{{Mangilli} {et~al.}(2015){Mangilli}, {Plaszczynski}, \&
  {Tristram}}]{mangilli15}
{Mangilli}, A., {Plaszczynski}, S., \& {Tristram}, M., {Large-scale cosmic
  microwave background temperature and polarization cross-spectra likelihoods}.
  2015, \mnras, 453, 3174, \eprint{1503.01347}

\bibitem[{{McQuinn}(2015)}]{mcquinn15}
{McQuinn}, M., {The Evolution of the Intergalactic Medium}. 2015, ArXiv
  e-prints, \eprint{1512.00086}

\bibitem[{{McQuinn} {et~al.}(2005){McQuinn}, {Furlanetto}, {Hernquist}, {Zahn},
  \& {Zaldarriaga}}]{mcquinn05}
{McQuinn}, M., {Furlanetto}, S.~R., {Hernquist}, L., {Zahn}, O., \&
  {Zaldarriaga}, M., {The Kinetic Sunyaev-Zel'dovich Effect from Reionization}.
  2005, \apj, 630, 643, \eprint{astro-ph/0504189}

\bibitem[{{Meiksin} \& {Madau}(1993)}]{Meiksin93}
{Meiksin}, A. \& {Madau}, P., {On the photoionization of the intergalactic
  medium by quasars at high redshift}. 1993, \apj, 412, 34

\bibitem[{Mesinger(2016)}]{RevMes2016}
Mesinger, A., ed. 2016, Astrophysics and Space Science Library, Vol. 423,
  {Understanding the Epoch of Cosmic Reionization} (Springer International
  Publishing)

\bibitem[{{Mesinger} \& {Furlanetto}(2008)}]{mesinger2008}
{Mesinger}, A. \& {Furlanetto}, S.~R., {Ly{$\alpha$} damping wing constraints
  on inhomogeneous reionization}. 2008, \mnras, 385, 1348, \eprint{0710.0371}

\bibitem[{{Mesinger} \& {Haiman}(2004)}]{mesinger2004}
{Mesinger}, A. \& {Haiman}, Z., {Evidence of a Cosmological Str{\"o}mgren
  Surface and of Significant Neutral Hydrogen Surrounding the Quasar SDSS
  J1030+0524}. 2004, \apjl, 611, L69, \eprint{astro-ph/0406188}

\bibitem[{{Mesinger} \& {Haiman}(2007)}]{mesinger2007}
{Mesinger}, A. \& {Haiman}, Z., {Constraints on Reionization and Source
  Properties from the Absorption Spectra of $z > 6.2$ Quasars}. 2007, \apj,
  660, 923, \eprint{astro-ph/0610258}

\bibitem[{{Mesinger} {et~al.}(2012){Mesinger}, {McQuinn}, \&
  {Spergel}}]{mesinger2012}
{Mesinger}, A., {McQuinn}, M., \& {Spergel}, D.~N., {The kinetic
  Sunyaev-Zel'dovich signal from inhomogeneous reionization: a parameter space
  study}. 2012, \mnras, 422, 1403, \eprint{1112.1820}

\bibitem[{{Miralda-Escude} \& {Ostriker}(1990)}]{Miralda90}
{Miralda-Escude}, J. \& {Ostriker}, J.~P., {What produces the ionizing
  background at large redshift?} 1990, \apj, 350, 1

\bibitem[{{Mitra} {et~al.}(2011){Mitra}, {Choudhury}, \& {Ferrara}}]{Mitra11}
{Mitra}, S., {Choudhury}, T.~R., \& {Ferrara}, A., {Reionization constraints
  using principal component analysis}. 2011, \mnras, 413, 1569,
  \eprint{1011.2213}

\bibitem[{{Mortlock} {et~al.}(2011){Mortlock}, {Warren}, {Venemans}, {Patel},
  {Hewett}, {McMahon}, {Simpson}, {Theuns}, {Gonz{\'a}les-Solares}, {Adamson},
  {Dye}, {Hambly}, {Hirst}, {Irwin}, {Kuiper}, {Lawrence}, \&
  {R{\"o}ttgering}}]{mortlock11}
{Mortlock}, D.~J., {Warren}, S.~J., {Venemans}, B.~P., {et~al.}, {A luminous
  quasar at a redshift of z = 7.085}. 2011, \nat, 474, 616, \eprint{1106.6088}

\bibitem[{{Mortonson} \& {Hu}(2008)}]{mortonson08}
{Mortonson}, M.~J. \& {Hu}, W., {Model-Independent Constraints on Reionization
  from Large-Scale Cosmic Microwave Background Polarization}. 2008, \apj, 672,
  737, \eprint{0705.1132}

\bibitem[{{Ostriker} \& {Vishniac}(1986)}]{ostriker86}
{Ostriker}, J.~P. \& {Vishniac}, E.~T., {Generation of microwave background
  fluctuations from nonlinear perturbations at the ERA of galaxy formation}.
  1986, \apjl, 306, L51

\bibitem[{{Page} {et~al.}(2007){Page}, {Hinshaw}, {Komatsu}, {Nolta},
  {Spergel}, {Bennett}, {Barnes}, {Bean}, {Dor{\'e}}, {Dunkley}, {Halpern},
  {Hill}, {Jarosik}, {Kogut}, {Limon}, {Meyer}, {Odegard}, {Peiris}, {Tucker},
  {Verde}, {Weiland}, {Wollack}, \& {Wright}}]{page2007}
{Page}, L., {Hinshaw}, G., {Komatsu}, E., {et~al.}, {Three-Year Wilkinson
  Microwave Anisotropy Probe (WMAP) Observations: Polarization Analysis}. 2007,
  \apjs, 170, 335, \eprint{astro-ph/0603450}

\bibitem[{{Pandolfi} {et~al.}(2011){Pandolfi}, {Ferrara}, {Choudhury},
  {Melchiorri}, \& {Mitra}}]{Pan11}
{Pandolfi}, S., {Ferrara}, A., {Choudhury}, T.~R., {Melchiorri}, A., \&
  {Mitra}, S., {Data-constrained reionization and its effects on cosmological
  parameters}. 2011, \prd, 84, 123522, \eprint{1111.3570}

\bibitem[{{Park} {et~al.}(2013){Park}, {Shapiro}, {Komatsu}, {Iliev}, {Ahn}, \&
  {Mellema}}]{park13}
{Park}, H., {Shapiro}, P.~R., {Komatsu}, E., {et~al.}, {The Kinetic
  Sunyaev-Zel'dovich Effect as a Probe of the Physics of Cosmic Reionization:
  The Effect of Self-regulated Reionization}. 2013, \apj, 769, 93,
  \eprint{1301.3607}

\bibitem[{{Patra} {et~al.}(2013){Patra}, {Subrahmanyan}, {Raghunathan}, \&
  {Udaya Shankar}}]{SARAS}
{Patra}, N., {Subrahmanyan}, R., {Raghunathan}, A., \& {Udaya Shankar}, N.,
  {SARAS: a precision system for measurement of the cosmic radio background and
  signatures from the epoch of reionization}. 2013, Experimental Astronomy, 36,
  319, \eprint{1211.3800}

\bibitem[{{Peebles}(1968)}]{Peebles1968}
{Peebles}, P.~J.~E., {Recombination of the Primeval Plasma}. 1968, \apj, 153, 1

\bibitem[{{\sorthelp{Planck Collaboration 2014O}}{Planck Collaboration
  XV}(2014)}]{planck2013-p08}
{\sorthelp{Planck Collaboration 2014O}}{Planck Collaboration XV},
  {\textit{Planck} 2013 results. XV. CMB power spectra and likelihood}. 2014,
  \aap, 571, A15, \eprint{1303.5075}

\bibitem[{{\sorthelp{Planck Collaboration 2014P}}{Planck Collaboration
  XVI}(2014)}]{planck2013-p11}
{\sorthelp{Planck Collaboration 2014P}}{Planck Collaboration XVI},
  {\textit{Planck} 2013 results. XVI. Cosmological parameters}. 2014, \aap,
  571, A16, \eprint{1303.5076}

\bibitem[{{\sorthelp{Planck Collaboration 2014T}}{Planck Collaboration
  XX}(2014)}]{planck2013-p15}
{\sorthelp{Planck Collaboration 2014T}}{Planck Collaboration XX},
  {\textit{Planck} 2013 results. XX. Cosmology from Sunyaev-Zeldovich cluster
  counts}. 2014, \aap, 571, A20, \eprint{1303.5080}

\bibitem[{{\sorthelp{Planck Collaboration 2015B}}{Planck Collaboration
  II}(2016)}]{planck2014-a03}
{\sorthelp{Planck Collaboration 2015B}}{Planck Collaboration II},
  {\textit{Planck} 2015 results. II. Low Frequency Instrument data processing}.
  2016, \aap, in press, \eprint{1502.01583}

\bibitem[{{\sorthelp{Planck Collaboration 2015G}}{Planck Collaboration
  VII}(2016)}]{planck2014-a08}
{\sorthelp{Planck Collaboration 2015G}}{Planck Collaboration VII},
  {\textit{Planck} 2015 results. VII. High Frequency Instrument data
  processing: Time-ordered information and beam processing}. 2016, \aap, in
  press, \eprint{1502.01586}

\bibitem[{{\sorthelp{Planck Collaboration 2015H}}{Planck Collaboration
  VIII}(2016)}]{planck2014-a09}
{\sorthelp{Planck Collaboration 2015H}}{Planck Collaboration VIII},
  {\textit{Planck} 2015 results. VIII. High Frequency Instrument data
  processing: Calibration and maps}. 2016, \aap, in press, \eprint{1502.01587}

\bibitem[{{\sorthelp{Planck Collaboration 2015I}}{Planck Collaboration
  IX}(2016)}]{planck2014-a11}
{\sorthelp{Planck Collaboration 2015I}}{Planck Collaboration IX},
  {\textit{Planck} 2015 results. IX. Diffuse component separation: CMB maps}.
  2016, \aap, in press, \eprint{1502.05956}

\bibitem[{{\sorthelp{Planck Collaboration 2015J}}{Planck Collaboration
  X}(2016)}]{planck2014-a12}
{\sorthelp{Planck Collaboration 2015J}}{Planck Collaboration X},
  {\textit{Planck} 2015 results. X. Diffuse component separation: Foreground
  maps}. 2016, \aap, in press, \eprint{1502.01588}

\bibitem[{{\sorthelp{Planck Collaboration 2015K}}{Planck Collaboration
  XI}(2016)}]{planck2014-a13}
{\sorthelp{Planck Collaboration 2015K}}{Planck Collaboration XI},
  {\textit{Planck} 2015 results. XI. CMB power spectra, likelihoods, and
  robustness of parameters}. 2016, \aap, in press, \eprint{1507.02704}

\bibitem[{{\sorthelp{Planck Collaboration 2015M}}{Planck Collaboration
  XIII}(2016)}]{planck2014-a15}
{\sorthelp{Planck Collaboration 2015M}}{Planck Collaboration XIII},
  {\textit{Planck} 2015 results. XIII. Cosmological parameters}. 2016, \aap, in
  press, \eprint{1502.01589}

\bibitem[{{\sorthelp{Planck Collaboration 2015O}}{Planck Collaboration
  XV}(2016)}]{planck2014-a17}
{\sorthelp{Planck Collaboration 2015O}}{Planck Collaboration XV},
  {\textit{Planck} 2015 results. XV. Gravitational lensing}. 2016, \aap, in
  press, \eprint{1502.01591}

\bibitem[{{\sorthelp{Planck Collaboration IntZU}}{Planck Collaboration Int.
  XLVI}(2016)}]{planck2014-a10}
{\sorthelp{Planck Collaboration IntZU}}{Planck Collaboration Int. XLVI},
  {\textit{Planck} intermediate results. XLVI. Reduction of large-scale
  systematic effects in HFI polarization maps and estimation of the
  reionization optical depth}. 2016, \aap, submitted, \eprint{1605.02985}

\bibitem[{{Pritchard} {et~al.}(2010){Pritchard}, {Loeb}, \& {Wyithe}}]{PLW10}
{Pritchard}, J.~R., {Loeb}, A., \& {Wyithe}, J.~S.~B., {Constraining
  reionization using 21-cm observations in combination with CMB and
  Ly{$\alpha$} forest data}. 2010, \mnras, 408, 57, \eprint{0908.3891}

\bibitem[{{Reichardt} {et~al.}(2012){Reichardt}, {Shaw}, {Zahn}, {Aird},
  {Benson}, {Bleem}, {Carlstrom}, {Chang}, {Cho}, {Crawford}, {Crites}, {de
  Haan}, {Dobbs}, {Dudley}, {George}, {Halverson}, {Holder}, {Holzapfel},
  {Hoover}, {Hou}, {Hrubes}, {Joy}, {Keisler}, {Knox}, {Lee}, {Leitch},
  {Lueker}, {Luong-Van}, {McMahon}, {Mehl}, {Meyer}, {Millea}, {Mohr},
  {Montroy}, {Natoli}, {Padin}, {Plagge}, {Pryke}, {Ruhl}, {Schaffer},
  {Shirokoff}, {Spieler}, {Staniszewski}, {Stark}, {Story}, {van Engelen},
  {Vanderlinde}, {Vieira}, \& {Williamson}}]{reichardt12}
{Reichardt}, C.~L., {Shaw}, L., {Zahn}, O., {et~al.}, {A Measurement of
  Secondary Cosmic Microwave Background Anisotropies with Two Years of South
  Pole Telescope Observations}. 2012, \apj, 755, 70, \eprint{1111.0932}

\bibitem[{{Robertson} {et~al.}(2010){Robertson}, {Ellis}, {Dunlop}, {McLure},
  \& {Stark}}]{robertson2010}
{Robertson}, B.~E., {Ellis}, R.~S., {Dunlop}, J.~S., {McLure}, R.~J., \&
  {Stark}, D.~P., {Early star-forming galaxies and the reionization of the
  Universe}. 2010, \nat, 468, 49, \eprint{1011.0727}

\bibitem[{{Robertson} {et~al.}(2015){Robertson}, {Ellis}, {Furlanetto}, \&
  {Dunlop}}]{Rob15}
{Robertson}, B.~E., {Ellis}, R.~S., {Furlanetto}, S.~R., \& {Dunlop}, J.~S.,
  {Cosmic Reionization and Early Star-forming Galaxies: A Joint Analysis of New
  Constraints from Planck and the Hubble Space Telescope}. 2015, \apjl, 802,
  L19, \eprint{1502.02024}

\bibitem[{{Robertson} {et~al.}(2013){Robertson}, {Furlanetto}, {Schneider},
  {Charlot}, {Ellis}, {Stark}, {McLure}, {Dunlop}, {Koekemoer}, {Schenker},
  {Ouchi}, {Ono}, {Curtis-Lake}, {Rogers}, {Bowler}, \&
  {Cirasuolo}}]{Robertson13}
{Robertson}, B.~E., {Furlanetto}, S.~R., {Schneider}, E., {et~al.}, {New
  Constraints on Cosmic Reionization from the 2012 Hubble Ultra Deep Field
  Campaign}. 2013, \apj, 768, 71, \eprint{1301.1228}

\bibitem[{{Schenker} {et~al.}(2014){Schenker}, {Ellis}, {Konidaris}, \&
  {Stark}}]{schenker14}
{Schenker}, M.~A., {Ellis}, R.~S., {Konidaris}, N.~P., \& {Stark}, D.~P.,
  {Line-emitting Galaxies beyond a Redshift of 7: An Improved Method for
  Estimating the Evolving Neutrality of the Intergalactic Medium}. 2014, \apj,
  795, 20, \eprint{1404.4632}

\bibitem[{{Schroeder} {et~al.}(2013){Schroeder}, {Mesinger}, \&
  {Haiman}}]{schroeder2013}
{Schroeder}, J., {Mesinger}, A., \& {Haiman}, Z., {Evidence of Gunn-Peterson
  damping wings in high-z quasar spectra: strengthening the case for incomplete
  reionization at $ z \sim$ 6-7}. 2013, \mnras, 428, 3058, \eprint{1204.2838}

\bibitem[{{Seager} {et~al.}(2000){Seager}, {Sasselov}, \& {Scott}}]{SSS2000}
{Seager}, S., {Sasselov}, D.~D., \& {Scott}, D., {How Exactly Did the Universe
  Become Neutral?} 2000, \apjs, 128, 407, \eprint{astro-ph/9912182}

\bibitem[{{Shaw} {et~al.}(2012){Shaw}, {Rudd}, \& {Nagai}}]{Shaw12}
{Shaw}, L.~D., {Rudd}, D.~H., \& {Nagai}, D., {Deconstructing the Kinetic SZ
  Power Spectrum}. 2012, \apj, 756, 15, \eprint{1109.0553}

\bibitem[{{Sokolowski} {et~al.}(2015){Sokolowski}, {Tremblay}, {Wayth},
  {Tingay}, {Clarke}, {Roberts}, {Waterson}, {Ekers}, {Hall}, {Lewis},
  {Mossammaparast}, {Padhi}, {Schlagenhaufer}, {Sutinjo}, \&
  {Tickner}}]{BIGHORNS}
{Sokolowski}, M., {Tremblay}, S.~E., {Wayth}, R.~B., {et~al.}, {BIGHORNS -
  Broadband Instrument for Global HydrOgen ReioNisation Signal}. 2015, \pasa,
  32, e004, \eprint{1501.02922}

\bibitem[{{Sunyaev} \& {Zeldovich}(1980)}]{SZ80}
{Sunyaev}, R.~A. \& {Zeldovich}, I.~B., {The velocity of clusters of galaxies
  relative to the microwave background - The possibility of its measurement}.
  1980, \mnras, 190, 413

\bibitem[{{Tilvi} {et~al.}(2014){Tilvi}, {Papovich}, {Finkelstein}, {Long},
  {Song}, {Dickinson}, {Ferguson}, {Koekemoer}, {Giavalisco}, \&
  {Mobasher}}]{til14}
{Tilvi}, V., {Papovich}, C., {Finkelstein}, S.~L., {et~al.}, {Rapid Decline of
  Ly{$\alpha$} Emission toward the Reionization Era}. 2014, \apj, 794, 5,
  \eprint{1405.4869}

\bibitem[{{Trac} {et~al.}(2011){Trac}, {Bode}, \& {Ostriker}}]{trac11}
{Trac}, H., {Bode}, P., \& {Ostriker}, J.~P., {Templates for the
  Sunyaev-Zel'dovich Angular Power Spectrum}. 2011, \apj, 727, 94,
  \eprint{1006.2828}

\bibitem[{{Tristram} {et~al.}(2005){Tristram}, {Mac{\'{\i}}as-P{\'e}rez},
  {Renault}, \& {Santos}}]{tristram2005}
{Tristram}, M., {Mac{\'{\i}}as-P{\'e}rez}, J.~F., {Renault}, C., \& {Santos},
  D., {XSPECT, estimation of the angular power spectrum by computing
  cross-power spectra with analytical error bars}. 2005, \mnras, 358, 833,
  \eprint{astro-ph/0405575}

\bibitem[{{Venemans} {et~al.}(2013){Venemans}, {Findlay}, {Sutherland}, {De
  Rosa}, {McMahon}, {Simcoe}, {Gonz{\'a}lez-Solares}, {Kuijken}, \&
  {Lewis}}]{venemans13}
{Venemans}, B.~P., {Findlay}, J.~R., {Sutherland}, W.~J., {et~al.}, {Discovery
  of Three z\textgreater 6.5 Quasars in the VISTA Kilo-Degree Infrared Galaxy
  (VIKING) Survey}. 2013, \apj, 779, 24, \eprint{1311.3666}

\bibitem[{Voytek {et~al.}(2014)Voytek, Natarajan, García, Peterson, \&
  López-Cruz}]{SCI-HI}
Voytek, T.~C., Natarajan, A., García, J. M.~J., Peterson, J.~B., \&
  López-Cruz, O., Probing the Dark Ages at z ~ 20: The SCI-HI 21 cm All-sky
  Spectrum Experiment. 2014, The Astrophysical Journal Letters, 782, L9

\bibitem[{{Willott} {et~al.}(2010){Willott}, {Delorme}, {Reyl{\'e}}, {Albert},
  {Bergeron}, {Crampton}, {Delfosse}, {Forveille}, {Hutchings}, {McLure},
  {Omont}, \& {Schade}}]{willott2010}
{Willott}, C.~J., {Delorme}, P., {Reyl{\'e}}, C., {et~al.}, {The Canada-France
  High-z Quasar Survey: Nine New Quasars and the Luminosity Function at
  Redshift 6}. 2010, \aj, 139, 906, \eprint{0912.0281}

\bibitem[{{Worseck} {et~al.}(2014){Worseck}, {Prochaska}, {Hennawi}, \&
  {McQuinn}}]{worseck14}
{Worseck}, G., {Prochaska}, J.~X., {Hennawi}, J.~F., \& {McQuinn}, M., {Early
  and Extended Helium Reionization Over More Than 600 Million Years of Cosmic
  Time}. 2014, ArXiv e-prints, \eprint{1405.7405}

\bibitem[{{Wyithe} \& {Loeb}(2004)}]{wyithe04}
{Wyithe}, J.~S.~B. \& {Loeb}, A., {A characteristic size of \~{}10Mpc for the
  ionized bubbles at the end of cosmic reionization}. 2004, \nat, 432, 194,
  \eprint{astro-ph/0409412}

\bibitem[{{Wyithe} {et~al.}(2005){Wyithe}, {Loeb}, \& {Carilli}}]{wyithe05}
{Wyithe}, J.~S.~B., {Loeb}, A., \& {Carilli}, C., {Improved Constraints on the
  Neutral Intergalactic Hydrogen surrounding Quasars at Redshifts $z > 6$}.
  2005, \apj, 628, 575, \eprint{astro-ph/0411625}

\bibitem[{{Zahn} {et~al.}(2012){Zahn}, {Reichardt}, {Shaw}, {Lidz}, {Aird},
  {Benson}, {Bleem}, {Carlstrom}, {Chang}, {Cho}, {Crawford}, {Crites}, {de
  Haan}, {Dobbs}, {Dor{\'e}}, {Dudley}, {George}, {Halverson}, {Holder},
  {Holzapfel}, {Hoover}, {Hou}, {Hrubes}, {Joy}, {Keisler}, {Knox}, {Lee},
  {Leitch}, {Lueker}, {Luong-Van}, {McMahon}, {Mehl}, {Meyer}, {Millea},
  {Mohr}, {Montroy}, {Natoli}, {Padin}, {Plagge}, {Pryke}, {Ruhl}, {Schaffer},
  {Shirokoff}, {Spieler}, {Staniszewski}, {Stark}, {Story}, {van Engelen},
  {Vanderlinde}, {Vieira}, \& {Williamson}}]{zahn12}
{Zahn}, O., {Reichardt}, C.~L., {Shaw}, L., {et~al.}, {Cosmic Microwave
  Background Constraints on the Duration and Timing of Reionization from the
  South Pole Telescope}. 2012, \apj, 756, 65, \eprint{1111.6386}

\bibitem[{{Zarka} {et~al.}(2012){Zarka}, {Girard}, {Tagger}, \&
  {Denis}}]{NenuFAR}
{Zarka}, P., {Girard}, J.~N., {Tagger}, M., \& {Denis}, L. 2012, in SF2A-2012:
  Proceedings of the Annual meeting of the French Society of Astronomy and
  Astrophysics, ed. S.~{Boissier}, P.~{de Laverny}, N.~{Nardetto}, R.~{Samadi},
  D.~{Valls-Gabaud}, \& H.~{Wozniak}, 687--694

\bibitem[{{Zel'dovich} {et~al.}(1969){Zel'dovich}, {Kurt}, \&
  {Syunyaev}}]{ZKS1969}
{Zel'dovich}, Y.~B., {Kurt}, V.~G., \& {Syunyaev}, R.~A., {Recombination of
  Hydrogen in the Hot Model of the Universe}. 1969, Soviet Journal of
  Experimental and Theoretical Physics, 28, 146

\end{thebibliography}

\appendix

\section{the Lollipop likelihood}
\label{sec:lollipop} 

\lollipop, the LOw-$\ell$ LIkelihood on POlarized Power-spectra, is a
likelihood function based on cross-power spectra for the low multipoles. 
The idea behind this approach is that the noise can be considered as
uncorrelated between maps and that systematics will be considerably reduced in
cross-correlation compared to auto-correlation. 

At low multipoles and for incomplete sky coverage, the $C_\ell$s are not
Gaussian distributed and are correlated between multipoles. \lollipop\ uses
the approximation presented in \citet{hamimeche08}, modified as described in
\citet{mangilli15} to apply to cross-power spectra. The idea is to apply a
change of variable $C_\ell \rightarrow X_\ell$ so that the new variable
$X_\ell$ is Gaussian.  Similarly to \citet{hamimeche08}, we define
\begin{equation}
  X_\ell = \sqrt{ C_\ell^{\rm f} + O_\ell} \,\,
    g{\left(\frac{\widetilde{C}_\ell + O_\ell}{C_\ell + O_\ell}\right)}
    \,\, \sqrt{ C_\ell^{\rm f} + O_\ell} ,
\end{equation}
where $g(x)=\sqrt{2(x-\ln(x)-1)}$, $\widetilde{C}_\ell$ are the measured
cross-power spectra, $C_\ell$ are the power-spectra of the model to evaluate,
$C_\ell^{\rm f}$ is a fiducial model, and $O_\ell$ are the offsets needed in
the case of cross-spectra. 
For multi-dimensional CMB modes (i.e., $T$, $E$, and $B$), $C_\ell$ is a
$3\times3$ matrix of power-spectra:
\begin{equation}
	C_\ell = 
	\left(
	\begin{array}{ccc}
		C^{TT} & C^{TE} & C^{TB} \\
		C^{ET} & C^{EE} & C^{EB} \\
		C^{BT} & C^{BE} & C^{BB}
	\end{array}
	\right)_\ell \,,
\end{equation}
and the $g$ function is applied to the eigenvalues of $C^{-1/2}_\ell
 \widetilde{C}_\ell C^{-1/2}_\ell$.

In the case of auto-spectra, the offsets are replaced by the noise bias
effectively present in the measured power-spectra. For cross-power spectra, the
noise bias is null and here we use the effective offsets defined from the
$C_\ell$ noise variance:
\begin{equation}
	\Delta C_\ell \equiv \sqrt{ \frac{2}{2\ell+1}} O_\ell .
\end{equation}

The distribution of the new variable $X$ can be approximated as Gaussian, with
a covariance given by the covariance of the $C_\ell$s.
The likelihood function of the $C_\ell$ given the data $\widetilde{C}_\ell$ is
then
\begin{equation}
  -2\ln P(C_\ell|\widetilde{C}_\ell)=\sum_{\ell \ell'} X^{\sf T}_\ell
    M^{-1}_{\ell \ell'} X_{\ell'},
\end{equation}
where the $C_\ell$ covariance matrix $M_{\ell\ell'}$ is estimated via Monte
Carlo simulations.

\begin{figure}[htbp!]
\center
\includegraphics[width=\columnwidth]{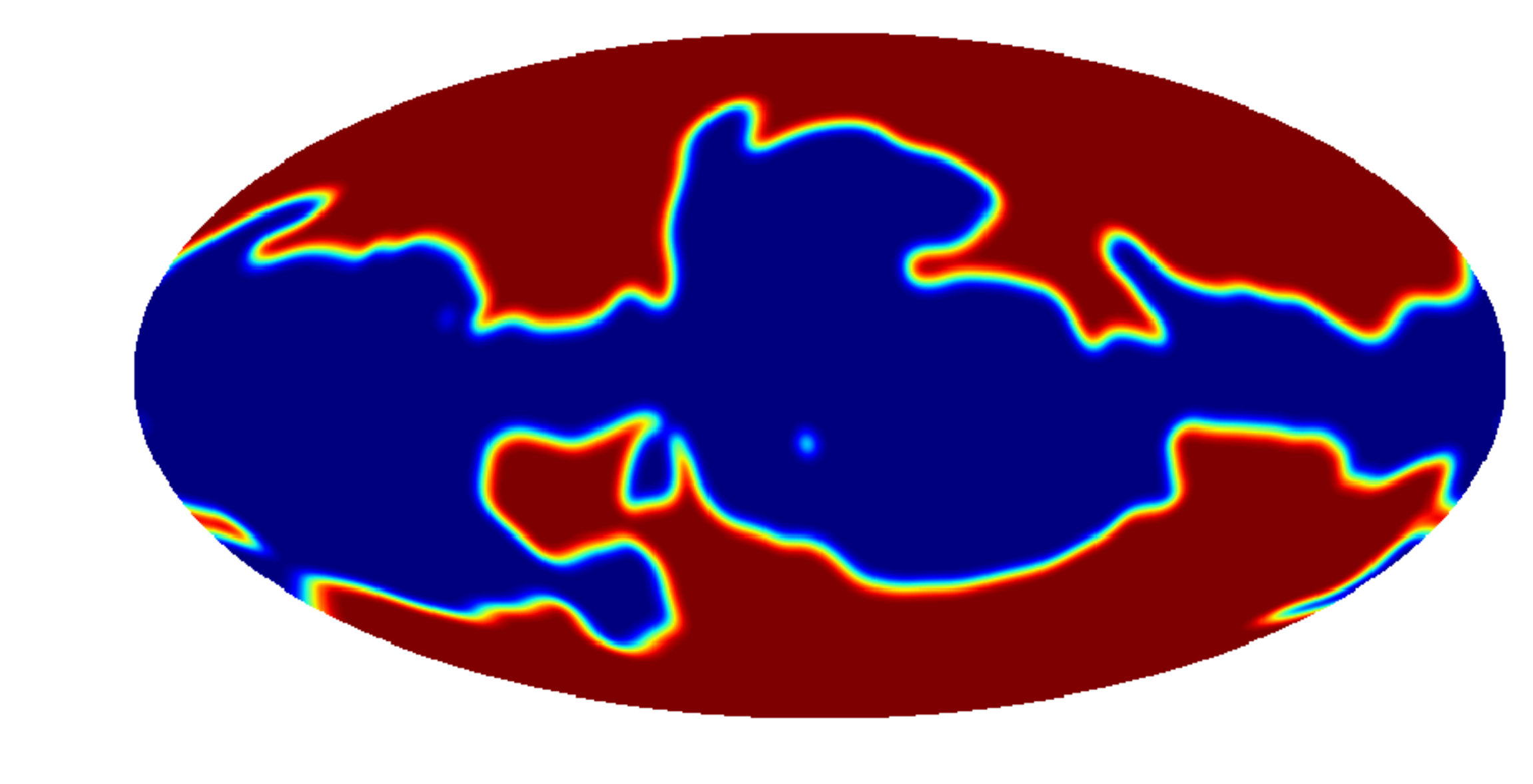}
\caption{Galactic mask used for the \lollipop\ likelihood, covering 50\,\% of
the sky.}
	\label{fig:mask_lollipop}
\end{figure}

In this paper, we restrict ourselves to the one-field approximation in order
to derive a likelihood function based only on the $EE$ power spectrum at very
low multipoles.  We use a conservative sky fraction including 50\,\% of the
sky, with a Galactic mask based on a threshold on the polarisation amplitude
measured in the 353\GHz\ \Planck\ channel, further apodized using a 4\deg\
Gaussian taper (see Fig.~\ref{fig:mask_lollipop}). 
We use {\tt Xpol} (a pseudo-$C_\ell$ estimator described in
\citealt{tristram2005} extended to polarisation) to derive cross-power spectra
between the 100 and 143\GHz\ channel maps from \Planck.
We also reject the first two multipoles ($\ell=2$ and 3), since they are more
subject to contamination by residual instrumental effects
\citep[see ][]{planck2014-a10}.

\begin{figure}[htbp!]
\center
\includegraphics[width=\columnwidth]{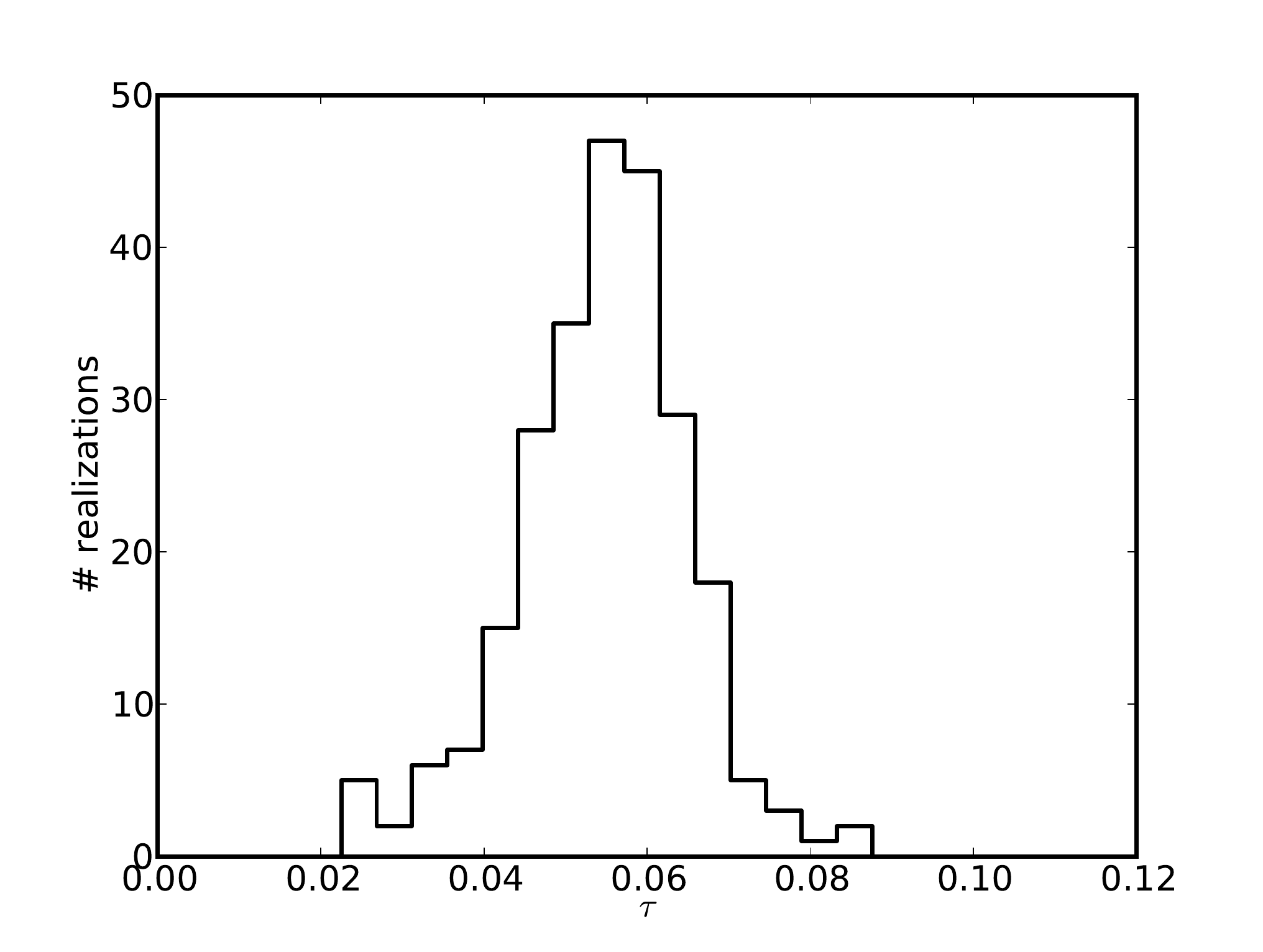}
\caption{Distribution of the peak value of the posterior distribution for
optical depth from end-to-end simulations including noise, systematic effects,
Galactic dust signal, and CMB with the fiducial value of $\tau = 0.06$.}
\label{fig:E2E_ohl}
\end{figure}

This likelihood has been tested on Monte Carlo simulations including signal
(CMB and foregrounds), realistic noise, and systematic effects. The simulated
maps are then foreground-subtracted, using the same procedure as for the data.
We constructed the $C_\ell$ covariance matrix $M_{\ell\ell'}$ using those
simulations.  Figure~\ref{fig:E2E_ohl} shows the distribution of the recovered
$\tau$ values for an input model with $\tau=0.06$, fixing all other
cosmological parameters to the \Planck\ 2015 best-fit values (including
$A_{\rm s} e^{-2\tau}$).

\begin{figure}[htbp!]
\center
\includegraphics[width=\columnwidth]{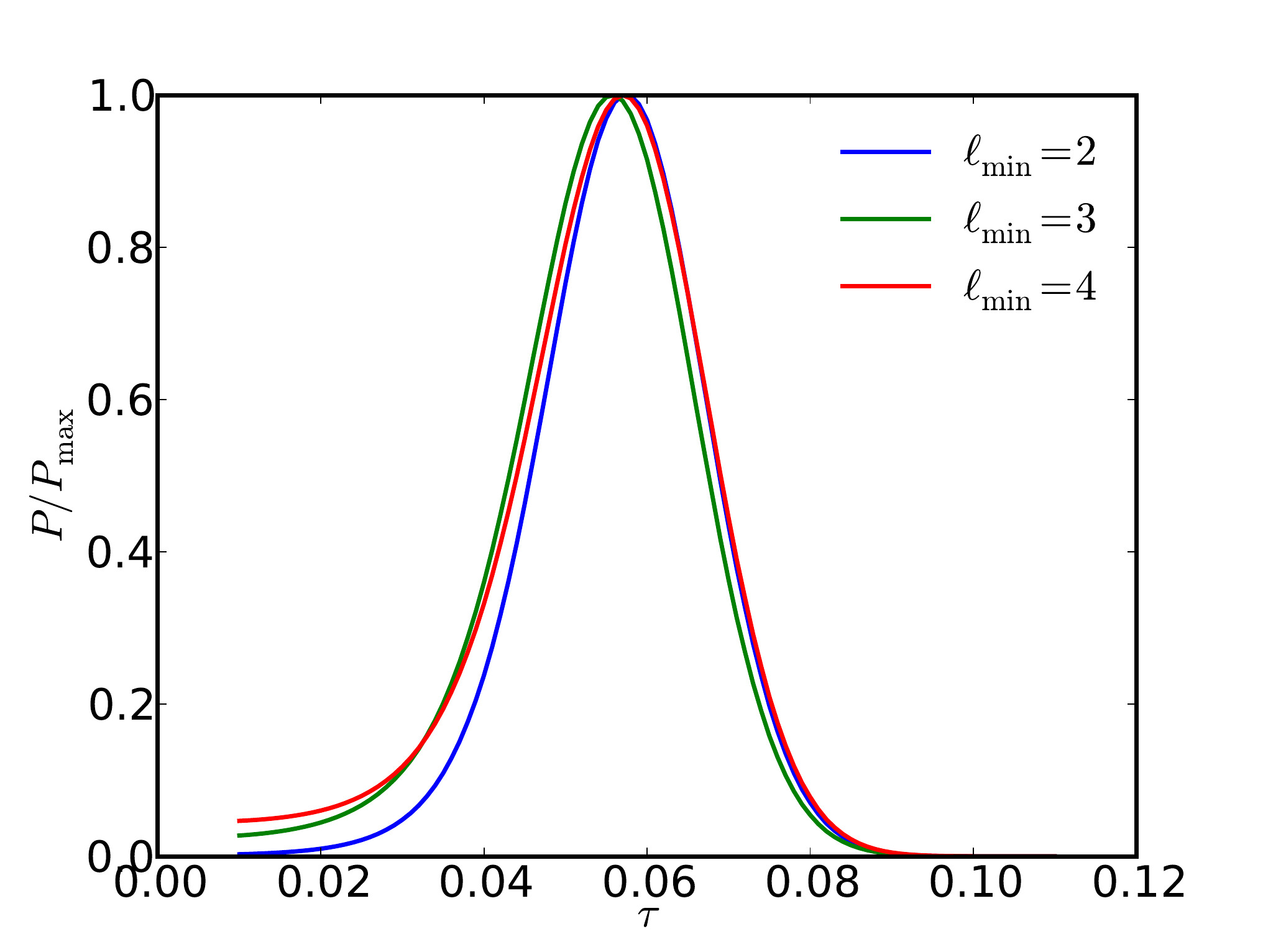}
\includegraphics[width=\columnwidth]{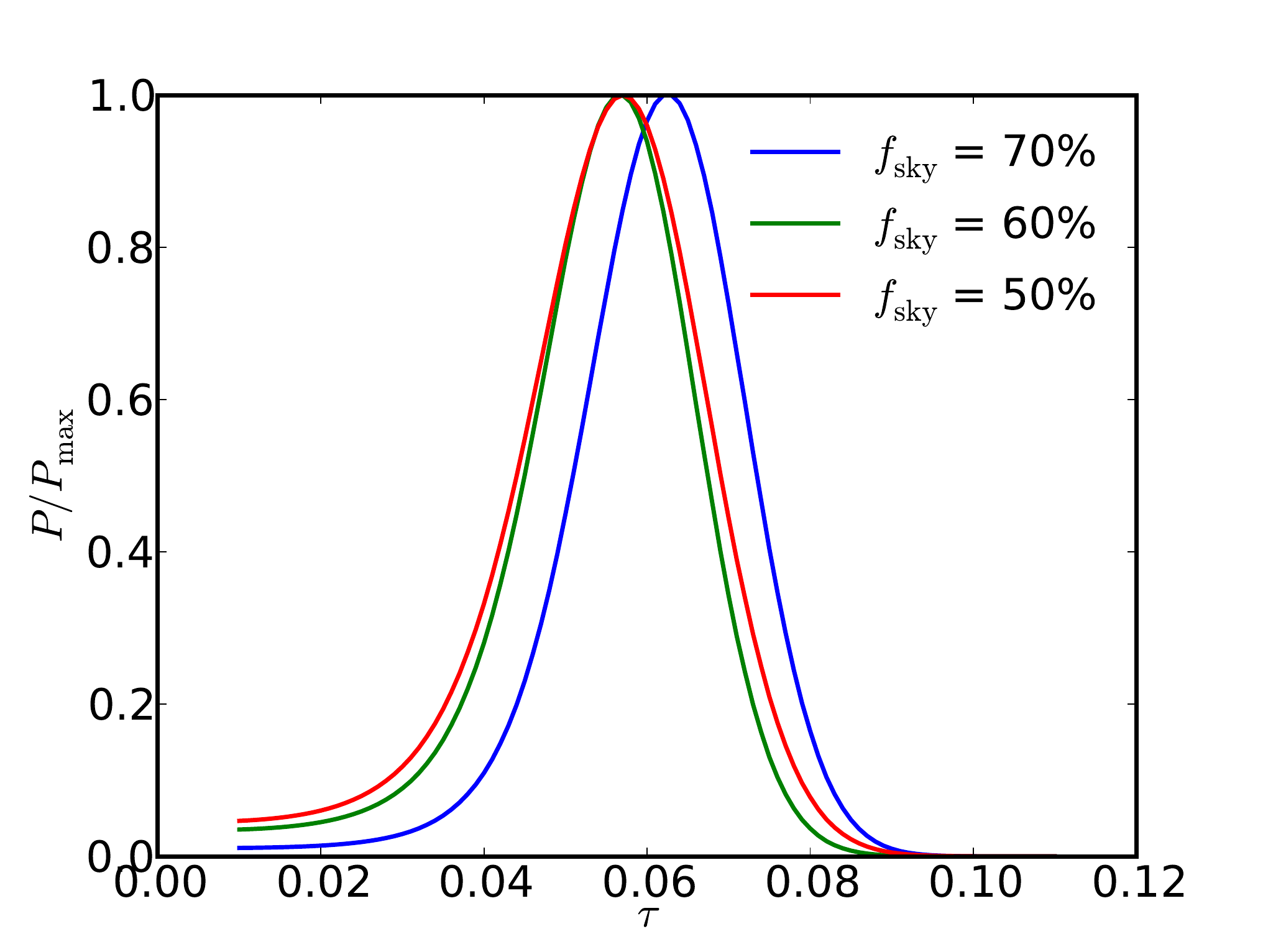}
\caption{Posterior distributions for optical depth, showing the effect of
  changing two of the choices made in our analysis.  {\it Top}: Different
  choices of minimum multipole.  {\it Bottom}: Different choices of sky
  fraction used.}
\label{fig:lollipop_consistency}
\end{figure}

To validate the choice of multipole and the stability of the result on $\tau$,
we performed several consistency checks on the \Planck\ data. Among them, we
varied the minimum multipole used (from $\ell=2$ to $\ell=4$) and allowed for
larger sky coverage (increasing to 60\,\% of the sky). The results are
summarized in Fig.~\ref{fig:lollipop_consistency}.

\section{Impact on \lcdm\ parameters}
\label{app:lcdm_impact}

\begin{figure*}[htbp!]
  \center
  \includegraphics[width=2\columnwidth]{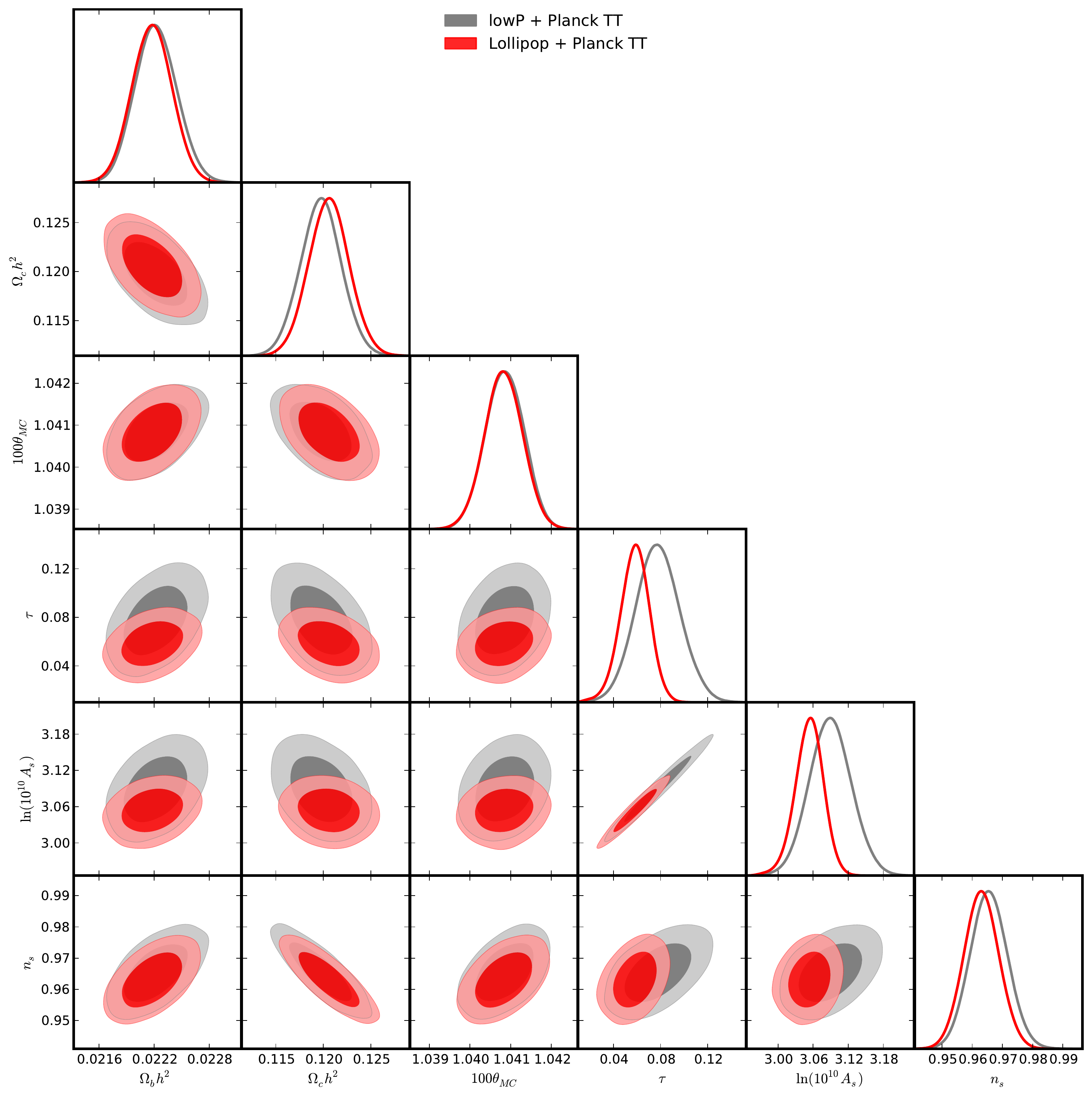}
  \caption{\lcdm\ parameters for \planckTT\ combined with the
    low-$\ell$ polarization likelihood from the \Planck\ 2015 release
    (lowP, in grey) and from this work (\lollipop, in red).}
  \label{fig:LCDM_triangle}
\end{figure*}
In addition to the restricted parameter set shown in
Fig.~\ref{fig:As_tau}, we describe here the impact of the \lollipop\
likelihood on \lcdm\ parameters in general.
Figure~\ref{fig:LCDM_triangle} compares results from
\lollipop+\planckTT\ with the lowP+\planckTT\ 2015.  The new
low-$\ell$ polarization results are sufficiently powerful that they
break the degeneracy between $n_{\rm s}$ and $\tau$.  The contours for
$\tau$ and $A_{\rm s}$, where the \lollipop\ likelihood dominates the
constraint, are significantly reduced.  The impact on other \lcdm\
parameters are small, typically below $0.3\,\sigma$.

\end{document}